\pdfoutput=1
\documentclass[12pt,a4paper]{article}    
\usepackage{jheppub}
\makeatletter
\renewcommand\@fpheader{\color{white}{`}}
\usepackage{amssymb,amsfonts}
\usepackage{fontenc}
\usepackage[utf8]{inputenc}
\usepackage{float}
\usepackage{units}
\usepackage{amsmath}
\usepackage{amssymb}
\usepackage{graphicx}
\usepackage{color}
\usepackage{esint}
\usepackage{youngtab}
\newcommand{\pd}[2]{\frac{\partial #1}{\partial #2}}

\DeclareRobustCommand{\DIEP}{\ensuremath{%
    \mathchoice{\includegraphics[height=2ex]{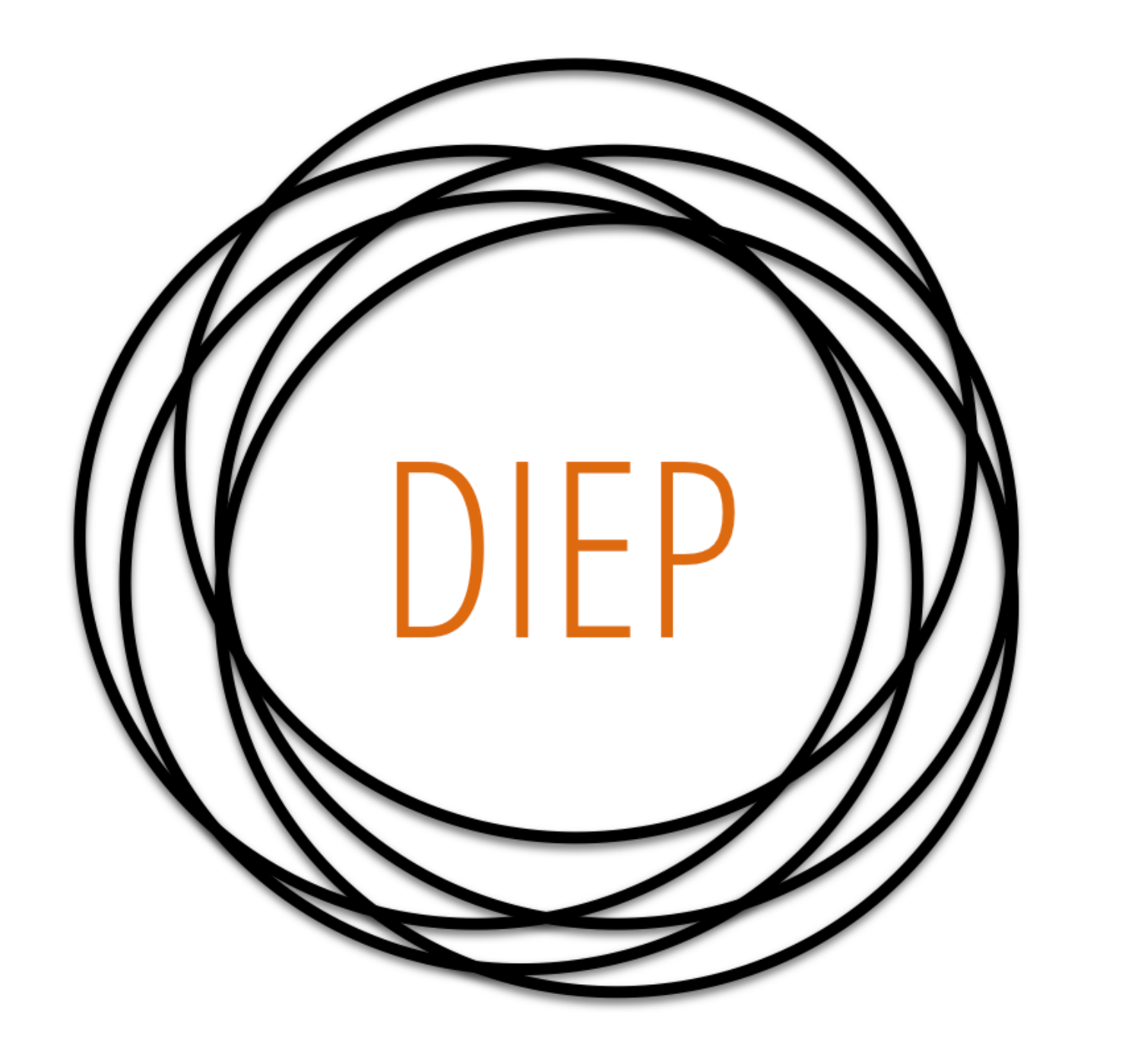}}
    {\includegraphics[height=2ex]{DIEPs.pdf}}
    {\includegraphics[height=1.5ex]{DIEPs.pdf}}
    {\includegraphics[height=1ex]{DIEPs.pdf}}
  }}

\usepackage{wrapfig}
\usepackage{slashed}
\usepackage{epsfig}
\usepackage{tensor}
\usepackage{subcaption}

\def\be{\begin{eqnarray}}
\def\ee{\end{eqnarray}}
\newcommand{\nn}{\nonumber}

\def\Dslash{\,\,{\raise.15ex\hbox{/}\mkern-12mu D}}
\def\Dbarslash{\,\,{\raise.15ex\hbox{/}\mkern-12mu {\bar D}}}
\def\delslash{\,\,{\raise.15ex\hbox{/}\mkern-9mu \partial}}
\def\delbarslash{\,\,{\raise.15ex\hbox{/}\mkern-9mu {\bar\partial}}}
\def\pslash{\,\,{\raise.15ex\hbox{/}\mkern-9mu p}}
\def\calDslash{\,\,{\raise.15ex\hbox{/}\mkern-12mu {\cal D}}}

\newcommand{\D}{{\partial}}
\newcommand{\dx}{{\text{d}x}}

\def\lae{\mathrel{\mathop{\smash{\lower .5 ex \hbox{$\stackrel<\sim$}}}}}
\def\lae{\mathrel{\mathop{\smash{\lower .5 ex \hbox{$\stackrel>\sim$}}}}}

\usepackage[useregional]{datetime2}

\title{\bf Newton--Cartan Submanifolds and Fluid Membranes}

\author[1,\DIEP\!]{Jay Armas,}
\author[2]{Jelle Hartong,}
\author[2]{Emil Have,}
\author[3]{Bjarke F. Nielsen,}
\author[4,3]{Niels A. Obers}

\affiliation[1]{Institute for Theoretical Physics, University of Amsterdam, 1090 GL Amsterdam, The Netherlands}
\affiliation[\DIEP]{Dutch Institute for Emergent Phenomena, 1090 GL Amsterdam, The Netherlands}
\affiliation[2]{School of Mathematics and Maxwell Institute for Mathematical Sciences,\\
 University of Edinburgh, Peter Guthrie Tait Road, Edinburgh EH9 3FD, UK}
\affiliation[3]{The Niels Bohr Institute, University of Copenhagen\\
Blegdamsvej 17, DK-2100 Copenhagen Ø, Denmark}
\affiliation[4]{Nordita, KTH Royal Institute of Technology and Stockholm University,\\
Roslagstullsbacken 23, SE-106 91 Stockholm, Sweden}

\emailAdd{j.armas@uva.nl}
\emailAdd{j.hartong@ed.ac.uk}
\emailAdd{emil.have@ed.ac.uk}
\emailAdd{bjarkenielsen@nbi.ku.dk}
\emailAdd{obers@nbi.ku.dk}

\abstract{We develop the geometric description of submanifolds in Newton--Cartan spacetime. This provides the necessary starting point for a covariant spacetime formulation of Galilean-invariant hydrodynamics on curved surfaces. We argue that this is the natural geometrical framework to study fluid membranes in thermal equilibrium and their dynamics out of equilibrium. A  simple model of fluid membranes that only depends on the surface tension is presented and, extracting the resulting stresses, we show that perturbations away from equilibrium yield the standard result for the dispersion of elastic waves. We also find a generalisation of the Canham--Helfrich bending energy for lipid vesicles that takes into account the requirements of thermal equilibrium.}

\usepackage[textsize=scriptsize]{todonotes}
\usepackage{mathrsfs}
\usepackage{amscd}
\newcommand{\diff}{\text{d}}

\usepackage{amsthm}

\theoremstyle{remark}

\usepackage{tcolorbox}
\usepackage{cancel}
\usepackage{xcolor}

\usepackage{tikz}

\DeclareFontFamily{U}{skulls}{}
\DeclareFontShape{U}{skulls}{m}{n}{ <-> skull }{}

\usepackage{pifont}
\begin{document}
\pagestyle{plain} \setcounter{page}{1}
\newcounter{bean}
\baselineskip16pt \setcounter{section}{0}
\notoc
\maketitle
\flushbottom

\section{Introduction}
The dynamics of surfaces and interfaces plays a prominent role in various instances of physical phenomena, ranging from fluid membranes in biological systems \cite{Canham1970,Helfrich:1973}, the interplay between liquid crystal geometry and hydrodynamics \cite{Keber1135} to surface/edge physics in condensed matter systems 
\cite{PhysRevLett.95.146802}. 
Fluid membranes comprised of lipid bilayers are essential in the physics of biological systems, and the characterisation of their geometric properties has been an active field of research for decades, as well as being key in understanding experimental outcomes (see e.g. 
\cite{doi:10.1080/00018739700101488, Tu:2014, Guckenberger_2017, Steigmann2018, Guven2018} for reviews). Hydrodynamics on curved surfaces has also recently received considerable attention, not only due to its relevance in embryonic processes \cite{streichan2017quantification} or cell migration \cite{Gut} where activity also plays a role, but also due to its relevance in understanding topological properties of wave dynamics such as Kelvin-Yanai waves on the Earth's equator \cite{Delplace1075}, flocking on a sphere \cite{PhysRevX.7.031039} or turbulence in active nematics \cite{PhysRevLett.122.168002, Henkes2018,Alaimo2017}. 

While the geometry and dynamics of surfaces in (pseudo)-Riemannian geometry has been deeply studied
in both physics and mathematics, a systematic treatment using covariant and geometrical structures 
has so far not been developed for Galilean-invariant systems.  In view of the relevance of such systems
in many branches of physics, and immediate applications in biophysical systems detailed below, the main
goal of this paper is to develop the theory of submanifolds in Newton-Cartan spacetime. This can be considered as the Galilean analogue of the (pseudo)-Riemannian case for which the geometry and its embeddings have local Euclidean (Poincar\'e) symmetry as opposed to Galilean symmetries. The formalism we develop allows for a covariant spacetime formulation of Galilean-invariant hydrodynamics on curved surfaces. 

As such it is thus the natural framework  to study fluid membranes in thermal equilibrium along with their dynamics
away from equilibrium. This includes in particular biophysical membranes such as lipid bilayers, which are membranes composed of lipid molecules that enclose the cytoplasm. The lipid molecules move as a fluid along the membrane surface, which itself behaves elastically when bent. It is well known that at mesoscopic scales, lipid bilayers can be approximated by thin surfaces whose equilibrium configurations are accurately described by geometrical degrees of freedom and a small set of material coefficients that encode the more microscopic biochemical details (see e.g. \cite{Guven2018}). The shapes of lipid bilayers, such as discoids characterising the morphology of red blood cells, are found by extremising the Canham-Helfrich (CH) free energy \cite{doi:10.1080/00018739700101488, Tu:2014}, which only depends on geometric properties. The stresses associated to such bilayers have received considerable attention \cite{Capovilla_2002, Guven2018} as well as deformations of the CH free energy away from equilibrium in order to identify stable deformations \cite{Capovilla_2004}. 

However, despite the CH free energy being taken to represent a system in thermodynamic equilibrium \cite{Terzi2018} (as well as its analogue in nematic liquid crystals - the Frank energy \cite{DF9582500019}), it disregards the basic lesson of equilibrium thermal field theory: that temperature and mass chemical potential (conjugate to particle number) also have a geometric interpretation. This results in the CH free energy giving rise to inaccurate stresses characterising the membrane, explicit by the fact that they do not describe the stresses intrinsic to a fluid, and neither do they yield elastic wave dispersion relations when deforming away from equilibrium. In this paper, we argue that the development of a spacetime covariant formulation of Galilean-invariant hydrodynamics using Newton--Cartan geometry is a more useful approach to understanding fluid dynamics on curved surfaces and the physics of equilibrium fluid membranes.

Newton--Cartan (NC) geometry was pioneered by Cartan in order to geometrise Newton's theory of gravity \cite{Cartan1,Cartan2}%
\footnote{See also \cite{Andringa:2010it} for a modern perspective and earlier references, and
the recent work \cite{Hansen:2018ofj} for an action principle for Newtonian gravity.}. As a non-dynamical
geometry its importance stems from the fact that it is the natural background geometry that non-relativistic field theories couple to  \cite{Jensen:2014aia,Hartong:2014pma}%
\footnote{In particular, the most general coupling requires a torsionful generalisation of NC geometry, called torsional
Newton--Cartan (TNC) geometry which was first observed as the boundary geometry in the context of Lifshitz holography \cite{Christensen:2013lma,Christensen:2013rfa,Hartong:2014oma}. TNC geometry also appears as the ambient space-time
for non-relativistic strings, see e.g. \cite{Harmark:2017rpg,Harmark:2018cdl,Harmark:2019upf}.}
and  thus provides a geometric and covariant formulation of many aspects of non-relativistic physics including broad classes of long-wavelength effective theories such as hydrodynamics. In particular, in the past few years NC geometry and variants have been applied to the formulation of Galilean-invariant fluid dynamics \cite{Jensen:2014ama, Banerjee:2015hra}, Lifshitz fluid dynamics \cite{Kiritsis:2015doa,Hartong:2016nyx} as well as hydrodynamics without boost symmetry 
\cite{deBoer:2017ing,deBoer:2017abi,Novak:2019wqg,deBoer:2019}%
\footnote{The boost non-invariant hydrodynamics of these papers is formulated in the regime where momentum is conserved, but may be generalised to include further breaking of translation symmetry, in which case it applies to 
flocking and active matter.}%
, which encapsulate the former as cases with extra symmetries. Furthermore, in the context of condensed matter systems, it was realised that NC geometry is the natural setting for developing an effective theory of the fractional quantum Hall effect \cite{Son:2013rqa,Geracie:2014nka,Gromov:2014vla,Geracie:2015dea}.  This body of work, together with previous work on Galilean superfluid droplets \cite{Armas:2016xxg} and connections between black holes and CH functionals \cite{Armas:2013hsa, Armas:2014bia}, suggests that NC geometry can also be useful in describing hydrodynamics on curved surfaces.

The development of submanifold calculus in (pseudo-)Riemannian/Euclidean geometry, written in multiple volumes (e.g \cite{aminov2014geometry}) and furthered in different contexts \cite{Carter:1993wy,Carter:1997pb,Carter:2000wv, Capovilla:1994bs, Armas:2017pvj}, is an essential pre-requisite for describing surfaces and hence for formulating and extremising the CH free energy. Therefore, the majority of the work presented in this paper, in particular sections \ref{sec:NCGeom} and \ref{sec:var} and appendix \ref{app:reduction}, consists of the novel development of submanifold calculus in Newton--Cartan geometry, the identification of geometrical properties describing surfaces, and the formulation of appropriate geometric functionals whose extrema are NC surfaces. Thus, the main part of the work presented here is foundational. However, in section \ref{sec:applications} we apply this machinery to different fluid membrane systems in order to show its usefulness and provide a generalised CH model that takes into account the requirements of thermodynamic equilibrium.  The work developed here will be the basis for a more detailed study of effective theories of fluid membranes, which takes into account a larger set of responses including viscosity, providing a more solid foundation for the physics of fluid membranes \cite{upcoming}.

\subsubsection*{Organisation of the paper}
A more detailed outline of the paper, including a brief summary of the main results is as follows. 

In Section \ref{sec:NCGeom}, after reviewing the geometric structure of a Newton-Cartan spacetime, we first define
what a submanifold structure is in such spacetimes. In particular, we develop the necessary geometric tools
to define an induced NC structure on the submanifold. We highlight in particular how the objects
transform under  local Galilean boosts, which is a key property for non-relativistic geometries.  
We then show, using the affine connection that is known for NC structures, how to construct a covariant
derivative along the surface directions, and give an expression for the corresponding surface torsion tensor. 
With this in hand, we discuss the exterior curvature and show how the (Riemannian) Weingarten identity
gets modified in this case. 

Section \ref{sec:var} develops the variational calculus for NC submanifolds, which is essential technology
in order to find equations of motion from effective actions. We consider first general variations of
the relevant quantities describing the embedding. Subsequently we obtain expressions for embedding map variations
as well as Lagrangian variations, which are diffeomorphisms in the ambient NC spactime that keep the embedding maps fixed. From the corresponding variations of the induced NC structures and the normal vectors we find in particular how the extrinsic curvature transforms under such variations. We subsequently use this technology to consider the dynamics
of submanifolds that arises from extremisation of an action. The resulting equations of motions for NC submanifolds
are thus obtained from the general response to varying the induced NC metric structure on the manifold
and the extrinsic curvature. These split up in 
a set of intrinsic equations, which are conservation equations of the worldvolume stress tensor and mass current 
accompanied by a set of extrinsic equations. We also analyse the boundary terms
that appear as a result of varying the general action functional and obtain the resulting boundary conditions. 

Then in section  \ref{sec:applications} we apply the action formalism presented in the previous section to describe equilibrium fluid membranes and lipid vesicles as well as their fluctuations.  We will show that employing 
NC geometry for such surfaces is not only natural but also provides a more complete description.  
First of all, it introduces (absolute) time and therefore fluctuations of the system can include temporal dynamics in a covariant form. Moreover, the symmetries of the problem are made manifest via the geometry of the 
submanifold and ambient spacetime. Even more important is the aspect that NC geometry allows to properly introduce thermal field theory of equilibrium fluid membranes. To illustrate all this we first
consider equilibrium fluid branes, i.e stationary fluid configurations on an arbitrary surface and the simplest example
with a free energy depending on surface tension only, for which we compute the resulting stresses. 
We then show that perturbations away from  equilibrium yield the standard result for the dispersion of elastic waves. 
We also briefly consider the case of a droplet, by adding  internal/external pressure to the previous case. 
Then we revisit the celebrated Canham-Helfrich model which describes equilbrium configurations of
biophysical membranes. We show how this model can be described using Newton-Cartan geometry and generalize
it by allowing its (material) parameters to depend on temperature and chemical potential. Finally, we review
the classic lipid vesicles using this framework. 

We end in section \ref{sec:Outlook} with a brief discussion and  description of further avenues of investigation.

A number of appendices are included containing further details.
Since it is known that torsional NC spacetimes can be obtained from Lorentzian spacetime using null reduction,
we show in appendix \ref{app:reduction} a complimentary perspective on NC submanifolds, by null reducing submanifolds of Lorentzian spacetimes. Appendix \ref{app:NCtypes} describes diffferent classes of NC spacetimes, depending on
properties of the torsion. In appendix \ref{App:ConnectionsAreImportant} we 
 find the relation between the NC connections of the ambient spacetime and the submanifold (described in section \ref{sec:CovDandExtK}). Finally, in appendix \ref{sec:GaussBonnet} we show how the Gauss--Bonnet theorem reduces the number of independent terms in an effective action for $(2+1)$-dimensional membranes that appear as closed co-dimension one surfaces embedded in flat $(3+1)$-dimensional Newton--Cartan geometry.

\section{The geometry of Newton--Cartan submanifolds}\label{sec:NCGeom}
This section is devoted to a proper geometrical treatment of surfaces (or embedded submanifolds) in NC geometry with the goal of subsequently applying it to the description of membrane elasticity and fluidity in later sections. To that aim, we begin by introducing the reader to the essential details of NC geometry. The basic structures that define a given NC geometry are then understood as background fields for the dynamical surfaces/objects, in direct analogy with embedding of surfaces in a (pseudo-)Riemannian geometry with background metric $g_{\mu\nu}$. This paves the way for defining the geometric structures that characterise non-relativistic surfaces.\footnote{Intuition originating from the description of surfaces in (pseudo-)Riemannian geometry suggests that geometric structures characterising surfaces in NC geometry would naively be constructed from pullbacks of NC ambient spacetime fields. It will turn out that this is only true for submanifolds of NC geometry provided we take the pullbacks of quantities that are invariant under the local Galilean boost transformations of the ambient NC geometry.} In appendix \ref{app:reduction}, we provide an alternative method for obtaining the theory of NC surfaces directly from the theory of surfaces in Lorentzian geometry. 

\subsection{Newton--Cartan geometry}\label{sec:NCreview}

Let $\mathcal{M}_{d+1}$ be a $(d+1)$-dimensional manifold endowed with a Newton--Cartan structure, which consists of the fields $(\tau_\mu,h_{\mu\nu},m_\mu)$. Here, the Greek indices denote spacetime indices such that $\mu,\nu,\dots = 0,\dots,d$. The tensor $h_{\mu\nu}$ is symmetric with rank $d$ and has signature $(0,1,1,...)$, while the nowhere vanishing 1-form $\tau_\mu$ is such that $-\tau_\mu\tau_\nu + h_{\mu\nu}$ has full rank. The field $m_\mu$ is the connection of an Abelian gauge symmetry that from the point of view of a Galilean field theory on a NC spacetime can be thought of as the symmetry underlying particle number conservation. Since the latter is a compact Abelian symmetry we refer to $m_\mu$ as the $U(1)$ gauge connection. It is useful to define an inverse NC structure $(v^\mu,h^{\mu\nu})$, where $v^\mu$ spans the kernel of $h_{\mu\nu}$ and $\tau_\mu$ spans the kernel of $h^{\mu\nu}$. The 1-form $\tau_\mu$ is sometimes called the \emph{clock 1-form}, while the vector $v^\mu$ is known as the \emph{Newton--Cartan velocity}. These structures satisfy the completeness relation and normalisation condition:
\begin{equation}
\delta^{\mu}_\nu = -v^\mu\tau_\nu + h^{\mu\rho}h_{\rho\nu}~~,\quad\text{so that}\quad~~v^\mu\tau_\mu=-1~~.
\end{equation}
It is occasionally useful to introduce vielbeins $e^{\underline a}_\mu$, $e^\mu_{\underline a}$ with $\underline{a},\underline{b},\dots = 1,\dots,d$ (that is, spatial tangent space indices are underlined lowercase Latin letters) such that
\begin{equation}
h_{\mu\nu} = \delta_{\underline{a}\hspace{0.5pt}\underline{b}}e^{\underline{a}}_\mu e^{\underline{b}}_\nu\,,\qquad h^{\mu\nu} = \delta^{\underline{a}\hspace{0.5pt}\underline{b}} e^\mu_{\underline{a}} e^\nu_{\underline{b}}\,,\label{eq:Metricsfromvielbein}
\end{equation} 
which furthermore satisfy the orthogonality relations
\begin{equation}
v^\mu e_\mu^{\underline{a}} = 0\,,\qquad \tau_\mu e^\mu_{\underline{a}}=0\,,\qquad e^\mu_{\underline{a}} e^{\underline{b}}_\mu = \delta^{\underline{b}}_{\underline{a}}~~.
\end{equation}

The Newton--Cartan structure on $\mathcal{M}_{d+1}$ in terms of the fields $(\tau_\mu,h_{\mu\nu},m_\mu)$ transforms under diffeomorphisms (coordinate transformations), $U(1)$ (mass) gauge transformations (akin to gauge transformations in Maxwell theory), local rotations and local Galilean boosts (also known as Milne boosts) in the following way:
\begin{equation}
\begin{split}
&\delta \tau_\mu=\pounds_\xi \tau_\mu\,,\qquad \delta e^{\underline{a}}_\mu =\pounds_\xi e^{\underline{a}}_{\mu}  + \lambda^{\underline{a}}{_{\underline{b}}}e_\mu^{\underline{b}}+\lambda^{\underline{a}}\tau_\mu\,,\qquad \delta m_\mu = \pounds_\xi m_\mu +\lambda_{\underline{a}} e^{\underline{a}}_\mu + \D_\mu\sigma\,,\\
&\delta v^\mu = \pounds_\xi v^\mu + \lambda^{\underline{a}} e^\mu_{\underline{a}}\,,\qquad \delta e^\mu_{\underline{a}} = \pounds_\xi e^\mu_{\underline{a}} + \lambda_{\underline{a}}{^{\underline{b}}} e^\mu_{\underline{b}}~~.
\label{eq:TNCgaugetrafos}
\end{split}
\end{equation}
Here $\xi^\mu$ is the generator of diffeomorphisms, $\sigma$ is the parameter of mass gauge transformations and $\lambda^{\underline{a}}$ is the parameter of local Galilean boosts. Finally, $\lambda_{\underline{a}}{^{\underline{b}}}=-\lambda_{\underline{b}}{^{\underline{a}}}$ corresponds to local $\mathfrak{so}(d)$ transformations.
When describing physical systems in NC geometry by means of a Lagrangian or action functional, one requires invariance under the gauge transformations \eqref{eq:TNCgaugetrafos}. In the restricted setting of a flat NC background (i.e. a spacetime with absolute time whose constant time slices are described by Euclidean geometry), which is the most relevant case in the context of biophysical membranes, invariance under \eqref{eq:TNCgaugetrafos} implies invariance under global Galilean symmetries centrally extended to include mass conservation. The centrally extended Galilei group is known as the Bargmann group. This implies that the geometry can be viewed as originating from `gauging' the Bargmann algebra as detailed in \cite{Andringa:2010it}.

\subsubsection{Galilean boost-invariant structures}
One may readily check that given \eqref{eq:TNCgaugetrafos}, the NC fields $h^{\mu\nu}$ and $h_{\mu\nu}$, which are constructed out of the vielbeins as in \eqref{eq:Metricsfromvielbein}, transform as
\begin{equation}
\delta h^{\mu\nu} = \pounds_\xi h^{\mu\nu}~~,~~ \delta h_{\mu\nu} = \pounds_\xi h_{\mu\nu}+ 2\lambda_{(\mu}\tau_{\nu)}~~,
\label{eq:hhtrans}
\end{equation}
where $\lambda_\mu = e^{\underline{a}}_\mu \lambda_{\underline{a}}$, immediately implying that $\lambda_\mu v^\mu = 0$. We conclude from this that $h^{\mu\nu}\partial_\mu\partial_\nu$ is an invariant of the geometry, a co-metric, while $h_{\mu\nu}dx^\mu dx^\nu$ is not an invariant because it transforms under the Galilean boosts. On the other hand $\tau_\mu dx^\mu$ is invariant. This means that NC geometry has a degenerate metric structure given by $\tau_\mu\tau_\nu$ and $h^{\mu\nu}$ and that $h_{\mu\nu}$ should not be viewed as a metric\footnote{We can fix diffeomorphisms such that $\tau_i=0$ where we split the spacetime coordinates $x^\mu=(t,x^i)$. In this restricted gauge the metric on slices of constant time $t$ is given by $h_{ij}dx^idx^j$ which is invariant under the diffeomorphisms that do not affect time. In this sense the constant time slices are described by standard Riemannian geometry. However when we include time into the formalism we have to abandon the notion of a metric and instead work with the NC triplet $(\tau_\mu,h_{\mu\nu},m_\mu)$. In this setting, in order to evaluate areas or volumes of given surfaces one can use the integration measure $e=\sqrt{-\text{det}\left(-\tau_\mu\tau_\nu+h_{\mu\nu}\right)}$, which is both Galilean boost- and $U(1)$-invariant.}.

Notice that while $h_{\mu\nu}$ transforms under Galilean boosts it does not transform under $U(1)$ gauge transformations. It is possible to define objects that have the opposite property, namely that they are Galilean boost invariant but not $U(1)$ invariant. We will often work with these fields and so we discuss their construction here. We can trade $U(1)$ gauge invariance for boost invariance by introducing the new set of fields
\begin{equation} \label{eq:barmunu}
\bar h_{\mu\nu} = h_{\mu\nu} - 2\tau_{(\mu}m_{\nu)}\,,\qquad \hat v^\mu = v^\mu - h^{\mu\nu}m_\nu~~,
\end{equation}
which transform as\footnote{Note that this is possible because the $U(1)$ connection $m_\mu$ also transforms under Galilean boosts. In this sense it is different from the Maxwell potential. The difference comes from the fact that the mass generator forms a central extension of the Galilei algebra whereas the charge $U(1)$ generator of Maxwell's theory forms a direct sum with in that case the Poincar\'e algebra. See \cite{Andringa:2010it,Festuccia:2016awg}) for more details.}
\begin{equation}
\delta \bar h_{\mu\nu} = \pounds_\xi \bar h_{\mu\nu}-2\tau_{(\mu}\D_{\nu)}\sigma\,,\qquad \delta \hat v^\mu = \pounds_\xi\hat v^\mu - h^{\mu\nu}\D_\nu \sigma~~,
\label{eq:sigmatrafo}
\end{equation}
and hence are manifestly Galilean boost-invariant. Additionally, it is also possible to construct a boost invariant scalar, which is the boost invariant counterpart of the Newtonian potential \cite{Bergshoeff:2014uea}, namely
\begin{equation}
\tilde{\Phi}=-v^\mu m_\mu+\frac{1}{2}h^{\mu\nu}m_\mu m_\nu~~.
\label{eq:NewtonPot}
\end{equation}
The Newtonian potential itself is just the time component of $m_\mu$. These quantities will be useful when discussing effective actions for fluid membranes in later sections.

\subsubsection{Covariant differentiation and affine connection}
NC geometry provides a way of formulating non-relativistic physics in curved backgrounds/substrates which has recently become an active research direction in soft matter \cite{Delplace1075, PhysRevX.7.031039, PhysRevLett.122.168002, Henkes2018,Alaimo2017}. Additionally, even in the traditional case of lipid membranes sitting in Euclidean space, it is useful to have explicit coordinate-independence as it can simplify many problems of interest. Therefore, it is important to introduce a covariant derivative adapted to curved backgrounds. However, in contrast to (pseudo-)\\Riemannian geometry without torsion, there is no unique metric-compatible connection in Newton--Cartan geometry. Rather, the analogue of metric compatibility in NC geometry is
\begin{equation}
\nabla_\mu \tau_\nu=0\,,\qquad \nabla_\mu h^{\nu\rho}=0~~,
\label{eq:MetricComp}
\end{equation}
where $\nabla$ is the covariant derivative with respect to the affine connection $\Gamma^\rho_{\mu\nu}$. It is possible to choose the affine connection as \cite{Bekaert:2014bwa, Hartong:2015zia}\footnote{As shown in \cite{Bekaert:2014bwa, Hartong:2015zia}, the most general affine connection satisfying \eqref{eq:MetricComp} takes the form $\bar\Gamma^\rho_{\mu\nu}={\Gamma}^\rho_{\mu\nu}+W^\rho_{\mu\nu}$ where $W^\rho_{\mu\nu}$ is the pseudo-contortion tensor, obeying $\tau_\rho W^\rho_{\mu\nu}=0$ and $W^\nu_{\mu\lambda}h^{\lambda\rho}+W^\rho_{\mu\lambda}h^{\nu\lambda}=0$. The choice \eqref{eq:TNCconnection} corresponds to $W^\rho_{\mu\nu}=0$. This choice is also the natural choice from the perspective of the Noether procedure \cite{Festuccia:2016awg}.}
\begin{equation}
{{\Gamma}^\rho_{\mu\nu} = -\hat{v}^\rho \D_\mu\tau_\nu + \frac{1}{2}h^{\rho\sigma}\left(\D_\mu \bar h_{\nu\sigma} + \D_\nu\bar h_{\mu\sigma} - \D_\sigma\bar h_{\mu\nu} \right)}~~.
\label{eq:TNCconnection}
\end{equation}
Given the connection $\Gamma$, covariant differentiation acts on an arbitrary vector $X^\mu$ in a similar manner as in (pseudo)-Riemannian geometry, that is
\begin{equation}
\nabla_\mu X^\nu = \D_\mu X^\nu +\Gamma^\nu_{\mu\rho}X^\rho~~.
\end{equation}

Notably, and in contradistinction to the Levi-Civita connection of (pseudo)-Riemannian geometry, the connection ${\Gamma}^\lambda_{\mu\nu}$ is generally torsionful. This is due to the condition $\nabla_\mu \tau_\nu=0$.  In particular, the affine connection has an anti-symmetric part given by
\begin{equation}
2\Gamma^\lambda_{[\mu\nu]} = -2\hat{v}^\lambda \D_{[\mu}\tau_{\nu]} = -\hat{v}^\lambda\tau_{\mu\nu}~~,
\end{equation}
where we defined the torsion 2-form 
\be  \label{eq:torsion2}
\tau_{\mu\nu}=2\D_{[\mu}\tau_{\nu]}~~.
\ee
For all physical systems studied in this paper, the torsion vanishes. However, when performing variational calculus (of the NC fields) it is required to keep variations of $\tau_\mu$ arbitrary\footnote{The condition that $\tau_\mu$ be unconstrained is not necessary when we perform variations of embedding scalars in a fixed ambient space geometry.}.

As written in \eqref{eq:TNCconnection} in terms of boost-invariant quantities, the affine connection does not transform under Galilean boosts. However, under the $U(1)$ gauge transformations \eqref{eq:sigmatrafo}, it transforms as
\begin{equation}\label{eq:gaugatrafoGamma}
\delta_\sigma\Gamma^\rho_{\mu\nu}=\frac{1}{2}h^{\rho\lambda}\left(\tau_{\mu\nu}\partial_\lambda\sigma+\tau_{\lambda\nu}\partial_\mu\sigma+\tau_{\lambda\mu}\partial_\nu\sigma\right)~~.
\end{equation}
In the absence of torsion, $\tau_{\mu\nu}=0$, the connection is invariant under such transformations.

\subsubsection{Absolute time and flat space} \label{sec:abs}
Depending on the conditions imposed on the clock 1-form $\tau_\mu$, there are different classes of NC geometries \cite{Christensen:2013rfa,Hartong:2015zia}. We refer the curious reader to appendix \ref{app:NCtypes}, which contains a classification of the different classes NC geometries, while in this section we focus on the most relevant case for the purposes of this work. If $\tau_\mu$ is exact, that is $\tau_\mu = \partial_\mu T$ for some scalar $T$, the torsion \eqref{eq:torsion2} vanishes and we are dealing with Newtonian absolute time. This is the simplest kind of Newton--Cartan geometry and the relevant one for the applications we consider in this work, namely lipid vesicles or fluid membranes. For example, for membrane geometries, which for each instant in time are embedded in three-dimensional Euclidean space, the ambient NC spacetime in Cartesian coordinates can be parametrised as 
\begin{equation} \label{eq:flatspace}
\tau_\mu=\delta^0_\mu~~,~~h_{\mu\nu}=\delta^i_\mu\delta^i_\nu~~,~~v^\mu = - \delta^\mu_0~~,~~h^{\mu\nu}=\delta_i^\mu\delta_i^\nu~~,~~m_\mu=0~~.
\end{equation}

In the context of non-relativistic physics in spatially curved backgrounds, the clock 1-form will still have the form $\tau_\mu=\delta^0_\mu$ but the tensor $h_{\mu\nu}$ can be non-trivial in the sense that it is not gauge equivalent to flat space. Thus for all practical applications, the first term in the affine connection \eqref{eq:TNCconnection} vanishes and the connection is purely spatial. However, while for physically relevant spacetimes we will always require that $\tau_\mu$ must be of the form $\tau_\mu = \partial_\mu T$, when we are dealing with $\tau_\mu$ as a background source in some action functional for matter fields, we need to require that it is unconstrained in order to be able to vary it freely.

\subsection{Submanifolds in Newton-Cartan geometry}\label{sec:SubmanifoldStruc}
In this section we formulate the theory of non-relativistic NC timelike\footnote{The submanifolds we consider are timelike in the sense that the normal vectors are required to be spacelike (see \eqref{eq:tauIzero}). The submanifolds will inherit a NC structure of their own.} surfaces (or submanifolds) embedded in arbitrary NC geometries. Following the literature that deals with the relativistic counterpart \cite{Armas:2017pvj}, we focus on the description of a single surface placed in an ambient NC spacetime and not on a foliation of such surfaces. In practice, this means that all geometric quantities, such as tangent and normal vectors, describing the surface are only well-defined on the surface and not away from it. In this section we introduce the necessary geometrical structures for dealing with a single surface in a NC spacetime. 

\subsubsection{Embedding map, tangent and normal vectors}
A $(p+1)$-dimensional Newton--Cartan submanifold $\Sigma_{p+1}$ of a $(d+1)$-dimensional Newton--Cartan manifold ${\mathcal{M}}_{d+1}$ is specified by the embedding map
\begin{equation}
X^\mu : \Sigma \rightarrow \mathcal{M},~~\mu=0,\ldots,d\,,
\end{equation}
which maps the coordinates $\sigma^a$ on $\Sigma_{p+1}$ to $X^\mu(\sigma^a)$ on $\mathcal{M}$ (lowercase Latin letters,  $a,b,\ldots = 0,\ldots,p$, denote submanifold spacetime indices). Concretely, the embedding map specifies the location of the surface as $x^\mu=X^\mu(\sigma^a)$ where $x^\mu$ are coordinates in $\mathcal{M}$. The manifold $\mathcal{M}$ into which the embedding scalars map is usually referred to as the target spacetime. The manifold described by the spacetime coordinates $x^\mu$ is the ambient spacetime. For simplicity, we will refer to both as ambient spacetime. 

Given the embedding map, the tangent vectors to the surface are explicitly defined via $u^\mu_a=\partial_a X^\mu$. In turn, the normal 1-forms $n_\mu ^I dx^\mu$ (where $I$ runs over the $d-p$ transverse directions) are implicitly defined via the relations
\begin{equation}
n^I_\mu u^\mu_a = 0~~,~~{h^{\mu\nu} n^I_\mu n^J_\nu = \delta^{IJ },}~~I=1,\dots,d-p ~~.
\label{eq:1formnormalization}
\end{equation}
This normalisation implies that in the normal directions we can use $\delta_{IJ}$ and $\delta^{IJ}$ to raise and lower transverse indices, meaning that we can write $Y_IY^I=Y^IY^I$ for some arbitrary vector $Y^I$. However, eq.~\eqref{eq:1formnormalization} does not fix the normal 1-forms uniquely. In fact, the 1-forms $n^I_\mu$ transform under local $SO(d-p)$ rotations such that
\begin{equation} \label{eq:normrotations}
n^I_\mu \rightarrow \mathcal{M}^I{}_J n^J_\mu~~,
\end{equation}
where $\mathcal{M}^I{}_J$ is an element of $SO(d-p)$. The transformation \eqref{eq:normrotations} leaves \eqref{eq:1formnormalization} invariant and hence expresses the freedom of choosing the normal 1-forms.\footnote{More formally, since the orientation of the normal 1-forms can be chosen freely as inward/outward pointing, $\mathcal{M}^I{}_J$ is a matrix in $O(d-p)$.}

We can furthermore introduce "inverse objects" $u^a_\mu$ and $n_I^\mu$ to the tangent vectors and normal 1-forms via the completeness relation
\begin{equation}
\delta^\mu_\nu = u^\mu_a u^a_\nu + n^I_\nu n^\mu_I~~,\label{eq:ProjectorCompleteness}
\end{equation}
which in turn satisfy the relations
\begin{equation} \label{eq:defstructure}
u_\mu^a n^\mu_I =0~~,~~u^\mu_a u^b_\mu = \delta^b_a~~,~~ n^\mu_I n_\mu^J = \delta^J_I~~.
\end{equation}
The tangent vectors, normal 1-forms and their inverses can be used to project any tensor tangentially or orthogonally to the surface. For instance, we may project some tensor $X^\mu{}_\nu{}_\rho{}^\lambda$ and denote the result as
\begin{equation}
X^{a}{_I}{}_b{}^J=u^a_\mu n_I^\nu u_b^\rho n_\lambda^J X^\mu{}_\nu{}_\rho{}^\lambda\,.\label{eq:inlinewith}
\end{equation}
It is also useful to define the tangential spacetime projector 
\begin{equation} \label{eq:tanprojector}
P^\mu_\nu = u^\mu_a u^a_\nu = \delta^\mu_\nu - n^\mu_In^I_\nu~~,
\end{equation}
which can be shown to be idempotent and of rank $p+1$. The object \eqref{eq:tanprojector} can be used to project arbitrary tensors onto tangential directions along the surface and satisfies $P^\mu_\nu n_\mu^I=0$.

\subsubsection{Timelike submanifolds and boost-invariance}
\begin{figure}
\centering
\includegraphics[width=0.55\linewidth]{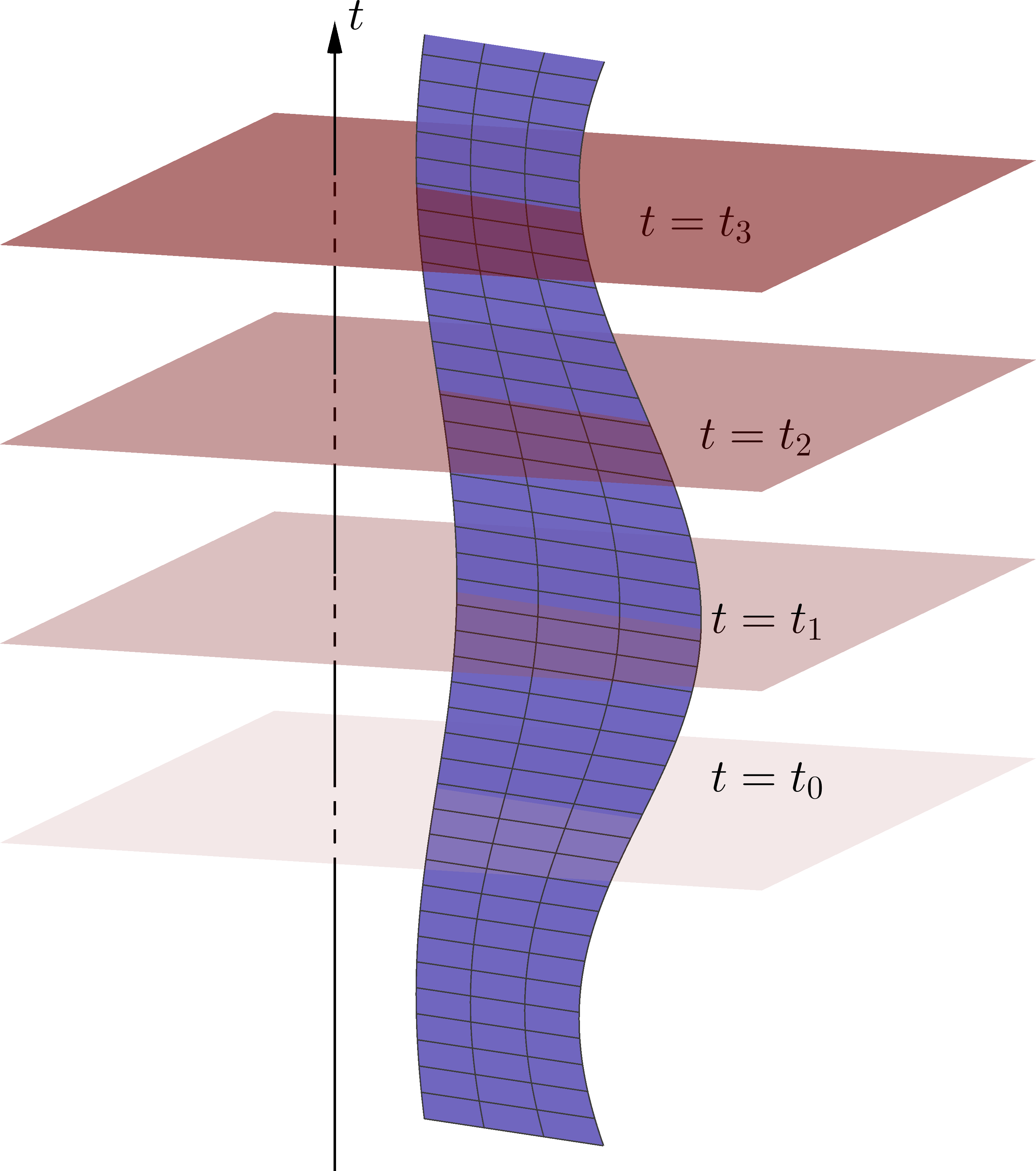}
\caption{Graphical depiction of the embedding of timelike Newton--Cartan submanifolds. The vertical direction represents the time $t$, while the spatial directions are in the plane orthogonal to the $t$--axis. The spatial hypersurfaces of constant time are denoted by their corresponding value of $t$. Note in particular that the condition \eqref{eq:tauIzero} implies that the submanifold does not “bend” away from the time direction in the ambient spacetime. }
\label{fig:wrinklync}
\end{figure}

Our goal is formulate a theory of non-relativistic submanifolds $\Sigma_{p+1}$ characterised by a Newton--Cartan structure that is inherited from the NC structure of the ambient spacetime. We introduce the submanifold clock 1-form as the pullback of the clock 1-form of the ambient spacetime such that
\begin{equation}\label{eq:subclock}
\tau_a = u^\mu_a \tau_\mu~~.
\end{equation}
As mentioned earlier, we focus on timelike submanifolds, by which we mean that the normal vectors $n_I^\mu$ satisfy
\begin{equation}
\tau_I= n_I^\mu \tau_\mu = 0~~,
\label{eq:tauIzero}
\end{equation}
and so $\tau_a$ is nowhere vanishing on $\Sigma_{p+1}$ (see figure \ref{fig:wrinklync} for an illustration of this condition). Then, taking 
\begin{equation}
{n^{\mu I} = h^{\mu\nu}n^I_\nu\,,}\label{eq:NormalVecDef}
\end{equation}
we make \eqref{eq:tauIzero} manifest. We note that these considerations imply that
\begin{eqnarray}
h^{IJ} & = & h^{\mu\nu}n^I_\mu n^J_\nu=\delta^{IJ}\,,\\
h^{aI} & = & h^{\mu\nu} u^a_\mu n^I_\nu = u^a_\mu n^{\mu I} = 0\,,\\\label{haizero}
h_{IJ} & = & h_{\mu\nu}n^\mu_I n^\nu_J = h_{\mu\nu}h^{\nu \rho}n^\mu_I n_{\rho J} = (\delta_\mu^\rho + v^\rho \tau_\mu)n^\mu_I n_{\rho J}  = \delta_{IJ}\,,\\\label{eq:hIJIndicesDown}
h_{a I} & = &  h_{\mu\nu} u^\mu_a n^\nu_I = h_{\mu\nu} u^\mu_a h^{\nu\rho}n_{\rho I} = v_I\tau_a\,,\label{eq:MixedLowerIdentity}
\end{eqnarray}
where $v_I = n_{I\mu}v^\mu$, which we will denote as the normal velocity. 

The description of submanifolds in NC geometry must be invariant under Galilean boosts, as these just express a choice of frame. This implies that the defining structure of NC submanifolds, namely \eqref{eq:1formnormalization} and \eqref{eq:defstructure}, must be invariant under local Galilean boost transformations. We start by noting that the embedding map does not transform under boosts, that is
\begin{equation} \label{eq:mapnoboost}
\delta_G X^\mu=0~~\Rightarrow~~\delta_G u^\mu_a=0~~,
\end{equation}
and hence the tangent vectors to the surface are boost-invariant.\footnote{Note that the embedding map specifies the location of the surface such that $x^\mu=X^\mu(\sigma^a)$. The spacetime coordinates $x^\mu$ do not transform under local Galilean boosts and hence neither does the embedding map $X^\mu(\sigma)$.} Specialising to timelike submanifolds, using \eqref{eq:NormalVecDef}, the variations of \eqref{eq:1formnormalization} and \eqref{eq:defstructure}, together with \eqref{eq:mapnoboost}, require
\begin{equation}  \label{eq:NoTransverseBoosts}
\begin{split}
u^{\mu}_a\delta_G n_\mu^I=&-n_\mu^I\delta_G u^\mu_a~~,~~n^{\mu J}\delta_G n_\mu^I=0~~\Rightarrow~~\delta_G n_\mu^I=0~~,\\
u_{\mu}^ah^{\mu\nu}\delta_G n_{\mu I}=&-n^\mu_I\delta_G u_\mu^a~~,~~u^{\mu}_a\delta_Gu^b_\mu=-u_\mu^b\delta_G u^\mu_a~~\Rightarrow~~\delta_G u^a_\mu=0~~,
\end{split}
\end{equation}
while $\delta_G n^\mu_I=0$ follows trivially from \eqref{eq:NormalVecDef}. Thus, Eq.~\eqref{eq:mapnoboost} ensures that the defining structure of timelike NC submanifolds is boost-invariant.\footnote{In particular, \eqref{eq:NoTransverseBoosts} implies that $\delta_G v^I = n_\mu^I\delta_G v^\mu = n_\mu^I h^{\mu\nu} \lambda_\nu =\lambda^I$. This is consistent with \eqref{eq:NoTransverseBoosts} since $n^I_\mu = n_{\underline{a}}^I e^{\underline{a}}_\mu - v^I \tau_\mu$, so that $n^I_{\underline{a}} = e^\mu_{\underline{a}} n^I_\mu$. Given that $\delta_G e^\mu_{\underline{a}} = \delta_G \tau_\mu=0$ and $\delta_G e^{\underline{a}}_\mu = \lambda^{\underline{a}} \tau_\mu$, we find that $\delta_G n^I_\mu = n^I_{\underline{a}} \lambda^{\underline{a}} \tau_\mu - \lambda^I\tau_\mu$ and  since $\lambda_\mu = n^I_\mu \lambda_I + u^a_\mu \lambda_a$, we get $n^I_{\underline{a}} \lambda^{\underline{a}} = n^I_{\underline{a}} e^{{\underline{a}}\mu}\lambda_\mu= \lambda^I$, thus confirming \eqref{eq:NoTransverseBoosts}.}

\subsubsection{Induced Newton--Cartan structures}
Besides the defining conditions \eqref{eq:1formnormalization} and \eqref{eq:defstructure}, NC submanifolds have other inherent geometric structures, such as induced tensors, that can be introduced via appropriate contractions of ambient tensors with any of the objects $u_\mu^a$ and $u^\mu_a$. We wish to identify the induced NC structures on the submanifold that have the same properties as the NC structures of the ambient spacetime. For instance, these induced structures should transform as in \eqref{eq:TNCgaugetrafos} and \eqref{eq:hhtrans} but now involving only tangential directions to the submanifold. 

The basic building blocks are the clock 1-form $\tau_a$ in eq.~\eqref{eq:subclock} and the normal velocity $v^I$ in eq.~\eqref{eq:MixedLowerIdentity} along with the pullbacks of the remaining ambient space fields
\begin{equation} \label{eq:pullbacks}
h_{ab} = u^\mu_a u^\nu_b h_{\mu\nu}\,, \qquad v^a = u^a_\mu v^\mu\,,\qquad h^{ab} = u^a_\mu u^b_\nu h^{\mu\nu}\,,\qquad m_a = u^\mu_a m_\mu~~.
\end{equation}
It is possible to see that these structures mimic many of the properties of the ambient NC structure. For instance we have $\tau_a h^{ab}=0$ and $v^a\tau_a=-1$ by virtue of \eqref{eq:tauIzero} and $\tau_\mu h^{\mu\nu} = 0$ as well as $v^\mu\tau_\mu=-1$. Additionally, they give rise to the completeness relation $h^{ac} h_{cb} = \delta^a_b + v^a \tau_b$, which in turn implies the relation $h^{\mu\nu}u_\mu^a=h^{ab}u_b^\nu$. However, using \eqref{eq:MixedLowerIdentity}, we find that
\begin{equation}
v^a h_{ab} = u^a_\mu v^\mu u^\rho _a u^\sigma_b h_{\rho\sigma} = -v^I h_{I b} = -v^I v^I \tau_b~~,\label{eq:FunnyContraction}
\end{equation}
which is non-zero, contrary to the corresponding ambient NC result $v^\mu h_{\mu\nu}=0$. Hence, the individual structures in \eqref{eq:pullbacks} do not form a NC geometry on the submanifold. Using \eqref{eq:FunnyContraction} we instead define 
\begin{equation}
{\check h_{ab} = h_{ab} - v^I v_I \tau_a\tau_b~~,}\label{eq:checkh}
\end{equation}
which leads to a completeness relation and satisfies the required orthogonality condition, that is
\begin{equation}
h^{ac}\check h_{cb} = \delta^a_b + \tau_b v^a~~,~~ v^a \check h_{ab} =0~~. 
\end{equation}

For $\check h_{ab}$ to be considered a NC structure on the submanifold, one must also ensure that it transforms under Galilean boosts as its ambient space counterpart $h_{\mu\nu}$ (cf. \eqref{eq:hhtrans}). Using \eqref{eq:TNCgaugetrafos}, \eqref{eq:hhtrans}, \eqref{eq:NoTransverseBoosts} and\footnote{This follows from the statement that $v^\mu\lambda_\mu = 0$.} $v^a\lambda_a = -v^I\lambda_I$, it can be shown that 
\begin{equation}
\begin{split}
\delta_G v^a &= h^{ab}\check\lambda_b~~,~~\delta_G (v^a h_{ab}) = -2\tau_b \lambda^I v_I~~,~~\delta_G h_{ab} = 2\tau_{(a}\lambda_{b)}~~,\\
\delta_G \check h_{ab} &= 2\tau_{(a}\lambda_{b)}-2\tau_a \tau_b v^I \lambda_I = 2\tau_{(a}\lambda_{b)}+2\tau_a\tau_b	v^c\lambda_c = 2\tau_{(a}\check\lambda_{b)}~~,
\end{split}
\end{equation}
where we have defined 
\begin{equation}
\check\lambda_a = \lambda_a+v^c\lambda_c\tau_a=\check h_{ab} h^{bc}\lambda_c~~,
\end{equation}
which satisfies $v^a \check\lambda_a =0$, analogously to the ambient orthogonality condition $v^\mu \lambda_{\mu}=0$. Thus $\check h_{ab}$ transforms under submanifold Galilean boosts in the same manner as $h_{\mu\nu}$ transforms under ambient Galilean boosts.

NC submanifolds admit boost-invariant structures similar to the ambient structures \eqref{eq:barmunu} and \eqref{eq:NewtonPot}. Given that the set of tangent and normal vectors is boost-invariant (see eq.~\eqref{eq:NoTransverseBoosts}), two of these structures are obtained by contractions of the corresponding ambient quantities, namely
\begin{equation}
\bar h_{ab} =u^\mu_au^\nu_b \bar h_{\mu\nu}= \check h_{ab} - 2\tau_{(a}\check m_{b)}~~,~~\hat{v}^a =u_\mu^a \hat v^\mu= v^a - h^{ab}\check m_b ~~,
\end{equation}
where we have defined the submanifold $U(1)$ connection
\begin{equation}
{\check m_a = m_a - \frac{1}{2}v^I v_I \tau_a\,,}\label{eq:checkm}
\end{equation}
which transforms under boosts as $\delta_G \check m_a =\check\lambda_a$, analogous to the boost transformation of the ambient connection $ m_\mu$. Given that in the ambient space we have the identity $\hat v^\nu \bar h_{\nu\mu} = 2\tilde{\Phi}\tau_\mu$ where $\tilde{\Phi}$ is defined in \eqref{eq:NewtonPot} we require an analogue condition of the form $\hat v^a\bar h_{ab}=2\check\Phi \tau_b$ for some scalar $\check \Phi$. Explicit manipulation shows that
\begin{equation}
\hat v^a \bar h_{ab} = u^a_\mu \hat{v}^\mu u^\nu_a u^\rho_b \bar h_{\nu\rho} = \hat v^\nu \bar h_{\nu\rho} u^\rho_b - n^I_\mu h^{\nu\sigma}n^I_{\sigma} \hat v^\mu u^\rho_b \bar h_{\nu\rho}= 2(\tilde{\Phi} - 1/2 \hat{v}^I \hat{v}^I)\tau_b~~,\label{eq:Recycle}
\end{equation}
which leads us to identify
\begin{equation}
\check\Phi = \tilde{\Phi} - \frac{1}{2}\hat{v}^I \hat{v}^I= -v^a \check m_a + \frac{1}{2}h^{ab} \check m_a\check m_b~~,
\label{eq:CheckPhi}
\end{equation}
thus taking the same form as its ambient counterpart \eqref{eq:NewtonPot} but now in terms of $\check m_a$. 

In summary, we define the induced Newton--Cartan structure on the submanifold $\Sigma_{p+1}$ to consist of the fields $(\tau_a, \check h_{ab},  \check m_a)$ and $(v^a,h^{ab})$ along with the boost invariant combinations $\hat v^a$, $\bar h_{ab}$ and $\check\Phi$, satisfying the relations
\begin{equation}
\delta^a_b=h^{ac}\check h_{cb}-\tau_b v^a~~,~~ \tau_a h^{ab}=0\,,\qquad v^a\check h_{ab}=0~~,
\end{equation}
as well as 
\begin{equation}
\hat v^a\bar h_{ab}=2\check\Phi \tau_b~~.
\end{equation}
These are related to the ambient Newton--Cartan structures in the following way
\begin{eqnarray}
\hspace{-1cm}&&\tau_a = u^\mu_a\tau_\mu\,,\qquad \check h_{ab}=u^\mu_a u^\nu_bh_{\mu\nu}-v^I v_I\tau_a\tau_b=h_{ab}-v^I v_I\tau_a\tau_b~~,\\
\hspace{-1cm}&&\check m_a =u^\mu_a m_\mu-\frac{1}{2}v^I v_I\tau_a = m_a -\frac{1}{2}v^I v_I\tau_a\,,\qquad v^a = u^a_\mu v^\mu,\qquad h^{ab}=u^a_\mu u^b_\nu h^{\mu\nu}~~,\\
\hspace{-1cm}&&\hat v^a = v^a - h^{ab}\check m_b=u^a_\mu \hat v^\mu\,,\qquad \bar h_{ab}=\check h_{ab} - 2\tau_{(a}\check m_{b)} = u^\mu_a u^\nu_b \bar h_{\mu\nu}~~,\\
\hspace{-1cm}&&\check \Phi = -v^a \check m_a + \frac{1}{2}h^{ab} \check m_a\check m_b=\tilde{\Phi} - \frac{1}{2}\hat{v}^I \hat{v}^I~~.
\end{eqnarray}
These structures transform according to 
\begin{eqnarray}
\hspace{-1cm}&&\delta\tau_a=\pounds_\zeta \tau_a~~,~~\delta\check h_{ab}=\pounds_\zeta \check h_{ab}+2\check\lambda_{(a}\tau_{b)}~~,~~\delta \check m_a = \pounds_\zeta \check m_a + \check\lambda_a + \D_a \sigma~~,\label{eq:InducedTrafo1}\\
\hspace{-1cm}&&\delta v^a =\pounds_\zeta v^a + h^{ab}\check\lambda_b~~,~~ \delta h^{ab}=\pounds_\zeta h^{ab}~~,\\
 \hspace{-1cm}&&\delta\hat v^a = \pounds_\zeta \hat v^a - h^{ab}\D_b\sigma~~,~~ \delta\bar h_{ab}=\pounds_\zeta \bar h_{ab}-2\tau_{(a}\D_{b)}\sigma~~,~~ \delta\check\Phi=\pounds_\zeta \check\Phi - \hat v^a\D_a\sigma~,\label{eq:InducedTrafo2}
\end{eqnarray}
under submanifold diffeomorphisms $\zeta^a$, Galilean boosts $\check\lambda_a$ (satisfying $v^a\check\lambda_a =0$) and $U(1)$ gauge transformations $\sigma$.

\subsubsection{The role of the transverse velocity $v^I$}\label{sec:BoostedBrane}
In order to elucidate the role of $v^I$, we consider for concreteness a co-dimension one submanifold $\Sigma$ moving with (constant) linear velocity $v^\mu_\Sigma = (0,0,0,\mathfrak{v})$ in the $z$-direction of a four-dimensional flat ambient Newton--Cartan spacetime, which was introduced in \eqref{eq:flatspace} and where $i$ runs only over spatial directions. Defining $\Sigma$ via the embedding equation
\begin{equation}
F(x,y,z-\mathfrak{v}t)=0~~,
\end{equation}
we can write the normal 1-form as
\begin{equation}
n = N\diff F = N\D_x F + N\D_y F + N\D_u F\diff z - \mathfrak{v}\D_u F\diff t~~,
\end{equation}
where we have defined $u = z - \mathfrak{v}t$ and where $N$ is fixed by the normalisation condition \eqref{eq:1formnormalization}. This means that 
\be 
v^\mu n_\mu = -n_0 = \mathfrak{v}N\D_u F,~~,~~ v^\mu_{\Sigma}n_\mu = \mathfrak{v} n_z = \mathfrak{v}N\D_u F~~,
\ee
leading us to conclude that $v^\mu n_\mu = v^\mu_{\Sigma}n_\mu$. Thus, the normal projection of the NC velocity is the same as the normal projection of the linear velocity vector $v^\mu_\Sigma$ of the submanifold $\Sigma$. 

To illustrate this in the simplest possible setting, we consider an infinitely extended moving flat membrane embedded in $(3+1)$-dimensional flat NC space, described by
\begin{equation}
u = z - \mathfrak{v}t = 0~~,
\end{equation}
leading to the normal 1-form
\begin{equation}
n_\mu = -\mathfrak{v} \delta^0_\mu + \delta^3_\mu~~\Rightarrow~~v^\mu n_\mu = \mathfrak{v}~~.
\label{eq:vperp_is_velocity}
\end{equation}
Therefore, for a flat brane, where the normal vector is the same everywhere, we see that the normal projection of the NC velocity vector is just the magnitude of the linear velocity of the plane.

\subsubsection{Covariant derivatives, extrinsic curvature and external rotation}\label{sec:CovDandExtK}
Since we are dealing with the description of a single surface, and not of a foliation, covariant differentiation of submanifold structures only has meaning along tangential directions to the surface. Analogously to Lorentzian surfaces (see e.g. \cite{Armas:2017pvj}), we define a covariant derivative along surface directions that is compatible both with the surface Newton--Cartan structure, $ D_a \tau_b = 0 =  D_a h^{bc}$, and the ambient Newton--Cartan structure, $ D_a \tau_\mu = 0 =  D_a h^{\mu\nu}$, that acts on an arbitrary mixed tensor $T^{b \mu}$ as
\begin{equation}
{ D_a T^{b \mu } = \D_a T^{b\mu}	+ \gamma^b_{ac} T^{c\mu} + u^\rho_a \Gamma^\mu_{\rho\lambda} T^{b\lambda}~~,}\label{eq:SurfaceCovD}
\end{equation}
where we have introduced the surface affine connection according to
\begin{equation} \label{eq:indcon}
{\gamma}^c_{ab} = -\hat{v}^c \D_a\tau_b + \frac{1}{2}h^{cd}\left(\D_a \bar h_{b d} + \D_b\bar h_{ad} - \D_d\bar h_{ab} \right)~~,
\end{equation}
in analogy with the the spacetime affine connection \eqref{eq:TNCconnection}.
Note in particular that $D_a$ does not act on transverse indices. The relation between ${\gamma}^c_{ab}$ and $\Gamma^\mu_{\rho\lambda}$ is obtained in appendix \ref{App:ConnectionsAreImportant} and is shown to be
\begin{equation}
\gamma^c_{ab} = \Gamma ^c_{ab} + u^c_\mu\D_a u^\mu_b=u^c_\mu u^\nu_a\nabla_\nu u^\mu_b~~,\label{eq:gamma}
\end{equation}
where the corresponding surface torsion tensor is
\begin{equation}
2 \gamma^c_{[ab]} = -\hat{v}^c \tau_{ab} =-\hat v^cu^\mu_a u^\nu_b \tau_{\mu\nu}~~,
\end{equation}
and where the last equality follows from the fact that exterior derivatives commute with pullbacks.\footnote{Alternatively, this conclusion can be reached via the relation $\partial_a u^\mu_b = \partial_a \partial_b X^\mu = \partial_b \partial_a X^\mu = \partial_b u_a^\mu$.}

It is also convenient to introduce a covariant derivative ${\mathfrak{D}}_a$ that acts on all indices, i.e. $\mu,a,I$ \cite{Armas:2017pvj}, and whose action on the normal 1-forms and tangent vectors allows for the Weingarten decomposition\footnote{The action of ${\mathfrak{D}}_a$ on some vector $T^I$ takes the form ${\mathfrak{D}}_a T^I= D_aT^I-\omega_a{}^I{}_J T^J$.}
\begin{equation}
\begin{split}
{\mathfrak{D}}_a n^I_\sigma &= \D_a n^I_\sigma - \Gamma^\lambda_{\mu\sigma} u^\mu_a n_\lambda^I - \omega_{a}{^I}{_J}n_\sigma^J  = - u^b_\sigma K_{ab}{^I}+\frac{1}{2}u^b_\sigma\hat v^I \tau_{ab}~~, \\
{\mathfrak{D}}_a u_b^\mu &= D_a u_b^\mu =  n^\mu_I K_{ab}{^I}-\frac{1}{2}n^\mu_I\hat v^I \tau_{ab}~~,
\end{split}
\label{eq:Weingart1}
\end{equation}
where we have defined the extrinsic curvature to the submanifold according to
\begin{equation}
K_{ab}{^I} = n_\mu^I D_a u^\mu_b + \frac{1}{2}\hat v^I \tau_{ab}=n_\mu^I \left(\partial_a u^\mu_b+u^\nu_a u^\rho_b\Gamma_{(\nu\rho)}^\mu\right)= -u^\mu_a u^\nu_b\nabla_{(\mu}n_{\nu)}^I~~.
\label{eq:ExtCurv}
\end{equation}
The extrinsic curvature tensor, when defined in this manner, is symmetric and invariant under Galilean boosts but transforms under $U(1)$ gauge transformations according to
\begin{equation}\label{eq:U1trafoK}
\delta_\sigma K_{ab}{}^I=\frac{1}{2}\tau_{Ia}\partial_b\sigma+\frac{1}{2}\tau_{Ib}\partial_a\sigma\,,
\end{equation}
where we used \eqref{eq:gaugatrafoGamma}. In \eqref{eq:Weingart1} we also introduced the external rotation tensor, which can be interpreted as a $SO(d-p)$ connection, defined as
\begin{equation}
\omega_a{^I}{_J} = n^\mu_J  D_a n^I_\mu~~,\label{eq:ExtRotTens}
\end{equation}
which is antisymmetric in $I,J$ indices and transforms under $U(1)$ gauge transformations as
\begin{equation}\label{eq:U1trafoW}
\delta_\sigma \omega_a{^I}{_J}=-\frac{1}{2}\left(\tau_{aJ}\partial^I\sigma+{\tau^I}_J \partial_a \sigma+{\tau^{I}}_a \partial_J \sigma\right)~~.
\end{equation}
If the submanifold is co-dimension one, the external rotation vanishes by definition. 

Both the extrinsic curvature tensor and the external rotation tensor introduced here are direct analogues of their Lorentzian counterparts \cite{Armas:2017pvj}. To see that $\omega_a{^I}{_J}$ transforms as a connection we examine what happens if we perform a local $SO(d-p)$ rotation of the normal vectors as in \eqref{eq:normrotations}. If we focus on an infinitesimal rotation ${\mathcal{M}^{I}}_J=\delta^I_J + \lambda^I{}_J$ where $\lambda^I{}_J=-\lambda^J{}_I$, the extrinsic curvature tensor and external rotation tensor transform as
\be \label{eq:deltaKundernormalvectrafo}
\delta_\lambda K_{ab}{^I} = \lambda^I{_J}K_{ab}{^J}~~,~~\delta_\lambda \omega_a{^I}{_J} = \partial_a\lambda^I{}_J+\lambda^I{_K}\omega_a{^K}{_J}+\lambda_J{^K}\omega_a{^I}{_K}~~.
\ee
In addition, under a change of sign of the normal vectors $n^\mu_I\to-n^\mu_I$, the extrinsic curvature changes sign.

\subsubsection{Integrability conditions}
Certain combinations of geometric structures of Lorentzian submanifolds are related to specific contractions of the Riemann tensor of the ambient space. These are known as integrability conditions. In this section we derive the analogous conditions in the context of NC submanifolds, which are known as the Codazzi--Mainardi, Gauss--Codazzi and Ricci--Voss equations. In order to do so, we note that in the presence of torsion, the Ricci identity takes the form
\begin{equation}
[ \nabla_\mu, \nabla_\nu] X_\sigma =  R_{\mu\nu\sigma}{^\rho}X_\rho- 2\Gamma^\rho_{[\mu\nu]} \nabla_\rho X_\sigma~~,\label{eq:defRiem}
\end{equation}
where the Riemann tensor $R_{\mu\nu\sigma}{^\rho}$ of the ambient space is given by
\begin{equation}
R_{\mu\nu\sigma}{^\rho} = -\D_\mu  \Gamma^\rho_{\nu\sigma}	+ \D_\nu \Gamma^\rho_{\mu\sigma} - \Gamma^\rho_{\mu\lambda} \Gamma^\lambda_{\nu\sigma} + \Gamma^\rho_{\nu\lambda}\Gamma^\lambda_{\mu\sigma}~~.\label{eq:Riemann}
\end{equation}

The integrability conditions to be derived below take a nice form if we work with an object that is closely related to the extrinsic curvature, namely
\begin{equation}
\tilde K_{ab}{^I} = n_\mu^I D_a u^\mu_b =K_{ab}{^I}-\frac{1}{2}\hat v^I \tau_{ab}~~, \label{eq:ExtCurv1}
\end{equation}	
which has a non-vanishing antisymmetric part $2\tilde{K}_{[ab]}{^J}=-\hat v^J\tau_{ab}$.

We begin by deriving the Codazzi--Mainardi equation (see e.g. \cite{Armas:2017pvj,aminov2014geometry}) by considering the quantity $ D_a \tilde{K}_{bc}{^I}  -  D_b \tilde{K}_{ac}{^I}$. We find
\begin{align}
 D_a \tilde{K}_{bc}{^I} = \tilde{K}_{ab}{^I}n_I^\rho ( \nabla_\rho u^\mu_c)n^I_\mu -\omega_b{^I}{_J}\tilde{K}_{ac}{^J}   - u^\mu_c u^\rho_a u^\sigma_b   \nabla_\rho\nabla_\sigma n_\mu^I~~,
\end{align}
where we used \eqref{eq:ExtRotTens}. From here, using \eqref{eq:defRiem} and the covariant derivative ${\mathfrak{D}}_a$ introduced in \eqref{eq:Weingart1} we derive the Codazzi--Mainardi equation
\begin{equation}
{\mathfrak{D}}_a \tilde{K}_{bc}{^I}  -  {\mathfrak{D}}_b \tilde{K}_{ac}{^I} = -   R_{abc}{^I} + \hat v^d\tau_{ab} \tilde{K}_{d c}{^I}~~.
\label{eq:Codazzi-Mainardi}
\end{equation}

In order to derive the Gauss--Codazzi equation, we let $\omega_c$ be any submanifold 1-form that is the pullback of $\omega_\mu$ whose normal components vanish, i.e. $\omega_\mu=u^c_\mu\omega_c$. The Ricci identity for the submanifold reads
\begin{equation}
[D_a,D_b]\omega_c = {\mathcal{R}}_{abc}{^d}\omega_d +\hat v^d\tau_{ab} D_d \omega_c~~,\label{eq:step1}
\end{equation}
where ${\mathcal{R}}_{abc}{^d}$ is the Riemann tensor of the submanifold and takes the same form as \eqref{eq:Riemann} but with the connection $\Gamma^\rho_{\nu\sigma}$ replaced by $\gamma^c_{ab}$ of \eqref{eq:indcon}. Using  $u_\mu^a D_b u_c^\mu=0$ (which follows from \eqref{eq:gamma}) and $n^\mu_I D_b u_\mu^d=h^{de}\tilde K_{be I}$, explicit manipulation leads to 
\begin{align}
{\mathcal{R}}_{abc}{^d}\omega_d +\hat v^d\tau_{ab} D_d \omega_c = &h^{ed}\tilde{K}_{ac}{^I}\tilde{K}_{beI}\omega_d -h^{ed}\tilde{K}_{bc}{^I}\tilde{K}_{aeI}\omega_d +  R_{abc}{^d}\omega_d\nn\\
& +\tau_{ab}\left( - \hat v^I n^\rho_I u^\mu_c \nabla_\rho \omega_\mu+\hat v^\nu u^\mu_c \nabla_\nu \omega_\mu\right)~~,
\end{align}
where we used \eqref{eq:defRiem}. In this expression, the terms proportional to $\tau_{ab}$ on both sides cancel and since it must be true for any one form $\omega_c$, the Gauss--Codazzi equation becomes
\begin{equation}
{{\mathcal{R}}_{abc}{^d}  = \tilde{K}_{ac}{^I}\tilde{K}_b{^d}{_I} - \tilde{K}_{bc}{^I}\tilde{K}_a{^d}{_I} +  R_{abc}{^d}\,,}\label{eq:GaussCodazzi}
\end{equation}
where $\tilde{K}_b{^d}{_I}=h^{dc}\tilde{K}_{bc}{_I}$.

Although we will not use it in this paper, we will briefly discuss the Ricci--Voss equation for completeness. This equation becomes useful for surfaces of co-dimension higher than one, where we can define the outer curvature in terms of the external rotation tensor \eqref{eq:ExtRotTens} as
\begin{equation}
\Omega^I{_{Jab} = 2\D_{[a}\omega_{b]}{^I}{_J} - 2\omega_{[a\vert}{^{I}}{}_K\omega_{\vert b]}{}^K{}_J~~.}
\end{equation}
In terms of this tensor, the Ricci--Voss equation for Newton--Cartan geometry can be shown to read
\begin{equation}
\Omega^I{}_{Jab} =   R_{ab J}{^I} -2h^{cd} \tilde{K}_{[a\vert c}{^I} \tilde{K}_{\vert b]d J} ~~.
\end{equation}
This completes the description of the geometric structures of NC submanifolds.


\section{Variations and dynamics of Newton--Cartan submanifolds} \label{sec:var}
In the previous section we defined timelike NC submanifolds and their characteristic geometric properties. In this section, closely following the Lorentzian case \cite{Armas:2017pvj}, we develop the variational calculus for NC submanifolds for the geometric structures of interest. These results are necessary to later introduce geometric action functionals capable of describing different types of soft matter systems, including the case of bending energies for lipid vesicles. 

\subsection{Variations of Newton--Cartan objects on the submanifold}\label{sec:Variations}

In the following, we consider two types of variations, namely embedding map variations, which are displacements of the submanifold, and Lagrangian variations which consist of the class of diffeomorphisms that displace the ambient space but keep the embedding map fixed (see e.g. \cite{Carter:1993wy,Carter:1997pb} and also \cite{Armas:2013hsa, Armas:2017pvj}). As in the Lorentzian case \cite{Armas:2017pvj}, the sum of these two types of variations yield the transformation properties of the submanifold structures under full ambient space diffeomorphisms. When considering action functionals that give dynamics to submanifolds, they are equivalent, up to normal rotations.\footnote{In the context of continuum mechanics, these two viewpoints are known as the Lagrangian and Eulerian descriptions.} 

\subsubsection{Embedding map variations}\label{sec:EmbedVars}
Before specialising to any of the two types of variations, it is useful to consider general variations of the normal vectors. In particular, we decompose the variation of the normal vectors as 
\begin{equation}
\delta n_\mu^I = u^a_\mu u^\nu_a \delta n^I_\nu + n^J_\mu n^\nu_J \delta n^I_\nu
= -u^a_\mu n^I_\nu \delta u^\nu_a + \frac{1}{2}n_{\mu J} (n^{\nu J} \delta n^I_\nu + n^{\nu I} \delta n^J_\nu )  +  \lambda^{I}{}_Jn_{\mu}^J~~,
\end{equation}
where
\begin{equation}
\lambda^{I}{}_J= \frac{1}{2}\left(n^{\nu}_J \delta n^I_\nu  - n^{\nu I} \delta n_{J\nu}\right)~~,
\end{equation}
is a local $\mathfrak{so}(d-p)$ transformation of the normal vectors. By varying the second relation in \eqref{eq:1formnormalization}, we find the relation $n^{\nu J} \delta n^I_\nu + n^{\nu I} \delta n^J_\nu = -n_\mu^I n_\nu^J \delta h^{\mu\nu}$ and hence
\begin{equation}
\delta n_\mu^I = -u^a_\mu n^I_\nu \delta u^\nu_a - \frac{1}{2}n_{\mu J} n_\nu^I n_\rho^J \delta h^{\nu\rho} +  \lambda^{I}{}_Jn_{\mu}^J~~.
\end{equation}
By varying the completeness relation \eqref{eq:ProjectorCompleteness} one may express variations of $\delta h^{\nu\rho}$ in terms of variations of $\tau_\nu$ and $h_{\nu\rho}$ such that $\delta h^{\mu\nu} = 2v^{(\mu}h^{\nu)\lambda}\delta\tau_\lambda - h^{\mu\rho}h^{\nu\sigma}\delta h_{\rho\sigma}$. This leads to
\begin{equation}
{\delta n_\mu^I =  - v^{(I}n^{J)\nu}n_{\mu J}\delta \tau_\nu + \frac{1}{2}n^{\rho J} n_{\mu J} n^{\nu I}\delta h_{\rho \nu}- {n}^I_\nu u^a_\mu\delta u^\nu_a +  \lambda^{I}{}_Jn_{\mu}^J~~,}\label{eq:NormalVar}
\end{equation}
which describes arbitrary infinitesimal variations of the normal vectors.

We now specialise to infinitesimal variations of the embedding map which we denote by
\begin{equation}
\delta X^\mu(\sigma) = - \xi^\mu(\sigma)~~,
\end{equation}
where $\xi^\mu(\sigma)$ is understood as being an infinitesimal first order variation. Under this variation, the ambient tensor structures evaluated at the surface (i.e. $\tau_\mu(X)$, $\bar h_{\mu\nu}(X)$) vary as
\begin{equation}
\delta_X \tau_\mu (X) = -\xi^\nu \D_\nu \tau_\mu~~,~~\delta_X \bar h_{\mu\nu} (X) = -\xi^\rho \D_\rho \bar h_{\mu\nu}~~,
\end{equation}
which follows from $\delta_X \tau_\mu(X) = \tau_\mu (X - \xi ) - \tau_\mu(X) = -\xi^\nu\D_\nu\tau_\mu + \mathcal{O}(\xi^2)$. In turn, the tangent vectors transform as
\begin{equation}
\delta_X u^\mu_a = \D_a \delta X^\mu = -\D_a \xi^\mu~~,
\end{equation}
while variations of the induced metric structures take the form
\begin{equation}
\delta_X \tau_a = -u^\mu_a \pounds_\xi\tau_\mu,\qquad \delta_X \bar h_{ab} = -u^\mu_a u^\nu_b\pounds_\xi \bar h_{\mu\nu}~~.\label{eq:LagVar?}
\end{equation} 
In other words, for these structures, performing embedding map variations is equivalent to performing a diffeomorphism in the space of embedding maps that keep $u^\mu_a$ fixed, i.e. they are diffeomorphisms that are independent of $\sigma^a$. Using \eqref{eq:NormalVar}, we can write the variation of the normal vector as
\begin{equation}
\delta_X n_\mu^I = -n_{\mu J}n_\rho^{(I}n^{J)\nu}\nabla_\nu\xi^\rho -n_{\mu J}\hat v^{(I}n^{J)\nu}\tau_{\nu\rho}\xi^\rho+ n^I_\rho\D_\mu \xi^\rho+ \widetilde \lambda^{IJ}n_{\mu J}~~,\label{eq:NormalVecTrafo9000}
\end{equation}
where the third term ensures that the orthogonality relation $u^\mu_a n^I_\mu = 0$ is obeyed after the variation while the last term is a local transverse rotation of the form $\widetilde \lambda^{IJ}=\lambda^{IJ}+ n_\rho^{[J}n^{I]\nu}\D_\nu\xi^\rho$.

For the purposes of this work, as mentioned in sec.~\ref{sec:abs}, we will be focusing on ambient NC geometries with absolute time, i.e. zero torsion. This extra assumption greatly simplifies many expressions after variation. We stress, however, that it is in general not possible to assume zero torsion before variation, as variation and setting torsion to zero do not always commute.\footnote{For instance, when considering equations of motion for surfaces via extremisation of a Lagrangian as in the next section, a term of the form $X^{\mu\nu}\tau_{\mu\nu}$ in the Lagrangian can give a nonzero contribution to the equation of motion of $\tau$ as neither $X^{\mu\nu}$ nor $\delta\tau_{\mu\nu}$ need to vanish on ambient spaces with zero torsion. }

However, specifically in the case of embedding map or Lagrangian variations, the variation of $\tau_{\mu\nu}$ is guaranteed to vanish when the torsion itself vanishes. This means that we can set torsion to zero in the Lagrangian \textit{if} all we are interested in are the equations of motion for $X^\mu$. For example $\delta_X\tau_{\mu\nu}(X)=-\xi^\rho\partial_\rho\tau_{\mu\nu}$, which vanishes when $\diff\tau=0$. Under the assumption of vanishing torsion, variations of the extrinsic curvature \eqref{eq:ExtCurv} take the form 
\begin{eqnarray}
\delta_X K_{ab}{^I}  &= &(\delta_X n^I_\mu) \D_a u^\mu_b - n^I_\mu \D_a\D_b \xi^\mu + (\delta_X n^I_\mu) u^\rho_a \Gamma^\mu_{\rho\lambda} u^\lambda_b\nn \\
&& - n^I_\mu (\D_a \xi^\rho) \Gamma^\mu_{\rho\lambda} u^\lambda_b - n^I_\mu u^\rho_a \xi^\kappa\D_\kappa( \Gamma^\mu_{\rho\lambda}) u^\lambda_b - n^I_\mu u^\rho_a  \Gamma^\mu_{\rho\lambda}\D_b\xi^\lambda\nn \\
&=& -n_\mu^I D_a  D_b\xi^\mu +\xi^\rho R_{\rho ab}{^I} + n_\rho^{[I}n^{J] \nu}\Gamma^\rho_{\nu\sigma}\xi^\sigma K_{ab J}~~,\label{eq:VarOfK}
\end{eqnarray}
where we have used \eqref{eq:NormalVecTrafo9000} as well as  $\delta_X  \Gamma^\mu_{\rho\lambda}(X) = - \xi^\kappa\D_\kappa  \Gamma^\mu_{\rho\lambda}$. The last term in \eqref{eq:VarOfK} denotes a local $\mathfrak{so}(d-p)$ transformation and we have explicitly ignored further rotations by setting $\lambda^{IJ}=0$ in \eqref{eq:NormalVar}. It is also straightforward to consider variations of the external rotation tensor \eqref{eq:ExtRotTens} but since we do not explicitly consider this structure in the dynamics of submanifolds, we will not dwell on this.

\subsubsection{Lagrangian variations}\label{sec:LagVars}
In the previous section we have described how to perform variations of the embedding map. In this section we focus on a particular class of diffeomorphisms $x^\mu\to x^\mu-\xi^\mu$ that only act on fields with support in the entire ambient spacetime, that is, they only act on the NC triplet $(\tau_\mu(x),h_{\mu\nu}(x),m_\mu(x))$. In general, diffeomorphisms also displace the embedding map according to $\delta_\xi X^\mu=-\xi^\mu$ where $\delta_\xi $ denotes an infinitesimal diffeomorphism variation. However, here we consider the case of Lagrangian variations for which $\delta_\xi X^\mu=0$ (see e.g. \cite{Carter:1993wy,Carter:1997pb,Armas:2013hsa,Armas:2017pvj}). In turn, this implies that the tangent vectors do not vary, that is\footnote{If we were working with foliations of surfaces instead of a single surface, we could define a set of vector fields $u^\mu_a(x)$ where $x$ is any point in the ambient spacetime. We could then require that the Lie brackets between these vector fields vanish so that their integral curves can be thought of as locally describing a set of curvilinear coordinates for the submanifold. In other words, the restriction of these vector fields to the submanifold obeys the condition that the $u^\mu_a$ are tangent vectors, i.e. $u^\mu_a(x)\vert_{x=X}=\partial_a X^\mu$. When we perform ambient diffeomorphisms within the context of a foliation, we must ensure that this condition is respected. This means that $\left(\xi^\rho(x)\partial_\rho u^\mu_a(x)-u^\rho_a(x)\partial_\rho\xi^\mu(x)\right)\vert_{x=X}=\pounds_{\xi}u^\mu_a=\delta_\xi u^\mu_a=0$. Lagrangian diffeomorphisms are thus generated by $\xi^\mu(x)$ such that \eqref{eq:nomap} is obeyed. See for instance \cite{Capovilla:1994bs}.}
\begin{equation}\label{eq:nomap}
\delta_\xi u^\mu_a=0~~.
\end{equation}

In the remainder of this section, we will explicitly work out Lagrangian variations of submanifold structures and compare them with embedding map variations, thereby extracting the transformation properties under full ambient diffeomorphisms. In particular, using \eqref{eq:nomap} and the fact that $\delta_\xi \tau_\mu=\pounds_\xi \tau_\mu$ and $\delta_\xi \bar h_{\mu\nu}=\pounds_\xi  \bar h_{\mu\nu}$ we find
\begin{equation}
\delta_\xi \tau_a = u^\mu_a\pounds_\xi \tau_\mu ~~,~~ \delta_\xi  \bar h_{ab } = u^\mu_a u^\nu_b\pounds_\xi  \bar h_{\mu\nu}~~.
\end{equation}
Comparing this with \eqref{eq:LagVar?}, it follows that for pullbacks of Newton--Cartan objects we have the relations
\begin{equation}
(\delta_\xi + \delta_X)\tau_a = (\delta_\xi + \delta_X)\bar h_{ab}=0~~,
\end{equation}
and thus these objects transform as scalars under ambient diffeomorphisms.  For later purposes, we rewrite these results as
\begin{eqnarray}
\delta_\xi \tau_a &=&  \tau_\rho  D_a \xi^\rho~~,\\
\delta_\xi  \bar h_{ab } &=& \bar h_{\rho b}  D_a \xi^\rho +  \bar h_{\rho a}D_b\xi^\rho - 2\tau_a \tau_b \xi^\sigma\D_\sigma\tilde{\Phi} - 2\xi^\sigma\tau_\sigma\tau_{(a}\D_{b)}\tilde{\Phi} - 2\tau_{(a}\bar{\mathcal{K}}_{b)\sigma}\xi^\sigma\,,
\end{eqnarray}
where we have used the relation (valid in the absence of torsion)
\begin{equation}
\nabla_\sigma \bar h_{\mu\nu} = - 2\tau_\mu \tau_\nu \D_\sigma\tilde{\Phi} - 2\tau_\sigma\tau_{(\mu}\D_{\nu)}\tilde{\Phi} - 2\tau_{(\mu}\bar{\mathcal{K}}_{\nu)\sigma}\,,\label{eq:CovDofbarh}
\end{equation}
as well as $\hat v^\lambda \bar h_{\lambda\mu} = 2\tau_\mu \tilde{\Phi}$ and where $\bar{\mathcal{K}}_{\mu\nu}=-\pounds_{\hat v}\bar h_{\mu\nu}/2$. 

Considering the normal one-forms, using \eqref{eq:NormalVar} we find that
\begin{equation}
\delta_\xi n_\mu^I =  - v^{(I}n^{J)\nu}n_{\mu J} \tau_\rho  \nabla_\nu \xi^\rho +n^{\lambda J} n_{\mu J} n^{\nu I}h_{\rho(\lambda} \nabla_{\nu)} \xi^\rho=n_{\mu J} n_\rho^{(I} n^{J)\nu}  \nabla_\nu \xi^\rho\label{eq:LagrangianVarOfNormal}\,,
\end{equation}
where we have used \eqref{eq:CovDofbarh} as well as the identity $n^\lambda_I h_{\rho\lambda} = h_{\rho I} = \tau_\rho v_I + n_{\rho I}$ and assumed vanishing torsion.
Comparing this to the embedding map variation \eqref{eq:NormalVecTrafo9000}, we  find that
\begin{equation}
(\delta_\xi + \delta_X)n_\mu^I=\tilde{\lambda}^I{}_Jn_\mu^J + n_\rho^I\D_\mu\xi^\rho~~,
\end{equation}
where $\tilde{\lambda}^{IJ}=-n_\rho^{[I}n^{J]\nu}\D_\nu \xi^\rho$ is a local $\mathfrak{so}(d-p)$ transformation and we have set $\lambda^{IJ}=0$ in \eqref{eq:NormalVar}. This implies that, up to a $SO(d-p)$ rotation, the normal one-forms $n_\mu^I$ transform as 1-forms under ambient diffeomorphisms. This is the expected result (and analogous to the Lorentzian case \cite{Armas:2017pvj}) as the 1-forms carry a spacetime index $\mu$. Repeating this procedure for the extrinsic curvature, we find that
\begin{equation}
\delta_\xi K_{ab}{^I} = K_{ab}{^\mu}\delta_\xi n^I_\mu + n^I_\mu u^\rho_a  u^\lambda_b\delta_\xi  \Gamma^\mu_{\rho\lambda}~~.\label{eq:variationofK}
\end{equation}
Since $\Gamma^\mu_{\rho\lambda}$ is an affine connection, it transforms in the following way under diffeomorphisms
\begin{equation}
{\delta_\xi \Gamma^\mu_{\lambda\nu} = \xi^\rho\D_\rho\Gamma^\mu_{\lambda\nu} - \Gamma^\rho_{\lambda\nu}\D_\rho \xi^\mu + \Gamma^\mu_{\rho\nu} \D_\lambda\xi^\rho + \Gamma^\mu_{\lambda\rho} \D_\nu \xi^\rho + \D_\lambda\D_\nu 	\xi^\mu= \nabla_{\lambda}  \nabla_{\nu} \xi^\mu - \xi^\rho  R_{\rho \lambda\nu}{^\mu}~~, }\label{eq:DiffTrans}
\end{equation}
where in the second equality we assumed vanishing torsion. This implies that
\begin{eqnarray}
\delta_\xi K_{ab}{^I} &=&  n_\mu^I D_{a} D_{b} \xi^\mu - \frac{1}{2}n^I_\mu K_{ab}{^\sigma} \nabla_\sigma \xi^\mu  + \frac{1}{2}K_{ab J}n_\rho^J n^{I\nu}\nabla_\nu \xi^\rho- n_\mu^I u^\lambda_a u^\nu_b \xi^\rho R_{\rho\lambda\nu}{^\mu}\nn\\
&=&  n_\mu^I D_{a} D_{b} \xi^\mu -  \xi^\rho R_{\rho ab}{^I} - K_{ab J} n_\rho^{[I}n^{J]\nu } \nabla_\nu \xi^\rho~~.
\end{eqnarray}
Comparing this to \eqref{eq:VarOfK}, we obtain
\begin{equation}
(\delta_X+\delta_\xi)K_{ab}{^I}=\tilde{\lambda}^{IJ}K_{ab J}~~,
\end{equation}
which, as in the Lorentzian case \cite{Armas:2017pvj}, states that the extrinsic curvature transforms as a scalar under ambient diffeomorphisms up to a transverse rotation.


\subsection{Action principle and equations of motion}\label{ref:Dynamics}

Equipped with the variational technology of the previous section, we consider the dynamics of submanifolds that arise via the extremisation of an action. In the context of soft matter systems this action can be interpreted as a free energy functional that depends on geometrical degrees of freedom. Examples of such systems are fluid membranes and lipid visicles, described by Canham-Helfrich type free energies. The equations of motion that arise from extremisation naturally split into tangential energy and mass-momentum conservation equations in addition to the shape equation (which describes the mechanical balance of forces in the normal directions), as well as constraints (Ward identities) arising from $SO(d-p)$ rotational invariance and boundary conditions.

\subsubsection{Equations of motion \textit{\&} rotational invariance}\label{sec:EoMs}
Following \cite{Armas:2017pvj}, we consider an action $S$ on a $(p+1)$-dimensional NC submanifold that is a functional of the metric data $\tau_a,\bar h_{ab}$ (this set contains all the fields $\tau_a, h_{ab}, m_a$ and is an equivalent choice of NC objects) as well as the extrinsic curvature, that is $S = S[\tau_a,\bar h_{ab},K_{ab}{^I}]$. The variation of this action takes the general form
\begin{equation} \label{eq:NCact}
\delta S = \int_\Sigma \diff^{p+1}\sigma e\left(\mathcal{T}^a \delta \tau_a + \frac{1}{2}\mathcal{T}^{ab}\delta \bar h_{ab} + \mathcal{D}^{ab}{_I}\delta K_{ab}{^I} \right)\,.
\end{equation}
Here $e$ is the integration measure given by $e=\sqrt{-\text{det}\,(-\tau_a\tau_b+h_{ab})}$ and invariant under local Galilean boosts and $U(1)$ gauge transformations. The response $\mathcal{T}^a$ is the energy current,\footnote{As mentioned throughout this paper, we have focused on the case of vanishing torsion $\tau_{\mu\nu}=0$, meaning that $\tau_a=\partial_a T$, where $T$ is some scalar. Therefore, varying $\tau_a$ is actually varying $T$ in \eqref{eq:NCact}, which in turn implies that we are not able to extract $\mathcal{T}^a$ from the action but only its divergence. This is sufficient for the purposes of this work.} while the response $\mathcal{T}^{ab}$ is the Cauchy stress-mass tensor \cite{Geracie:2016dpu}. Finally, $\mathcal{D}^{ab}{_I}$ is the bending moment, encoding elastic responses, and typically takes the form of an elasticity tensor contracted with the extrinsic curvature (strain) \cite{Armas:2013hsa,Armas:2017pvj}. Both $\mathcal{T}^{ab}$ and $\mathcal{D}^{ab}{_I}$ are symmetric as they inherit the symmetry properties of $\bar h_{ab}$ and $K_{ab}{^I}$. The temporal projection of the Cauchy stress-mass tensor, $\tau_b\mathcal{T}^{ab}$, is the mass current. 

We require the action \eqref{eq:NCact} to be invariant under $U(1)$ gauge transformations for which $\delta_\sigma \bar h_{ab}= -2\tau_{(a}\D_{b)}\sigma$ and invariant under $SO(d-p)$ rotations for which the extrinsic curvature transforms according to \eqref{eq:deltaKundernormalvectrafo}. Ignoring boundary terms, to be dealt with in section \ref{sec:bdarycond}, this leads to mass conservation and a constraint on the bending moment, respectively
\begin{equation}
\label{eq:RotWI}
 D_b\left(\mathcal{T}^{ab}\tau_a\right)=0~~,~~\mathcal{D}^{ab[I}K_{ab}{^{J]}}=0~~.
\end{equation}
In particular, the latter condition takes exactly the same form as in the Lorentzian context \cite{Armas:2013hsa,Armas:2017pvj} and can also be obtained by performing a Lagrangian variation of \eqref{eq:NCact} as we shall see. In order to obtain the equations of motion arising from \eqref{eq:NCact}, we can perform a Lagrangian variation as originally considered in \cite{Carter:1993wy,Carter:1997pb} and developed further in \cite{Armas:2017pvj}.\footnote{Alternatively, we may perform embedding map variations.} Under a Lagrangian variation, using section \ref{sec:LagVars}, the action \eqref{eq:NCact} varies according to
\begin{eqnarray}
\delta_\xi S &=& \int_\Sigma \diff^{p+1}\sigma~e\xi^\rho\bigg[  -\tau_\rho D_a\mathcal{T}^a - D_{a}\left(\bar{{h}}_{\rho b} \mathcal{T}^{ab} \right) - \mathcal{T}^{ab} \left\{ \tau_a\tau_b\D_\rho \tilde{\Phi} + \tau_\rho \tau_a\D_b \tilde{\Phi} + \tau_a\bar{\mathcal{K}}_{b\rho} \right\}\nn\\
&& +  D_a D_b \left(\mathcal{D}^{ab}{_I}n^I_\rho \right) - \mathcal{D}^{ab}{_I} R_{\rho ab}{^I} \bigg] \nn\\
&& + \int_\Sigma \diff^{p+1}\sigma~e  D_a \left[\mathcal{T}^a \tau_\rho \xi^\rho + \mathcal{T}^{ab} \bar h_{\rho b}\xi^\rho + \mathcal{D}^{ab}{_I}n^I_\rho  D_b \xi^\rho -  D_b\left(\mathcal{D}^{ab}{_I}n^I_\rho \right)\xi^\rho  \right]\nn\\
&& +\int_\Sigma \diff^{p+1}\sigma~e \mathcal{D}^{ab}{_I}K_{ab J}n_\rho^{[I}n^{J]\sigma} \nabla_\sigma \xi^\rho~~.\label{eq:varaction}
\end{eqnarray}
In this equation, the second integral gives rise to a boundary term which we consider in section \eqref{sec:bdarycond}. The last integral vanishes due to the requirement of rotational invariance \eqref{eq:RotWI}. However, even if \eqref{eq:RotWI} was not imposed, given that the last term involves a normal derivative of $\xi_\mu$, it cannot be integrated out and hence must vanish independently giving again rise to the second condition in \eqref{eq:RotWI}, as in the Lorentzian case \cite{Armas:2017pvj}. 

The first integral in \eqref{eq:varaction} must vanish for an arbitrary vector field $\xi^\mu$ and hence it gives rise to the equation of motion
\begin{eqnarray}
 &&-\tau_\rho D_a\mathcal{T}^a - \bar h_{\rho b} D_a \mathcal{T}^{ab} - \mathcal{T}^{ab}\bar h_{\rho\sigma} K_{ab}{^\sigma} + 2\tau_\rho \mathcal{T}^{ab}\tau_a \D_b\tilde{\Phi} + \tau_\rho \bar{\mathcal{K}}_{ab}\mathcal{T}^{ab} \nn\\
&&+  D_a  D_b (\mathcal{D}^{ab}{_I}n^I_\rho) - \mathcal{D}^{ab}{_I} R_{\rho ab}{^I}=0~~,\label{eq:FullEoM}
\end{eqnarray}
where we have used \eqref{eq:CovDofbarh}. In appendix \ref{app:reduction} we provide the relation \eqref{eq:UsefulRel9000} between $\bar{\mathcal{K}}_{ab}$, which is the pullback of $\bar{\mathcal{K}}_{\mu\nu}$, and $\bar{\mathcal{K}}^\Sigma_{ab}=-\pounds_{\hat v}^\Sigma \bar h_{ab}/2$ which is the actual surface-equivalent of $\bar{\mathcal{K}}_{\mu\nu}$. Here $\pounds_{\hat v}^\Sigma$ denotes the surface Lie derivative along $\hat v^a$. Using this relation, as well as \eqref{eq:CheckPhi}, which relates the Newtonian potential on the submanifold $\check\Phi$ to its ambient spacetime counterpart $\tilde{\Phi}$, the equation of motion \eqref{eq:FullEoM} can be written as
\begin{eqnarray}
 &&\tau_\rho  D_a\mathcal{T}^a + \bar h_{\rho b} D_a \mathcal{T}^{ab} + \mathcal{T}^{ab}\bar h_{\rho\sigma} K_{ab}{^\sigma} - 2\tau_\rho \mathcal{T}^{ab}\tau_a \D_b\check{\Phi} - \tau_\rho \bar{\mathcal{K}}^\Sigma_{ab}\mathcal{T}^{ab}\label{eq:FullEoM2}\nn\\
 &&-\tau_\rho \hat v^I K_{ab}{^I}\mathcal{T}^{ab}-  D_a D_b\left(\mathcal{D}^{ab}{_I}n^I_\rho\right) + \mathcal{D}^{ab}{_I} R_{\rho ab}{^I}=0~~.
\end{eqnarray}
The equation of motion \eqref{eq:FullEoM2} can be projected tangentially or orthogonally to $\Sigma$, yielding two independent equations. The tangential projection, known as the intrinsic equation of motion, is given by
\begin{equation}
\tau_c\left[ D_a\left(\mathcal{T}^a - 2\check{\Phi}\mathcal{T}^{ab}\tau_b \right) - \mathcal{T}^{ab} \bar{\mathcal{K}}^\Sigma_{ab} \right] + \bar h_{bc}  D_a\mathcal{T}^{ab} + 2 D_a\left(K_{bc}{^I}\mathcal{D}^{ab}{_I}\right) - \mathcal{D}^{ab}{_I} D_c K_{ab}{^I}=0~~,
\label{eq:intrinsicEoM}
\end{equation}
where we have used the Codazzi--Mainardi equation \eqref{eq:Codazzi-Mainardi}, assuming vanishing torsion, in order to eliminate contractions with the Riemann tensor. Eq.~\eqref{eq:intrinsicEoM} can be further projected along $h^{cd}$ and $\hat v^c$, which again yields two independent equations. These projections can be simplified by defining $\mathcal{T}_{\text{m}}^{ad}=\mathcal{T}^{ad} + 2\mathcal{D}^{b(a}{_I}h^{d)c}K_{bc}{^I}$ and $\mathcal{T}^a_{\text{m}}=\mathcal{T}^a - 2\hat v^c K_{bc}{^I}\mathcal{D}^{ab}{_I}$. 
In particular, the spatial projection using $h^{cd}$ gives rise to mass and momentum conservation
\begin{equation}
D_a\mathcal{T}_{\text{m}}^{ad} + 2 D_a\left(\mathcal{D}^{b[a}{_I}h^{d]c}K_{bc}{^I}\right) - h^{cd}\mathcal{D}^{ab}{_I} D_c K_{ab}{^I}=0~~,\label{eq:intrinsicEoM1}
\end{equation}
where we have used invariance under $U(1)$ gauge transformations (the first condition in \eqref{eq:RotWI}). In turn, the projection along $\hat v^c$ leads to energy conservation
\begin{equation}
D_a\mathcal{T}_m^a - \mathcal{T}^{ab}_{\text{m}} \bar{\mathcal{K}}^\Sigma_{ab}  - 2 \mathcal{T}^{ab}_{\text{m}}\tau_b \D_a \check\Phi + \mathcal{D}^{ab}{_I} \hat{v}^c  D_c K_{ab}{^I}=0~~,\label{eq:TemporalProj}
\end{equation}
where we have used the identity $ D_a \hat v^c = -h^{cd}\left(\bar{\mathcal{K}}^\Sigma_{ad} + \tau_a\D_d\check\Phi \right)$ as well as the first condition in \eqref{eq:RotWI}.

The intrinsic equations \eqref{eq:intrinsicEoM1} and \eqref{eq:TemporalProj} result from diffeomorphism invariance along the tangential directions $\xi^a=u_\mu^a\xi^\mu$ or, equivalently, from tangential reparametrisation invariance $\delta X^\mu=u^\mu_a\delta X^a$. Since the action only depends on the NC objects $\tau_a$, $\bar h_{ab}$ and ${K_{ab}}^I$, the intrinsic equations are nothing but Bianchi identities that result from the diffeomorphism invariance of the action and hence are identically satisfied.

Finally, the transverse projection of \eqref{eq:FullEoM2} is usually referred to as the \textit{shape equation} and it is given by
\begin{equation}
\mathcal{T}^{ab}K_{ab}{^I}={\mathfrak D}_a {\mathfrak D}_b \mathcal{D}^{ab I} - \mathcal{D}^{ab}{_J} K_{ac}{^I}K_{bd}{^J}h^{cd} -\mathcal{D}^{ab}{_J}{R}_{Iab}{^J}\,,\label{eq:ShapeEq1}
\end{equation}
where we have used the covariant derivative $\mathfrak{D}_a$ introduced in \eqref{eq:Weingart1}. Eq.~\eqref{eq:ShapeEq1} is valid in the absence of torsion and takes the exact same form as its Lorentzian counterpart \cite{Armas:2013hsa,Armas:2017pvj} and it is a non-trivial dynamical equation that determines the set of embedding functions $n^I_\mu X^\mu$. This equation, which is one of the main results of the paper, appears extensively in the context of lipid vesicles (see e.g. \cite{Guven2018}) but without time-components. 


\subsubsection{Boundary conditions}\label{sec:bdarycond}
In the previous section we considered the equations of motion arising from \eqref{eq:NCact} on $\Sigma$. In this section we consider the possibility of such submanifolds having a boundary. In such cases, the second integral in \eqref{eq:varaction} is non-trivial and gives rise to a non-trivial boundary term that must vanish, namely 
\begin{equation}
\int_{\D\Sigma }\diff^{p}y~e_\D \eta_a \left[\left(\mathcal{T}^a \tau_\rho + \mathcal{T}^{ab} \bar h_{\rho b}  - D_b\mathcal{D}^{ab}{_\rho} - \mathcal{D}^{ab}{_I}   D_b n^I_\rho\right)\xi^\rho + \mathcal{D}^{ab}{_I} D_b \xi^I \right]=0~~,
\end{equation}
where $\eta_a$ is a normal co-vector to the boundary while $e_\D$ is the integration measure on $\D\Sigma$ (parameterised by $y$). With the help of the boundary completeness relation $\Pi^c_b = \delta^c_b - \eta_b\eta^c$
where $\eta^c=h^{cd}\eta_d$, the boundary term can be rewritten as 
\begin{align}
&\int_{\D\Sigma}\diff^p y~e_\D \eta_a \eta_b \mathcal{D}^{ab}{_I}\eta^c\D_c \xi^I \nonumber \\
&+ \int_{\D\Sigma}\diff^p y~ e_\D \eta_a\left[\left(\mathcal{T}^a \tau_\rho + \mathcal{T}^{ab} \bar h_{\rho b}  - D_b\left(\mathcal{D}^{ab}{_I}n^I_\rho\right) - \mathcal{D}^{ab}{_I}  D_b n^I_\rho \right)\xi^\rho +\Pi^c_b\mathcal{D}^{ab}{_I} \D_c \xi^I\right]=0~~.
\label{eq:lastterm}
\end{align}

As in the case of the bulk equations of motion on $\Sigma$, normal derivatives to the boundary of the form $\eta^c\D_c\xi^I$ cannot be integrated out. Hence the above equation splits into two independent conditions
\begin{eqnarray}
\label{eq:bdy1}
\hspace{-.5cm}\eta_a \eta_b \mathcal{D}^{ab}{_I}\big\vert_{\D\Sigma}&=&0~~,\label{eq:FirstBdaryCond}~~~\\
\hspace{-.5cm}\left[\eta_a\left(\mathcal{T}^a \tau_\rho + \mathcal{T}^{ab} \bar h_{\rho b}  -   D_b\left(\mathcal{D}^{ab}{_I}n^I_\rho\right) - \mathcal{D}^{ab}{_I}  D_b n^I_\rho\right) - n^I_\rho  \Pi^d_c D_d\left(\eta_a\mathcal{D}^{ab}{_I}\Pi^c_b\right)\right]\!\!\big\vert_{\D\Sigma}&=&0~~~.
\end{eqnarray}
The first boundary condition in \eqref{eq:bdy1} is a consequence of $SO(d-p)$ invariance of the action and can also be derived by keeping track of boundary terms when using \eqref{eq:deltaKundernormalvectrafo} in \eqref{eq:NCact}. The second of these conditions can be projected tangentially and transversely to $\Sigma$, yielding respectively
\begin{equation}\label{eq:bdrycond}
\eta_a\left[\mathcal{T}^a\tau_c +\mathcal{T}^{ab}\bar h_{bc} + 2\mathcal{D}^{ab}{_I}K_{bc}{^I}\right]\!\!\big\vert_{\D\Sigma}=0~~,~~\left[\mathcal{D}^{ab}{_J\Pi^c_b} D_c\eta_a -2{\mathfrak{D}}_b(\eta_a\mathcal{D}^{ab}{_J}) \right]\!\!\big\vert_{\D\Sigma}=0~~,
\end{equation}
where we have used the first boundary condition \eqref{eq:FirstBdaryCond} as well as $\eta_a\tau_b\mathcal{T}^{ab}\big\vert_{\D\Sigma}=0$ which is a consequence of the $U(1)$ invariance of \eqref{eq:NCact}.
These boundary conditions can be further projected along $h^{cd}$ and $\hat v^c$, leading to
\begin{equation}
\left[\eta_a\mathcal{T}^{ad}_m+2\eta_a \mathcal{D}^{b[a}{}_I h^{d]c}K_{bc}{}^I\right]\bigg\vert_{\D\Sigma}=0~~,~~\eta_a\mathcal{T}^a_m\bigg\vert_{\D\Sigma}= 0~~,
\end{equation}
where $\mathcal{T}_m^{ad}$ and $\mathcal{T}_m^a$ were introduced in \eqref{eq:intrinsicEoM1} and \eqref{eq:TemporalProj}, respectively. This completes the analysis of the equations of motion and its boundary conditions. In the specific examples below, however, we will not consider the presence of boundaries.


\section{Applications to soft matter systems} \label{sec:applications}
In this section we apply the action formalism in order to describe equilibrium fluid membranes and lipid vesicles as well as their fluctuations. These systems are such that their deformations, at mesoscopic scales, are described by purely geometric degrees of freedom (see e.g. \cite{Guven2018}) and few material/transport coefficients, such as the bending modulus $\kappa$. The development of Newton-Cartan geometry for surfaces in the previous sections brings several advantages to the description of these systems. Firstly, it introduces absolute time and therefore fluctuations of the system can include temporal dynamics in a covariant form. Secondly, the symmetries of the problem are manifested via the geometry of the submanifold/ambient spacetime.\footnote{This point is reminiscent of the strategy adopted by Son \textit{et al.} in \cite{Son:2013rqa,Geracie:2014nka,Geracie:2014zha} where the authors take advantage of the fact that Newton--Cartan geometry is the natural geometric arena for the effective description of the fractional quantum Hall effect. In this way, by coupling a suitable field theory to Newton--Cartan geometry, information about correlation functions involving mass, energy and momentum currents can be extracted via geometric considerations.} 

More importantly, however, is perhaps the fact that NC geometry allows to properly introduce thermal field theory of equilibrium fluid membranes. Material coefficients such as $\kappa$ are functions of the temperature $T$ (see e.g. ~\cite{Terzi2018}) but also of the mass density $\mu$. However, the fact that $T$ and $\mu$ can be given a geometric interpretation, via the hydrostatic partition function approach, in which case they are associated with the existence of a background isometry (or timelike Killing vector field), is disregarded in all models of lipid vesicles. However this approach is required in order to understand the correct equations that describe fluctuations. We begin with a simple fluid membrane with only surface tension in order to elucidate these fundamental aspects and end with a generalisation of the Canham-Helfrich model.


\subsection{Fluid membranes}\label{sec:Fluids}
In this section we consider equilibrium fluid membranes, by which we mean stationary fluid configurations that live on some arbitrary surface.\footnote{We follow previous constructions of relativistic \cite{PartitionFunction1,PartitionFunction2, Armas:2013hsa, Armas:2015ssd} and non-relativistic fluids \cite{Jensen:2014ama, Banerjee:2015hra}.} As mentioned above, equilibrium requires the existence of an ambient timelike Killing vector field $k^\mu$ such that the fluid configuration is time-independent. In general, since we wish to describe fluids that are rotating or boosted along some directions, equilibrium requires the existence of a set of symmetry parameters $K=(k^\mu, \lambda_\mu^K, \Lambda^K)$ such that the transformation on the NC triplet (cf. eqs. \eqref{eq:TNCgaugetrafos} and \eqref{eq:hhtrans}) vanishes, that is
\be \label{eq:t1}
\pounds_k \tau_\mu=0~~,~~\pounds_k \bar h_{\mu\nu}=2\tau_{(\mu}\pounds_k m_{\nu)}+2\tau_{(\mu}\partial_{\nu)}\Lambda^K~~,~~\pounds_k m_\mu+\lambda^K_\mu+\partial_\mu\Lambda^K=0~~,
\ee
and whose pullback $k^a=u^a_\mu k^\mu$ is also a submanifold Killing vector field satisfying the relations
\be \label{eq:t2}
\pounds_k \tau_a=0~~,~~\pounds_k \bar h_{ab}=2\tau_{(a}\pounds_k \check m_{b)}+2\tau_{(a}\partial_{b)}\Lambda^K~~,~~\pounds_k \check m_a+\check\lambda^K_a+\partial_a\Lambda^K=0~~.
\ee
These relations make sure that the space in which the fluid lives does not depend on time.

The simplest example of $k^\mu$ in flat NC space \eqref{eq:flatspace} is the case of a static Killing vector where $k^\mu=\delta^\mu_t$.\footnote{Specific surfaces where the fluid lives, besides a timelike isometry, may have additional translational or rotational isometries. In such situations the Killing vector $k^\mu$ can have components along those spatial directions. The chemical potential $\mu$ introduced in \eqref{eq:Tmu} captures the spatial norm of the Killing vector, which is associated with the presence of linear or angular momenta.}  Since the fluid is in equilibrium, it is straightforward to construct an Euclidean free energy\footnote{This is also referred to as hydrostatic partition function $-i\ln \mathcal{Z}=T_0\mathcal{F}$ \cite{PartitionFunction1,PartitionFunction2}.} from the action $S$ by Wick rotation $t\to i t$, compactification of $t$ with period $1/T_0$ and integration over the time circle, where $T_0$ is the constant global temperature. This means that the Euclidean free energy $\mathcal{F}$ is given by
\be \label{eq:free}
\mathcal{F}[\tau_a,\bar h_{ab},{K_{ab}}^{I}]=T_0 S_{t\to it}~~.
\ee

Given the transformations \eqref{eq:t1}--\eqref{eq:t2}, the free energy can depend on two scalars, namely the local temperature $T$ and chemical potential $\mu$ (associated with particle number conservation) defined in terms of the symmetry parameters as
\be \label{eq:Tmu}
T=\frac{T_0}{k^a\tau_a}~~,~~\frac{\mu}{T}=\frac{\Lambda^K}{T_0}+\frac{1}{2T}\bar h_{ab}u^au^b~~,~~u^b=\frac{k^b}{k^a\tau_a}~~,
\ee
where $u^\mu$ is the fluid velocity.\footnote{The free energy considered here only depends on geometric quantities such as $T$ and $\mu$, where the Killing vector $K^\mu$ and the gauge parameter $\Lambda^K$ solve \eqref{eq:t2}. It is possible to promote the free energy to an effective action that does not require time-independence by treating $S$ as also being dependent on a arbitrary vector $\beta^\mu$ and gauge parameter $\Lambda$ (see \cite{Haehl:2015pja}). } We will now look at different cases.

\subsubsection{Surface tension}\label{sec:Fluidtension}
The simplest example of a fluid membrane is one in which the action only depends on the surface tension $\chi(T,\mu)$. Such an action describes, for instance, soap films. Thus the free energy \eqref{eq:free} takes the form
\be \label{eq:freesurface}
\mathcal{F}=\int_{\Sigma_s} d^{p}\sigma e_s~\chi (T,\mu)~~,
\ee
where $\Sigma_s$ and $e_s$ denote the spatial part of $\Sigma$ and the volume form $e$, respectively, due to integration over the time direction. We can now use \eqref{eq:NCact} to extract the currents at fixed symmetry parameters. It is useful to explicitly evaluate the variations
\be
\delta T=-T u^{a}\delta \tau_a~~,~~\delta \mu=\frac{\Lambda^K}{T_0}\delta T +\frac{1}{2}u^{a}u^{b}\delta \bar h_{ab} + \bar u^2\frac{\delta T}{T}~~,
\ee
where we have defined $\bar u^2=\bar h_{ab}u^{a}u^{b}$. This allows us to derive the variation of the surface tension as 
\be \label{eq:thermo1}
\delta\chi=s \delta T+n \delta \mu=-\left(Ts+n\mu+\frac{n}{2}\bar u^2\right)u^a\delta \tau_a+\frac{n}{2} u^{a}u^{b}\bar h_{ab}~~,
\ee
where we have defined the surface entropy density and surface particle number density (mass density) as
\be
s=\left(\frac{\partial \chi}{\partial T}\right)_{\mu}~~,~~n=\left(\frac{\partial \chi}{\partial \mu}\right)_{T}~~.
\ee
From \eqref{eq:thermo1} we also directly extract the Gibbs-Duhem relation $d\chi=s dT+nd\mu$. Using \eqref{eq:thermo1} we also determine the currents
\be\label{eq:currents1}
\mathcal{T}^a=-\chi \hat v^a-\left(\varepsilon+\chi+\frac{n}{2}\bar u^2\right)u^a~~,~~\mathcal{T}^{ab}=\chi h^{ab}+nu^{a}u^{b}~~,
\ee
where we have defined the internal energy $\varepsilon$ via the Euler relation $\varepsilon+\chi=Ts+n\mu$. This defines the constitutive relations of a Galilean fluid living on a submanifold in an ambient NC spacetime. Using the stress-mass tensor in \eqref{eq:currents1}, the non-trivial shape equation \eqref{eq:ShapeEq1} in the absence of bending moment becomes
\be \label{eq:shapetension}
\mathcal{T}^{ab}{K}_{ab}^{I}=0~~\Rightarrow~~\chi K^I+ n u^{a}u^{b}K_{ab}^I=0~~.
\ee

Physically relevant fluid membranes are co-dimension one and so we can omit the transverse index $I$. The shape equation \eqref{eq:shapetension} expresses the balance of forces between the surface tension $\chi K$ (normal stress) and the normal acceleration $n u^{a}u^{b}K_{ab}$ of the fluid.\footnote{Using the definition of extrinsic curvature \eqref{eq:ExtCurv}, we can rewrite $u^{a}u^{b}K_{ab}^I=n_\mu^I u^\nu \nabla_\nu u^\mu$. Hence the second term in \eqref{eq:shapetension} is in fact the normal component of the acceleration of the fluid $u^\nu \nabla_\nu u^\mu$ where $u^\mu = u^\mu_a u^a$. If the fluid is rotating along the surface, this term gives rise to centrifugal acceleration. } If we would consider a surface tension with no dependence on the temperature and chemical potential, then $n=0$ and the shape equation reduces to the equation of a minimal surface. To complete the thermodynamic interpretation of \eqref{eq:freesurface}, we note that varying the free energy with respect to the global temperature $T_0$ gives rise to the global entropy
\be
\mathcal{S}=\frac{\partial\mathcal{F}}{\partial T_0}=\int_{\Sigma_s} d^{p}\sigma e_s~\frac{s}{k^a\tau_a}=\int_{\Sigma_s} d^{p}\sigma e_s~su^a t_a~~,
\ee
where we have defined the timelike vector $t_a=\tau_a/(k^b\tau_b)$, and where $s u^a$ is the entropy current.

\subsubsection{Surface fluctuations: Elastic waves}\label{sec:elastic waves}
The shape equation \eqref{eq:shapetension} describes equilibrium configurations of fluid membranes in the absence of any bending moment. We consider a fluid at rest in the simplest scenario of a surface with 2 spatial dimensions embedded in a NC spacetime with 3 spatial dimensions such that $ \tau_a=\delta_a^t$ where $a=t,1,2$. The fluid thus has a velocity $u^a=(1,0,0)$. Such a trivial time embedding, $ \tau_a=\delta_a^t$, is typically the most physically relevant setting for soft matter applications. In this context, we have that $u^{a}u^{b}K_{ab}=0$ since $K_{tb}=0$ trivially. Thus, the second term in \eqref{eq:shapetension} does not contribute in equilibrium and it is acceptable to simply ignore the fact that the surface tension depends on the temperature and chemical potential. However, if one is interested in fluctuations away from equilibrium, the second term in \eqref{eq:shapetension} cannot be ignored. Here we consider the simplest case where the surface is flat and hence also trivially embedded in space such that
\be
h_{ab}=\delta_a^i\delta_b^i~~,~~m_a=0~~,~~n_\mu = \delta_\mu^3~~.
\ee
This is an equilibrium configuration that trivially solves \eqref{eq:shapetension} since $K_{ab}=0$. 

We now consider a small fluctuation of the embedding map along the normal direction $X^3=X^\perp$. Using \eqref{eq:VarOfK} we find
\be \label{eq:wave}
\delta_X\mathcal{T}^{ab}{K}_{ab} + \mathcal{T}^{ab}\delta_X{K}_{ab} = \left(\chi h^{ab}+nu^{a}u^{b}\right)\partial_a\partial_b\xi^\perp=0~~,
\ee
where we have used that $K_{ab}=0$ to eliminate the first term and converted ${\mathfrak D}_a\to\partial_a$ as we are dealing with a flat surface in a flat ambient space. Eq.~\eqref{eq:wave} is a wave equation, and considering wave-like solutions of the form $\xi^\perp \sim e^{-i\omega t +i( k_1 \sigma_1+k_2\sigma_2)}$ one finds the linear dispersion relation
\be \label{eq:dispersion}
\omega=\pm\sqrt{\frac{-\chi}{n}}k~~,
\ee
where $\omega$ is the frequency, $k_1, k_2$ are wavenumbers and $k^2=k_1^2+k_2^2$.\footnote{Note that in order to match conventions with the classical literature one should redefine $\chi\to-\chi$.} This is the classical answer for the oscillations of uniform elastic sheets (see e.g. \cite{landau1989theory}). 

This result shows the importance of considering NC geometry in the theory of fluid membranes, since omitting the dependence of the surface tension on the temperature and chemical potential would not have allowed for the derivation of \eqref{eq:dispersion}. We note that the result \eqref{eq:dispersion} is valid for any type of elastic membrane with mass density and does not require any "flow" on the membrane, in particular the initial equilibrium configuration was static $u^a=(1,0,0)$.\footnote{If one was describing an elastic material, the surface tension would also be dependent on the Goldstone modes of broken translations and hence on intrinsic elastic moduli.} In a future publication, we will consider a more general analysis of fluctuations of fluid membranes which will also include the Canham-Helfrich model \cite{upcoming}.

\subsubsection{Droplets}\label{sec:FluidsOnSurfaces}
Here we briefly consider the case of a droplet (or soap bubble) in which the fluid membrane encloses some volume with uniform internal pressure $P_{\text{int}}$ separating it from an exterior medium with uniform external pressure $P_{\text{ext}}$.
In order to describe these situations we augment the action with the bulk pieces
\begin{equation}
S_{\text{bulk}} =  \int_{\text{int}(\Sigma)}\diff^{d+1}x~e_{\text{b}}P_{\text{int}} + \int_{\text{ext}(\Sigma)}\diff^{d+1}x~e_{\text{b}}P_{\text{ext}}~~,\label{eq:Underpressure}
\end{equation}
where $e_{\text{b}}$ is the bulk measure, $\text{int}(\Sigma)$ is the interior of the closed surface $\Sigma$ \footnote{By a closed surface we mean a NC submanifold whose constant time slices are closed.}, whereas $\text{ext}(\Sigma)$ is the exterior region of the bulk outside the surface. The variation of the density $e_{\text{b}}$ with respect to a bulk (or ambient spacetime) diffeomorphism reads
\begin{equation}
\delta_\xi e_{\text{b}}	=\D_\mu(e_{\text{b}}\xi^\mu)~~,
\end{equation}
which, using Stokes theorem, implies that the variation takes the form
\begin{equation}
\delta_\xi S_{\text{bulk}}=-\Delta p \int_\Sigma\diff^d\sigma \,n_\mu \xi^\mu~~,\label{eq:pressurevariation}
\end{equation}
where $\Delta p = P_{\text{ext}}-P_{\text{int}}$ is the constant pressure difference across the surface $\Sigma$.\footnote{In order to describe gases or fluids in the interior/exterior, one should consider the dependence of internal/external pressures on bulk temperature and chemical potential as in \cite{Armas:2016xxg}.} In a biophysical context, where the pressure difference is attributable to two different chemical solutions separated by a semi-permeable membrane, this pressure is the \textit{osmotic pressure} \cite{PhysRevE.54.2816}. 

From \eqref{eq:pressurevariation}, we deduce that $S_{\text{bulk}}$ does not contribute to the intrinsic equations of motion, while it adds the constant term $-\Delta p$ to the shape equation \eqref{eq:shapetension} such that 
\begin{equation}
\mathcal{T}^{ab}K_{ab}=\chi K+n u^a u^bK_{ab}=-\Delta p~~.
\end{equation}
This is a generalisation of the Young-Laplace equation, which includes the possibility of the fluid having non-trivial acceleration, and was first derived in \cite{Armas:2016xxg} in the context of null reduction.


\subsection{The Canham--Helfrich model revisited}\label{sec:EquilibriumSurfaces}
In this section we consider a more elaborate case of fluid membranes, namely that of the Canham-Helfrich model \cite{Canham1970, Helfrich:1973}. This model describes equilibrium configurations of biophysical membranes (see e.g.~\cite{Tu:2014}) comprised of a phospolipid bilayer \cite{CellMembranes}, and captures several shapes of biophysical interest \cite{Tu:2014}, namely the sphere (corresponding to spherical vesicles such as liposomes), the torus (toroidal vesicles) and the biconcave discoid (the red blood cell or \textit{erythrocyte}). This model includes, besides the presence of a surface tension $\chi$, also the bending modulus $\kappa$ that incorporates the bending energy of the membrane. We show how to describe this model within Newton--Cartan geometry and generalise it by allowing the material parameters to be functions of $T,\mu$. We also review the family of classical lipid vesicles (spherical, toroidal, discoid) within this framework. We leave a more detailed analysis of this model and its generalisations to a future publication \cite{upcoming}.

\subsubsection{Generalised Canham--Helfrich model}\label{sec:Helfrich}
The Canham-Helfrich model contains quadratic terms in the extrinsic curvature and a set of material coefficients. It describes lipid vesicles in thermal equilibrium. As in the previous section, a proper description of such systems requires taking into account the dependence of the material coefficients on the temperature and chemical potential. As a starting point we take the more general free energy
\begin{equation}
\mathcal{F}_{\text{CH}}= \int_{\Sigma_s} \diff^{p}\sigma {e}_s\big[a_0(T,\mu) + a_1(T,\mu)  K + a_2(T,\mu) K^2+ a_3(T,\mu)K\cdot K \big]~~,\label{eq:model}
\end{equation}
where $\{a_0,a_1,a_2,a_3\}$ is a set of material coefficients characterising the phenomenological specifics of the biophysical system under scrutiny. In the expression above, we have defined $K\cdot K = h^{ac}h^{bc}K_{ab}K_{cd}$. 

It is well known that the last term in \eqref{eq:model} can usually be ignored due to the Gauss-Codazzi equation \eqref{eq:GaussCodazzi} in flat ambient space, as it can be related to the Gaussian curvature of the membrane and hence integrated out for two-dimensional surfaces (see app.~\ref{sec:GaussBonnet} for details). However, this is only possible if $a_3$ is treated as a constant. Since a proper geometric and thermodynamic treatment requires promoting $a_3$ to a non-trivial function of $T,\mu$ this implies that new non-trivial contributions to the equations of motion will appear. Additionally, based solely on effective field theory reasoning, it is possible to augment \eqref{eq:model} with further terms involving the fluid velocity (see \cite{Armas:2013hsa} for the relativistic case). We will leave a thorough analysis of this for the future \cite{upcoming}. Here we focus on extracting the stresses on the membrane using \eqref{eq:NCact}. 

We find the energy current
\be\label{eq:T1}
\mathcal{T}^a=\!-\!\left(a_0+a_1K+a_2K^2+a_3K\cdot K\right)\hat v^a\!\!-\!\! \left(L_0+L_1K+L_2K^2+L_3K\cdot K\right)u^a~,~~~~
\ee
where we have defined the thermodynamic parameters
\be
L_i=Ts_i+n_i\mu+\frac{n_i}{2}\bar u^2~~,~~s_i=\left(\frac{\partial a_i}{\partial T}\right)_\mu~~,~~n_i=\left(\frac{\partial a_i}{\partial \mu}\right)_T~~.
\ee
Similarly, we extract the Cauchy stress-mass tensor
\begin{eqnarray}
\mathcal{T}^{ab} & = & h^{ab}\left(a_0 + a_1 K + a_2 K^2 + a_3 K\cdot K \right) - 2h^{ac}h^{bd}K_{cd}(a_1 + 2a_2 K) \nn\\
&&- 4a_3h^{f d}h^{c a}h^{e b}K_{cd}K_{ef}+\left(n_0+n_1K+n_2K^2+n_3K\cdot K\right)u^{a}u^{b}~~.\label{eq:Explicit1}
\end{eqnarray}
As this model contains terms involving the extrinsic curvature, it has a bending moment of the form
\begin{equation}
\mathcal{D}^{ab}
= a_1 h^{ab} + \mathcal{Y}^{abcd}K_{cd}~~,~~\mathcal{Y}^{abcd}=2a_2 h^{ab}h^{cd} + 2a_3 h^{a(c}h^{d)b}~~,\label{eq:Explicit2}
\end{equation}
where $\mathcal{Y}^{abcd}$ is the Young modulus of the membrane and has the usual symmetries of a classical elasticity tensor.\footnote{This was first introduced in an effective theory for relativistic fluids in Ref.~\cite{Armas:2013hsa}. The Young modulus tensor also appears when considering finite size effects in the dynamics of black branes \cite{Armas:2011uf}.} 
Eqs.~\eqref{eq:Explicit1}--\eqref{eq:Explicit2} tell us that if $a_3$ is a non-trivial function of $T,\mu$, then it will contribute non-trivially to the shape equation \eqref{eq:ShapeEq1}. 

Let us be a bit more precise about the role of $a_3$. First of all we redefine the coefficient $a_2$ as $a_2=\tilde a_2-a_3$ so that $a_3$ now multiplies the integrand of the Gauss--Bonnet term, the Gaussian curvature. All terms proportional to $a_3$ in the shape equation can be shown to cancel identically using a set of identities such as the Codazzi--Mainardi and Gauss--Codazzi equations (i.e. \eqref{eq:Codazzi-Mainardi} and \eqref{eq:GaussCodazzi} suitably adapted to the case of a co-dimension one submanifold) as well as the identity \eqref{eq:no2dEinstein} which expresses the fact that the Einstein tensor of the Riemannian geometry on constant time slices vanishes in two dimensions. This means that $a_3$ will contribute only to the shape equation through its derivatives that we denoted by $s_3$ and $n_3$. There are only two such terms, namely $n_3 K\cdot K u^a u^b K_{ab}$ and $\left(h^{ac}h^{bd}K_{cd}-h^{ab}K\right)D_a D_b a_3$. In particular the latter is interesting since it will make a contribution to the shape equation even in the case of a static fluid.

We now show how the model \eqref{eq:model} recovers the standard Canham-Helfrich model.

\subsubsection{The standard Canham-Helfrich model}\label{sec:cCH}
We focus on three-dimensional flat spacetime \eqref{eq:flatspace} and surfaces with two spatial dimensions. We also assume that the functions $\{a_0,a_1,a_2,a_3\}$ are constant. In this case, as explained above and detailed in app.~\ref{sec:GaussBonnet}, we can set $a_3=0$. Additionally, we require the free energy \eqref{eq:model} to be invariant under a change of the inwards/outwards orientation of normal vectors, that is, invariant under $n^\mu \to  -n^\mu$. This leads to
\begin{equation} \label{eq:reducedmodel}
\mathcal{F}_{\text{CH}} = \int_{\Sigma_s} \diff^{2}\sigma~{e}_s\big[\chi + \kappa (K+ c_0)^2\big] ~~,
\end{equation}
where we have redefined the coefficients such that
\begin{equation}
a_0=\chi+\kappa c_0^2~~,~~ a_1=2\kappa c_0~~,~~a_2=\kappa~~,
\end{equation}
and where $c_0$ changes sign under $n^\mu \to  -n^\mu$.
This is the direct analog of the Canham--Helfrich model of lipid bilayer membranes \cite{Helfrich:1973}. The constant $c_0$ is the \textit{spontaneous curvature}, which reflects a preference to adopt a specific curvature due to e.g. different aqueous environments or lipid densities on the two sides of the bilayer \cite{Rozycki:2015}. The parameter $\chi$ is the surface tension and the parameter $\kappa$ is the \textit{bending modulus} \cite{Tu:2014}. In this case, $s_i=n_i=0$ and the shape equation \eqref{eq:ShapeEq1} upon using \eqref{eq:Explicit1} and \eqref{eq:Explicit2} becomes
\begin{eqnarray}
-a_0 K - a_1K^2 -a_2K^3 +a_1K\cdot K +2 a_2 K(K\cdot K)+2a_2 h^{ab} D_a  D_b K - \Delta p=0~,~~~~
\end{eqnarray}
where we have added the contribution from constant interior/exterior pressures as in sec.~\ref{sec:FluidsOnSurfaces}. We will now review particular solutions to this model

\subsubsection{Biophysical solutions: axisymmetric vesicles}\label{sec:Solutions}
\begin{figure}
\begin{subfigure}{0.31\textwidth}
\includegraphics[width=\linewidth]{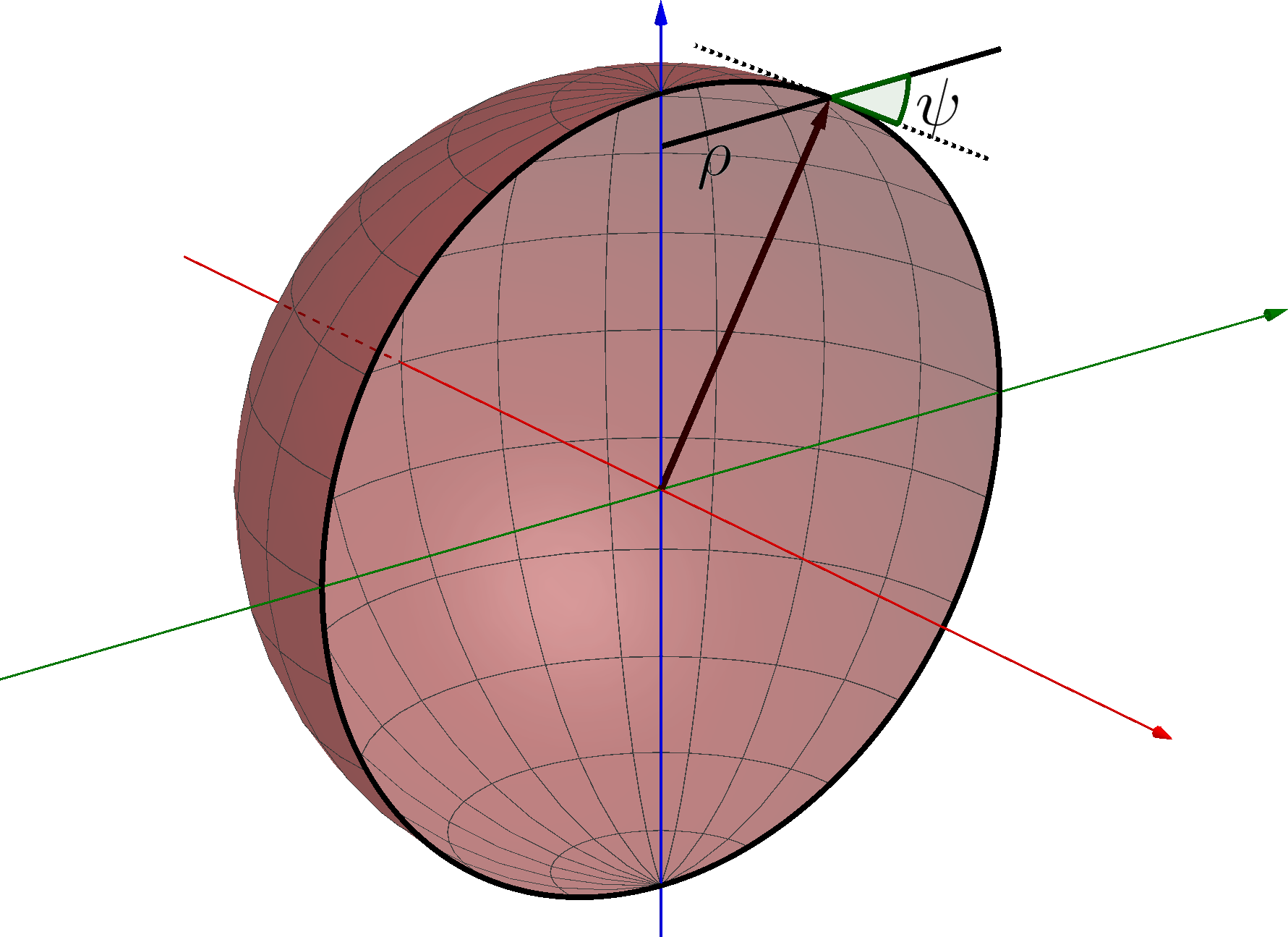}
\caption{Sphere} \label{fig:1a}
\end{subfigure}
\hspace*{\fill} 
\begin{subfigure}{0.31\textwidth}
\includegraphics[width=\linewidth]{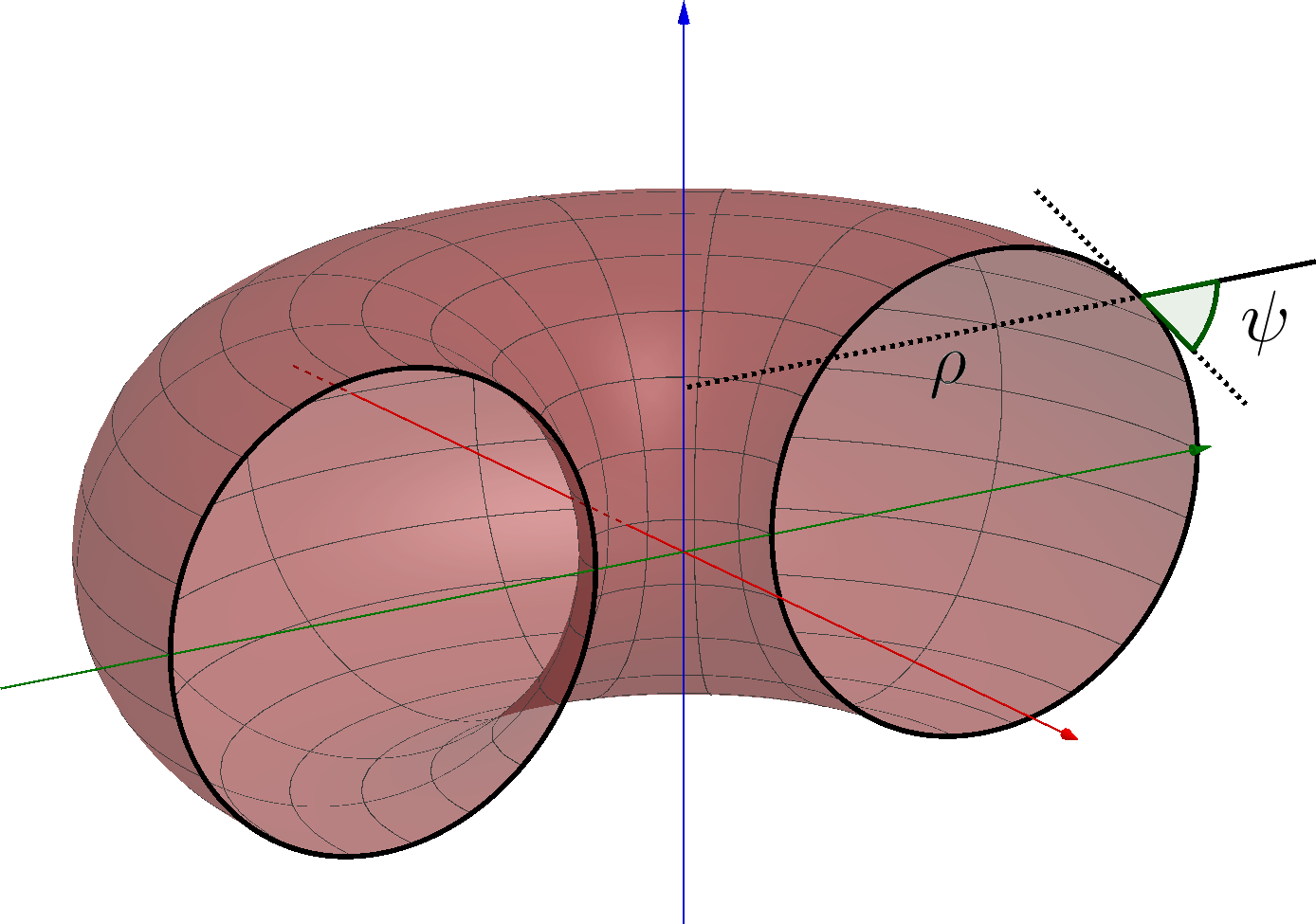}
\caption{Torus} \label{fig:1b}
\end{subfigure}
\hspace*{\fill} 
\begin{subfigure}{0.31\textwidth}
\includegraphics[width=\linewidth]{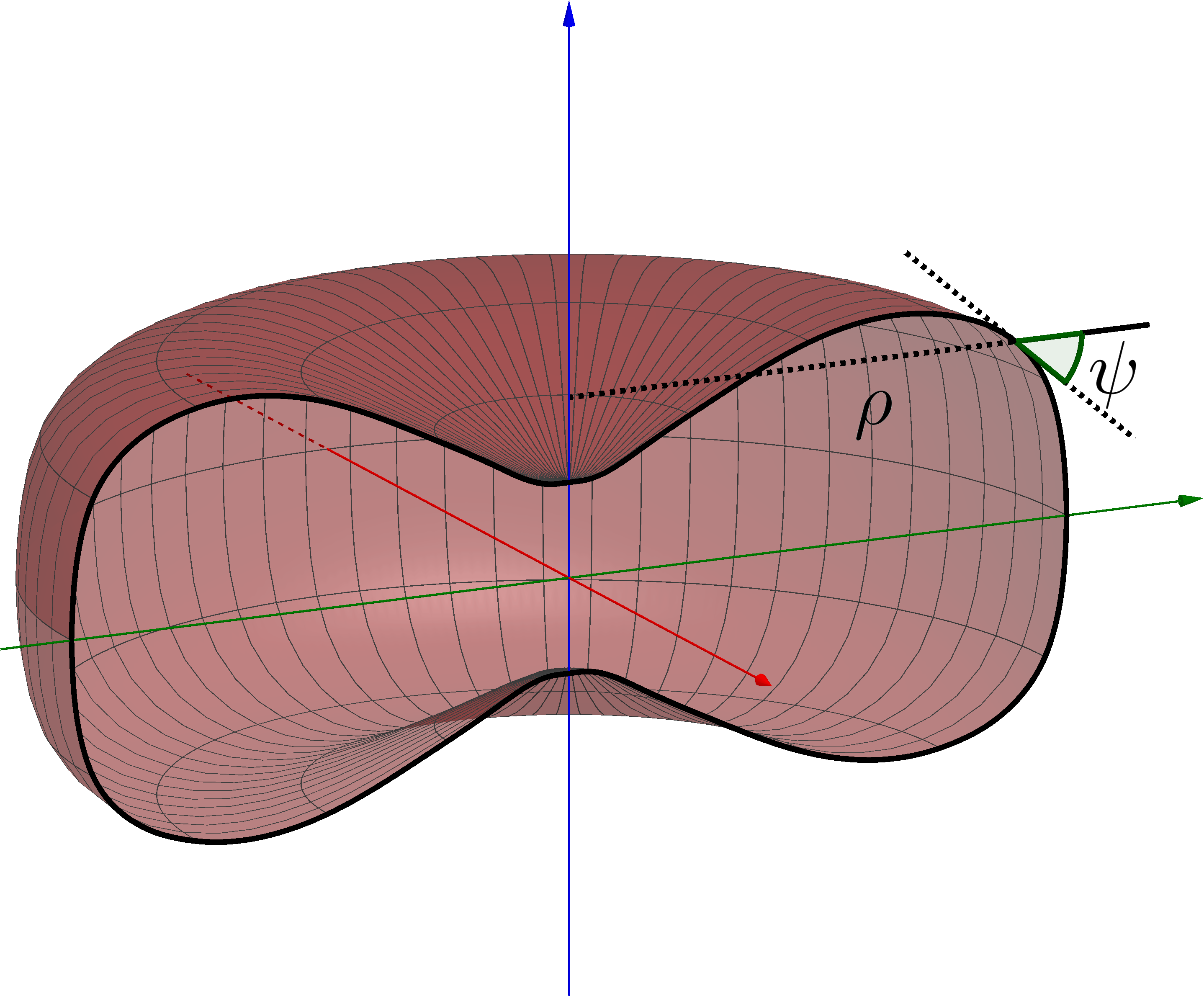}
\caption{Biconcave discoid} \label{fig:1c}
\end{subfigure}
\caption{The three axisymmetric biophysical solutions to the Canham--Helfrich model and how they arise as surfaces of revolution. The coordinate $\rho$ measures the perpendicular distance to the $z$-axis (blue), while $\psi$ is the angle between the tangent of the contour and the $\rho$-axis.} \label{fig:1}
\end{figure}

Here we discuss three well known axisymmetric solutions of the Canham-Helfrich model \cite{Tu:2014} (the spherical vesicle, the toroidal vesicle and the red blood cell) and how they are described within this approach. These surfaces arise as surfaces of revolution and therefore a particularly convenient way of parametrising these is to consider a “cross-sectional contour” described by the perpendicular distance $\rho$ to the symmetry axis (which we will take to be the $z$-axis) and the angle $\psi$, which is the angle between the tangent of the contour and the $\rho$-axis (see figure \ref{fig:1} for a graphical depiction). This gives us the relation $\tan\psi(\rho)=\frac{\diff z}{\diff \rho}$. The entire surface is then obtained by rotating this contour such that
\begin{equation}
X^\mu=\begin{pmatrix}
t\\
\rho\cos\phi\\
\rho\sin\phi\\
z_0+\int_{0}^{\rho}\diff\tilde{\rho}\tan\psi(\tilde{\rho})
\end{pmatrix}~~,
\end{equation}
which in turn gives rise to
\begin{eqnarray}
K =  -\frac{\sin\psi(\rho)}{\rho}-\cos\psi(\rho)\psi'(\rho)~~,~~K\cdot K& = & \frac{\sin^2\psi(\rho)}{\rho^2}+\cos^2\psi(\rho)(\psi'(\rho))^2~~.~~~~~
\end{eqnarray}

\paragraph{Spherical vesicle:}
A sphere of radius $R$ (see figure \ref{fig:1a}) is described by
\begin{equation}
\sin\psi(\rho)=\frac{\rho}{R}~~,
\end{equation}
which gives rise to the equation
\begin{equation}
0 = \Delta p R^2 + 4 c_0 \kappa + 2 c_0^2 R \kappa + 2 R \lambda~~.
\end{equation}
As was also pointed out in \cite{Tu:2014}, this has two solutions when viewed as an equation for the radius, provided that $\Delta p <0$ and $-4 c_0 \kappa \Delta p + (\kappa c_0^2 + \chi)^2 > 0$. The first condition reflects the fact that the internal pressure must be greater than the external pressure to stabilise the structure.

\paragraph{Torus:}
The torus can also be obtained as a surface of revolution (figure \ref{fig:1b}). This is achieved via
\begin{equation}
\sin\psi(\rho)=\frac{1}{r}\rho + \frac{R}{r}~~,
\end{equation}
where $R$ is the major axis and $r$ the minor axis. From this, we get the shape equation
\begin{eqnarray}
0 & = & \left(-\kappa  R^3+2 \kappa  r^2 R\right)+\rho ^2 \left(r^2 R \left(-\kappa {c_0}^2-\chi \right)-4 \alpha  {c_0} r
   R\right)\nn\\
   &&+\rho ^3 \left(-2 r^2 \left(\kappa  {c_0}^2+\chi \right)+4 \kappa  {c_0} r+\Delta p r^3\right)~~.
\end{eqnarray}
Each coefficient of $\{\rho^0,\rho^2,\rho^3 \}$ must vanish independently, giving us three equations
\begin{equation}
R=\sqrt{2}r~~,~~ \chi=\frac{\kappa c_0 \left(4-c_0 r\right)}{r}~~,~~ \Delta p=\frac{4 \kappa  c_0}{r^2}~~.
\end{equation} 
The first of these predicts a universal ratio between the major and minor axes. Theoretically predicted in \cite{TheoreticalToroidalVesicle}, this ratio was observed experimentally in \cite{ObsOfToroidalVesicle} with high precision. 

\paragraph{Biconcave discoid:} 
The biconcave discoid (figure \ref{fig:1c}) is the shape of the red blood cell. This axisymmetric vesicle is described by
\begin{equation}
\sin\psi(\rho)=a\rho(\log\rho+b)~~,
\end{equation}
where $a,b$ are parameters that are related to the characteristics of the discoid\footnote{For example, the radius of the discoid, i.e. the maximum value of $\rho=\rho_R$, is implicitly given by $1=a\rho_R(\log\rho_R+b)$, since $\psi(\rho_R)=\pi/2$ (see also \cite{PhysRevE.54.2816}).}. The resulting equation of motion is 
\begin{align}
0=&\left(\kappa  a^3-2 \kappa  a^2 b-4 \kappa  a b^2+4 \kappa  a b c_0-a \left(\kappa  c_0^2+\chi \right)+4 \kappa  b^2 c_0-2 b \left(\kappa  c_0^2+\chi \right)+\Delta p\right)\nn\\
&+\log \rho  \left(-2 \kappa  a^3-8 \kappa  a^2 b+4 \kappa
    a^2 c_0+8 \kappa  a b c_0-2 a \left(\kappa c_0^2+\chi \right)\right)\nn\\
&+\log ^2\rho  \left(-4 a^3 \kappa +4 a^2 \kappa  c_0\right)\,,
\end{align}
which again gives three equations. These equations yield 
\begin{equation}
a=c_0~~,~~\chi= \Delta p=0~~.
\end{equation}
Thus, we recover the result that the biconcave shape of the red blood cell relies on isotonicity, i.e. that the pressures on each side of the membrane are equal \cite{,PhysRevE.54.2816} (see also \cite{tanford1979hydrostatic}).

\section{Discussion \textit{\&} outlook}\label{sec:Outlook}
The majority of the work presented here was of a foundational nature. In order to describe the physical properties of fluid membranes in thermodynamic equilibrium, we developed the submanifold calculus for Newton--Cartan geometry. This parallels how the submanifold calculus of (pseudo-)Riemannian/Euclidean geometry is a pre-requisite for formulating and varying the standard Canham-Helfrich bending energy.
We identified the geometric structures characterising timelike submanifolds in NC geometry\footnote{The case of spacelike submanifolds is also interesting to pursue as it can be useful for understanding entanglement entropy in non-relativistic field theories \cite{Solodukhin:2009sk}} and obtained the associated integrability conditions. Deriving expressions for the infinitesimal variations and transformation properties of the basic objects allowed us to formulate a generic extremisation problem for broad classes of NC surfaces, including fluid membranes whose equilibrium configurations only depend on geometric properties.\footnote{It would be interesting to understand the connection between this work and other recently considered constructions involving extended objects embedded in Newton-Cartan spacetime (or related geometries), such as  non-relativistic strings  \cite{Harmark:2017rpg,Bergshoeff:2018yvt,Harmark:2018cdl,Harmark:2019upf}, non-relativistic D-branes \cite{Kluson:2019avy},  and Newton--Cartan $p$-branes \cite{Pereniguez:2019eoq}. It would also be interesting to connect this work to \cite{Gromov:2015fda}, where the boundary description of quantum Hall states involves a notion of Newton--Cartan submanifolds.}

In section \ref{sec:applications}, we applied this newly developed toolbox to the description of fluid membranes in thermodynamic equilibrium. The novel aspect of these applications is that the dependence on temperature and chemical potential of material coefficients, such as surface tension and bending modulus, is critical for the emergence of wave excitations. This relied on the fact that temperature and chemical potential have a geometric interpretation related to the existence of a timelike isometry in the ambient spacetime.
Standard examples of free energies such as the Canham-Helfrich bending energy are straightforwardly generalised by taking into account the geometric interpretation of thermodynamic variables. The resulting free energies are still purely geometric but the derived stresses on the membrane are different than standard results found in the literature. In particular, the Gaussian bending modulus can play a role in the shape of lipid vesicles since the Gaussian curvature cannot be integrated out when material coefficients are not constant. The resulting stresses produce elastic waves when perturbing away from equilibrium thus providing the correct dynamics of fluid membranes. 

This paves the way for tackling several open questions, which we plan to address in a future publication \cite{upcoming}:

\begin{itemize}
\item The fact that the Gaussian curvature cannot be integrated out in thermal equilibrium suggests that the family of closed lipid vesicles reviewed in section \ref{sec:Solutions} should be revisited and the effects of the Gaussian bending modulus should be considered (i.e. $a_3$ in \eqref{eq:model}), including the effects on deviations away from equilibrium.
\item The lipid vesicle solutions in section \ref{sec:Solutions} are static solutions, in which $u^{a}=(1,0,0)$. However, in principle such solutions can sustain rotation along the direction $\phi$. The question is thus: is it possible to obtain lipid vesicles with stationary flows?
\item From an effective field theory point of view, the Canham--Helfrich bending energy \eqref{eq:model} does not contain all possible responses that take into account thermal equilibrium. For instance a term quadratic in the extrinsic curvature of the form $u^au^bh^{cd}{K_{bc}}K_{ad}$ involving the fluid velocity can be added to \eqref{eq:model} (similarly to its relativistic counterpart \cite{Armas:2013hsa}). However, there are further couplings that involve derivatives of $u_{a}$ such as the square of the fluid acceleration $(u^a\mathfrak{D}_au^b)^2$ or the square of the vorticity. Some of these terms are related to the Gaussian curvature and thus, by the Gauss-Codazzi equation \eqref{eq:GaussCodazzi}, to combinations of squares of the extrinsic curvature. Therefore, from an effective theory point of view, they cannot be ignored a priori.
\item We have shown in section \ref{sec:elastic waves} that taking into account the geometric definitions of temperature and mass chemical potential in equilibrium gives rise to the correct dispersion relation for an elastic membrane when perturbing away from equilibrium. It would now be interesting to consider perturbations away from equilibrium solutions of the Canham--Helfrich model \eqref{eq:model} using the stresses \eqref{eq:T1}-\eqref{eq:Explicit2}. This would shed light on the stability of lipid vesicles.
\item The construction of effective actions or free energies in the manner described in this work is appropriate to describe equilibrium configurations. However, including different types of dissipation \cite{napoli2016hydrodynamic}, either due to viscous flows or diffusion of embedded proteins is of interest \cite{Steigmann2018}.  In order to include dissipation from an effective action point of view one could consider the more elaborate Schwinger-Keldysh framework \cite{Haehl:2015foa, Jensen:2017kzi, Glorioso:2018wxw} and adapt it to non-relativistic systems. Alternatively, one may construct the effective theory in a long-wavelength hydrodynamic expansion by classifying potential terms appearing in the currents $\mathcal{T}^a$ and $\mathcal{T}^{ab}$ and obtaining constitutive relations (see e.g. \cite{Kovtun:2012rj, Armas:2013goa}). We plan on addressing this in the near future. 
\item We focused on extrinsic curvature terms in effective actions \eqref{eq:NCact} but it would also be interesting to consider the effect of the external rotation tensor \eqref{eq:ExtRotTens}. In the (pseudo-)Riemannian/Euclidean setting, this corresponds to spinning point particles/membranes \cite{Guven:2006tc, Armas:2013hsa, Armas:2014rva, Armas:2017pvj} and are directly related to the Frenet curvature and Euler elastica (see e.g. ~\cite{guven2019conformal, guven2019conformal1, guven2019conformal2} for a recent discussion).
\item In sections \ref{sec:NCGeom} and \ref{sec:var} we formulated the description of a single surface in Newton--Cartan geometry for which the scalars $X^\mu$ can be seen as Goldstone modes of spontaneous broken translations at the location of the surface. It would be interesting to extend this further to the case of a foliation of surfaces, in which case the scalars $X^\mu$ form a lattice and can be used to describe viscoelasticity as in \cite{Armas:2019sbe}.
\end{itemize}

In this work we considered Newton--Cartan geometry but there are many other types of non-Lorentzian geometries 
depending on the space-time symmetry group, which can be e.g. Lifshitz, Schr\"odinger or Aristotelian, which have direct applications for the hydrodynamics of strongly correlated electron systems as well as for the hydrodynamics of flocking behaviour and active matter  \cite{deBoer:2017ing, deBoer:2017abi, Poovuttikul:2019ckt, Novak:2019wqg}. In these contexts, it is required to develop the mathematical description of submanifolds within these different types of ambient spacetimes. The description of surfaces within these geometries will be of interest for surface/edge physics in hard condensed matter.

\section*{Acknowledgements}

We thank L. Giomi and R. S. Green for useful discussions.
JA is partly supported by the Netherlands Organization for Scientific Research (NWO).
The work of JH is supported by the Royal Society University Research Fellowship ``Non-Lorentzian Geometry in Holography'' (grant number UF160197).
The work of EH is supported by the Royal Society Research Grant for Research Fellows 2017 ``A Universal Theory for Fluid Dynamics'' (grant number RGF$\backslash$R1$\backslash$180017).
The work of NO is supported in part by the project ``Towards a deeper understanding of  black holes with non-relativistic holography'' of the Independent Research Fund Denmark (grant number DFF-6108-00340) and 
by the Villum Foundation Experiment project 00023086.

\appendix

\section{Null reduction of Riemannian surfaces and perfect fluids}\label{app:reduction}
In this appendix we provide a completely different approach to formulating the theory of surfaces and fluid membranes in Newton--Cartan geometry. This approach consists in starting from relativistic surfaces and fluid membranes and performing a null reduction so as to obtain results in NC geometry. The purpose of this technical appendix is to provide a non-trivial check of the main results in the core of this paper.

\subsection{Submanifolds from null reduction}
It is well known that any Newton--Cartan geometry can be obtained as the null reduction of a Lorentzian manifold in one dimension higher equipped with a null killing vector \cite{Julia:1994bs,Christensen:2013rfa,Hartong:2016nyx}. Therefore, if we choose a timelike submanifold in a Lorentzian geometry such that the null Killing vector is tangent to the submanifold, its null reduction provides us with a Newton--Cartan submanifold embedded in a Newton--Cartan ambient spacetime. We illustrate this in the commuting diagram below:
\begin{equation}
\begin{CD}
(\widehat{\Sigma}_{p+2},\hat{\gamma}) @<{\hat{u}}^{\hat\mu}_{\hat{a}}<< (\widehat{\mathcal{M}}_{d+2},\hat{g})\\
@V\text{null red.}VV @VV\text{null red.}V\\
(\Sigma_{p+1},\{\tau\vert_\Sigma,\check h, \check m \}) @<<u^\mu_a< (\mathcal{M}_{d+1},\{\tau,h,m\})
\end{CD}
\end{equation}
In section \ref{sec:SubmanifoldStruc}, we described how to go from the NC manifold $(\mathcal{M}_{d+1},\{\tau,h,m\})$ to the NC  submanifold $(\Sigma_{p+1},\{\tau\vert_\Sigma,\check h, \check m \})$, while passing from the Lorentzian manifold $(\widehat{\mathcal{M}}_{d+2},\hat{g})$ to the Newton--Cartan manifold $(\mathcal{M}_{d+1},\{\tau,h,m\})$ is achieved by null reduction.

In this appendix, we will traverse the other route: our goal is to go from $(\widehat{\mathcal{M}}_{d+2}, \hat g)$ to $(\Sigma_{p+1},\{\tau\vert_\Sigma,\check h, \check m \})$ via $(\widehat{\Sigma}_{p+2},\hat{\gamma})$. The procedure to go from $(\widehat{\mathcal{M}}_{d+2},\hat g)$ to $(\widehat{\Sigma}_{p+2},\hat{\gamma})$ is nothing but the theory of submanifolds in Lorentzian geometry and is well known (see e.g. \cite{Armas:2013hsa,Armas:2017pvj}). We coordinatise $\widehat{\mathcal{M}}_{d+2}$ with $x^{\hat{\mu}}=(u,x^\mu)$ and $\widehat{\Sigma}_{p+2}$ with $\hat\sigma^{\hat{a}}=(w,\sigma^a)$. The metric on $\widehat{\mathcal{M}}_{d+2}$ can -- by assumption -- be written in null reduction form
\begin{equation}
ds_{\widehat{\mathcal{M}}_{d+2}}^2 = \hat{g}_{\hat\mu\hat\nu}\dx^{\hat\mu} \dx^{\hat\nu} = 2\tau_\mu \dx^\mu(\diff u - m_\nu\dx^\nu) + h_{\mu\nu}\dx^\mu\dx^\nu~~.\label{eq:nullred}
\end{equation}
This line element is invariant under the Newton--Cartan gauge transformations \eqref{eq:TNCgaugetrafos} and conversely all gauge invariance of this line element are of the form \eqref{eq:TNCgaugetrafos}. The invariance under the $U(1)$ transformation with parameter $\sigma (x^\mu)$ requires that we vary the higher-dimensional coordinate $u$ as $\delta u=\sigma$. From the higher-dimensional perspective this corresponds to a diffeomorphism that leaves the $x^\mu$ unaffected but that shifts $u$ by some function of $x^\mu$. 

The Lorentzian submanifold is defined via a set of embedding maps $\hat X^{\hat\mu}(\sigma^{\hat{a}})$ in the usual way. We define the projector
\begin{equation}
\hat{P}^{\hat\mu}_{\hat\nu} = \hat u^{\hat\mu}_{\hat{a}} \hat u_{\hat\nu}^{\hat a} = \delta^{\hat\mu}_{\hat\nu} - \hat{n}^I_{\hat{\rho}} \hat{n}^J_{\hat\nu}\delta_{IJ}\hat{g}^{\hat\rho\hat\mu}~~,
\end{equation}
where $\hat{n}^I_{\hat{\rho}}$ are the normal one-forms to $\widehat{\Sigma}_{p+2}$ and where $\hat u^{\hat\mu}_{\hat a}=\D_{\hat a}\hat X^{\hat\mu}$. We require that the null direction is shared between $\widehat{\mathcal{M}}_{d+2}$ and $\widehat{\Sigma}_{p+2}$, which can be expressed as the requirements
\begin{equation}
\hat{u}^u_w = 1~~,~~ \hat{u}^u_a = 0~~,\label{eq:Requirements}
\end{equation}
where the null direction on the submanifold is described by $w$. Further, we want to impose a null reduction analogue of the timelike requirement \eqref{eq:tauIzero}. To this end, we introduce a vector $U^{\hat\mu} = \left(\pd{}{u}\right)^{\hat\mu} = \delta^{\hat\mu}_u$ so that $U_{\hat\mu} = (0,\tau_\mu)$. Requiring that the null Killing vector field is tangential to the submanifold $\hat{n}_u^I = U^{\hat\mu} \hat n^I_{\hat\mu} = U_{\hat\mu} \hat{n}^{\hat\mu I} = 0$
for all $I $ implies the desired relation $\tau_\mu n^\mu_I = 0$ where we have identified $\hat n^\mu_I = n^\mu_I $. This further implies that $n^{\mu I} = \hat{g}^{\mu \hat\nu} \hat{n}^I_{\hat\nu} = h^{\mu\nu}n^I_\nu$
in agreement with the timelike constraint. This also implies that $\hat{P}^\mu_\nu = P^\mu_\nu$, as well as the normalisation $\hat{g}^{\hat\mu\hat\nu}\hat{n}^I_{\hat\mu}\hat{n}^J_{\hat\nu} = h^{\mu\nu}n^I_\mu n^J_\nu =\delta^{IJ}$.
Further, the above considerations lead us to conclude that
\begin{equation}
\hat n^{uI} = \hat g^{u\hat\mu}\hat n^I_{\hat\mu} = -\hat v^\mu n^I_\mu =-\hat v^I\,.
\end{equation}

The metric on  $ \widehat{\Sigma}_{p+2} $ can also be written in null reduction form
\begin{eqnarray}
ds_{\widehat{\Sigma}_{p+2}}^2 = \hat{\gamma}_{\hat a\hat b}\dx^{\hat{a}} \dx^{\hat{b}}& =& 2\tau_a\dx^a( \diff w - m_b\dx^b) + h_{ab}\dx^a\dx^b\nn\\
& =& 2\tau_a\dx^a( \diff w - \check m_b\dx^b) + \check h_{ab}\dx^a\dx^b~~,
\end{eqnarray}
where we recall the definitions of $\check h_{ab}$ and $\check m_a$ in \eqref{eq:checkh} and \eqref{eq:checkm}, respectively. As manifested in the equations above, the null reduction form of the metric is Galilean boost-invariant and does not distinguish between \emph{checked} and \emph{un-checked} metric data. In turn, the Lorentzian metric $\hat{\gamma}$ on $\widehat{\Sigma}_{p+2}$ is the pullback of the metric $\hat{g}$ on $\widehat{\mathcal{M}}_{d+2}$, that is
\begin{equation}
\hat{\gamma}_{\hat a\hat b} = \hat{u}^{\hat\mu}_{\hat{a}} \hat{u}^{\hat\nu}_{\hat{b}} \hat{g}_{\hat\mu\hat\nu}~~,
\end{equation}
which implies that
\begin{equation}
\tau_a = \hat{\gamma}_{a w} =  \hat{u}^{\hat\mu}_a \hat{u}^{\hat\nu}_w \hat{g}_{\hat\mu\hat\nu} =
  \hat{u}^\mu_a \hat{u}^u_w \hat{g}_{\mu u} +  \hat{u}^\mu_a \hat{u}^\nu_w \hat{g}_{\mu \nu} = \hat{u}^\mu_a \tau_\mu +  \hat{u}^\mu_a \hat{u}^\nu_w \bar{h}_{\mu \nu}~~.
\end{equation}
Thus, taking 
\be
\hat{u}^\mu_w = 0~~,\label{eq:ThisIsZero}
\ee
and identifying $\hat{u}^\mu_a = {u}^\mu_a$ we get the desired relation between the two clock 1-forms, namely $\tau_a = u^\mu_a\tau_\mu$. Next, we consider 
\begin{equation}
\bar h_{ab} = \hat{\gamma}_{ab} = \hat{u}^{\hat\mu}_a \hat{u}^{\hat\nu}_b \hat{g}_{\hat\mu\hat\nu} = \hat{u}^\mu_a \hat{u}^\nu_b \hat{g}_{\mu \nu} + \hat{u}^\mu_a \hat{u}^u_b \hat{g}_{\mu u} + \hat{u}^u_a \hat{u}^\nu_b \hat{g}_{u\nu } = u^\mu_a u^\nu_b \bar h_{\mu\nu}~~,
\end{equation}
where we have used \eqref{eq:Requirements}, which again agrees with the results of section \ref{sec:SubmanifoldStruc}. The relation $\hat{u}^{\hat\mu}_w\hat{u}^w_{\hat\mu} = \hat{u}^w_u=1$ where we used \eqref{eq:ThisIsZero}, fixes  $\hat{u}^w_u = 1$. To determine $\hat u^w_\mu$ we bring into play the orthogonality requirement $\hat u^w_{\hat\mu} \hat n^{\hat\mu I} = \hat g^{\hat\mu\hat\nu}\hat u^w_{\hat\mu} \hat n^I_{\hat\nu}=0$, which translates into the relation
\begin{equation}
\hat v^\mu n^I_\mu = \hat u^w_\mu n^{\mu I}~~,\label{eq:NotWeirdRel}
\end{equation}
where we have used that $\hat u^w_u = 1$ and $n^\mu_I = \delta_{IJ}h^{\mu\nu}n^J_\nu$. This is only possible if 
\begin{equation}
{ \hat u^w_\mu = \hat{v}^I n^I_\mu~~.}\label{eq:GoodRel}
\end{equation} 
The null reduction of the ambient inverse metric is 
\begin{equation}
\hat g^{uu} = 2\tilde{\Phi}~~,~~ \hat g^{u\mu} = -\hat{v}^\mu~~,~~ \hat g^{\mu\nu} = h^{\mu\nu}~~,
\end{equation}
while the relation between $\hat{g}^{-1}$ and $\hat{\gamma}^{-1}$ is given by $\hat{\gamma}^{\hat a\hat b} = \hat u^{\hat{a}}_{\hat\mu}\hat u^{\hat b}_{\hat\nu} \hat g^{\hat\mu\hat\nu}$. In turn, the relation $\hat{\gamma}^{ab}=h^{ab}$ requires that $\hat{u}^a_u = 0$. Using this, we can write
\begin{equation}
\hat{\gamma}^{wa}=\hat{u}^w_{\hat{\mu}}\hat{u}^a_{\hat{\nu}}\hat{g}^{\hat\mu\hat\nu} = \hat{v}^I n^I_\mu \hat{u}^a_\nu h^{\mu\nu}+\hat{u}^a_\nu \hat{g}^{u\nu}\,,
\end{equation}
where we have used \eqref{eq:GoodRel}, which leads us to identify $\hat{u}^a_\mu = u^a_\mu$ and, by the orthogonality relation \eqref{eq:NormalVecDef}, leads to $\hat v^a = u^a_\mu \hat v^\mu$
as desired. The relation \eqref{eq:GoodRel} furthermore implies that
\begin{equation}
\hat{\gamma}^{ww} = \hat u^w_{\hat\mu}\hat u^w_{\hat\nu}\hat g^{\hat\mu\hat\nu} 
= 2\tilde{\Phi}-\hat{v}^I\hat v_I=\check \Phi~~.
\end{equation}

In summary, the Lorentzian objects arrange themselves under submanifold null reduction according to
\begin{align}
\hat{u}^{\hat{\mu}}_{\hat{a}}&\overset{\text{null red.}}{\longrightarrow}\qquad \hat{u}^\mu_a=u^\mu_a~~,~~\hat{u}^u_w=1~~,~~ \hat{u}^\mu_w=0~~,~~ \hat{u}^u_a=0~~,\label{eq:NullRred1}\\
\hat{u}_{\hat{\mu}}^{\hat{a}}&\overset{\text{null red.}}{\longrightarrow}\qquad \hat{u}^a_\mu = u^a_\mu~~,~~ \hat{u}^w_u=1~~,~~ \hat{u}^w_\mu=\hat{v}^I n^I_\mu~~,~~ \hat{u}^a_u=0~~,\\
\hat{n}^{I}_{\hat{\mu}}&\overset{\text{null red.}}{\longrightarrow}\qquad \hat{n}^I_\mu = n^I_\mu~~,~~ \hat{n}^I_u=0~~,\label{eq:NullRred3}\\
\hat{n}_{I}^{\hat{\mu}}&\overset{\text{null red.}}{\longrightarrow}\qquad \hat{n}_I^\mu = n_I^\mu~~,~~ \hat{n}_I^u=-\hat v_I~~.
\end{align}
The metric on $\widehat{\Sigma}_{p+2}$ is
\begin{equation}
ds_{\widehat{\Sigma}_{p+2}}^2=2\tau_a\dx^a( \diff w - \check m_b\dx^b) + \check h_{ab}\dx^a\dx^b~~,
\end{equation}
while the components of the inverse metric on $\widehat{\Sigma}_{p+2}$ are
\begin{equation}
{\hat{\gamma}^{ww} = 2\check \Phi  = 2\tilde\Phi - \hat v^I \hat v^I,\qquad \hat{\gamma}^{wa} = -\hat v^a,\qquad \hat\gamma^{ab} = h^{ab}~~.}\label{eq:CheckNullRed?}
\end{equation}

\subsubsection{Null reduction of the connection \textit{\&} the extrinsic curvature} \label{sec:NullRed}
We now consider the null-reduction of the Lorentzian connection. The non-zero components of the higher-dimensional Christoffel symbols are
\begin{eqnarray}
\hat\Gamma_{\mu\nu}^\rho &=& \bar{\Gamma}^\rho_{(\mu\nu)}=\bar{\Gamma}^\rho_{\mu\nu}+ \frac{1}{2}\hat v^\rho\tau_{\mu\nu}~~,\\
\hat{\Gamma}_{\mu\nu}^u &=& -\bar {\mathcal{K}}_{\mu\nu} - 2\tau_{(\mu}\D_{\nu)}\tilde{\Phi}~~,\\
\hat{\Gamma}_{u\mu}^\rho &=& \frac{1}{2}h^{\rho\sigma}\tau_{\mu\sigma}~~,\\
\hat{\Gamma}_{u\mu}^u &=& \frac{1}{2}a_\mu~~,
\end{eqnarray}
where
\begin{eqnarray}
\bar{ \mathcal{K}}_{\mu\nu} = -\frac{1}{2}\pounds_{\hat{v}}\bar h_{\mu\nu}\label{eq:TNCexcurv} ~~,~~a_\mu = \pounds_{\hat{v}}\tau_\mu = \hat v^\rho \tau_{\rho\mu}~~.\label{eq:Defofa}
\end{eqnarray}
The NC extrinsic curvature $\bar{\cal K}_{\mu\nu}$ should not be confused with the submanifold extrinsic curvature $K_{ab}{}^I$.
The pullback of the ambient TNC extrinsic curvature, $\bar{\mathcal{K}}_{ab}=u^\mu_a u^\nu_b \bar{\mathcal{K}}_{\mu\nu}$, is related to the TNC extrinsic curvature on the submanifold $\Sigma_{p+1}$, 
\be
\bar{\mathcal{K}}^{\Sigma}_{ab}=-\frac{1}{2}\pounds^{\Sigma}_{\hat v}\bar h_{ab}~~,
\ee
where $\pounds^{\Sigma}_{\hat v}$ denotes the Lie derivative along $\hat v^a$ on $\Sigma_{p+1}$, in the following way
\begin{equation}
\bar{\mathcal{K}}_{ab}=\bar{\mathcal{K}}^{\Sigma}_{ab}-\tau_{(a}\partial_{b)}(\hat v^I\hat v^I)+\hat v^I K_{ab}{^I}~~.\label{eq:UsefulRel9000}
\end{equation}
This can be shown by starting with $\bar{\mathcal{K}}_{ab}=u^\mu_a u^\nu_b \bar{ \mathcal{K}}_{\mu\nu}$ and using $\hat v^\rho=\hat v^c u_c^\rho+\hat v^I n^\rho_I$ in \eqref{eq:TNCexcurv}. The following identity
\begin{equation}
\pounds_{n^I}\bar h_{\mu\nu}=2\nabla_{(\mu}n_{\nu)}^I+2\tau_{(\mu}\partial_{\nu)}\hat v^I\,,
\end{equation}
together with equation \eqref{eq:ExtCurv} can then be used to derive \eqref{eq:UsefulRel9000}.

The higher-dimensional extrinsic curvature $\hat{K}_{\hat{a}\hat{b}}{^I}$ is determined in terms of the higher-dimensional analogue of the surface covariant derivative of \eqref{eq:SurfaceCovD}, which we will call $\hat{D}_{\hat{a}}$. It acts on a mixed tensor $\hat T^{\hat{b}\hat{\mu}}$ according to
\begin{equation}
\hat D_{\hat{a}} \hat{T}^{\hat{b}\hat{\mu}}=\D_{\hat{a}} \hat{T}^{\hat{b}\hat{\mu}}+\hat\gamma^{\hat{b}}_{\hat a\hat c} \hat{T}^{\hat{c}\hat{\mu}} + \hat{u}^{\hat{\nu}}_{\hat{a}}\hat{\Gamma}^{\hat{\mu}}_{\hat{\nu}\hat{\lambda}} \hat{T}^{\hat{b}\hat{\lambda}}~~,
\end{equation}
where $\hat\gamma^{\hat{b}}_{\hat a\hat c}$ is the Levi-Civita connection of $\hat{\gamma}$, while $\hat{\Gamma}^{\hat{\mu}}_{\hat{\nu}\hat{\lambda}} $ is the Levi-Civita connection of $\hat{g}$. The higher-dimensional extrinsic curvature is 
\be 
\hat K_{\hat a\hat b}{^I}=\hat n_{\hat\mu}^I\hat D_{\hat a}\hat u^{\hat\mu}_{\hat b} = \hat n_{\hat\mu}^I\left(\D_{\hat a}\hat u^{\hat\mu}_{\hat b} + \hat u^{\hat \nu}_{\hat a}\hat\Gamma^{\hat\mu}_{\hat\nu\hat\lambda}\hat u^{\hat\lambda}_{\hat b}\right)~~,
\ee
which using \eqref{eq:NullRred1} and \eqref{eq:NullRred3} means that
\begin{equation}
\hat K_{ab}{^I} = n_\mu^I D_a u^\mu_b + \frac{1}{2}\hat v^I{\tau_{ab}} =K_{ab}{^I}\,,
\end{equation}
where we have recognised the extrinsic curvature of \eqref{eq:ExtCurv}. This is invariant under both gauge transformations and Galilean boosts. The other non-zero components of the higher-dimensional extrinsic curvature are $\hat K_{w b}{}^I=-\frac{1}{2}\tau_{Ib}$.

Below equation \eqref{eq:nullred}, we have shown that the $U(1)$ gauge transformation is a specific diffeomorphism in the higher-dimensional description. This is a useful way to find out how various objects transform under the $\sigma$ gauge transformation. This also applies to tensors defined on the submanifold $\Sigma_{p+1}$, since they descend from the Lorentzian manifold $\widehat{\Sigma}_{p+2}$. A diffeomorphism of a generic tensor $X_{\hat a}{}^{\hat b}$ is given by
\begin{equation}
\delta X_{\hat a}{}^{\hat b}=\hat\xi^{\hat c}\partial_{\hat c}X_{\hat a}{}^{\hat b}+X_{\hat c}{}^{\hat b}\partial_{\hat a}\hat\xi^{\hat c}-X_{\hat a}{}^{\hat c}\partial_{\hat c}\hat\xi^{\hat b}~~.
\end{equation}
In order to find the $U(1)$ transformation, we need to choose a diffeomorphism for which $\hat \xi^{\hat a}=-\sigma\delta^{\hat a}_w$. Since all objects are independent of $u$ we find that a one-form $X_{a}$ in this case transforms as
\begin{equation}
\delta X_{a}=-X_{w}\partial_{a}\sigma~~,
\end{equation}
while a vector $X^b$ is $U(1)$ invariant. Applying this to the extrinsic curvature $\hat K_{ab}{^I}$ we find
\begin{equation}
\delta_\sigma\hat K_{ab}{^I} =-\hat K_{aw}{^I} \partial_b\sigma-\hat K_{wb}{^I} \partial_a\sigma~~.
\end{equation}
Using that $\hat K_{w b}{}^I=-\frac{1}{2}\tau_{Ib}$ we recover the transformation rule \eqref{eq:U1trafoK}.

\subsubsection{Variations from null reduction}
Here we obtain some of the results of section \ref{sec:Variations} using null reduction. We begin with the variations of the normal 1-forms. In the relativistic case, the normal 1-forms can be shown to transform as \cite{Armas:2017pvj}
\begin{equation}
\delta \hat n^I_{\hat\mu} = \frac{1}{2}\hat n^{\hat\nu}_J\hat n^J_{\hat\mu}\hat n^{{\hat\rho}I} \delta \hat g_{\hat\nu\hat\rho} - \hat n^I_{\hat\nu} \hat u^{\hat{a}}_{\hat\mu} \delta\hat u^{\hat\nu}_{\hat a} + \frac{1}{2}\hat{n}_{\hat\mu J}(\hat{n}^{\hat\nu J} \delta \hat{n}^I_{\hat\nu} - \hat{n}^{\hat\nu I}\delta \hat{n}^J_{\hat\nu})~~.
\end{equation}
Restricting to $\hat\mu = \mu$, the last term simply reduces to $\lambda^{I}{}_J n_{\mu}^J$. This follows from demanding that $\hat n^I _u = 0$ is preserved under transformations, implying that $\delta \hat n^I_u=0$. Ignoring rotations of the normal one-forms, we get
\begin{equation}
\delta n^I_\mu = -\frac{1}{2}\hat v^\rho n^J_\rho n_{\mu J}n^{\nu I} \delta\tau_\nu -\frac{1}{2}n^{\nu J} n_{\mu J}  \hat v^\rho n^I_\rho \delta \tau_\nu  +\frac{1}{2}{n}^{\rho J}n_{\mu J} n^{\nu I} \delta \bar h_{\rho \nu}  - {n}^I_\nu u^a_\mu\delta u^\nu_a~~,\label{eq:deltan}
\end{equation}
where we have used that $\hat n^{u I} = -\hat v^\mu n^I_\mu$. Using the definitions of $\hat v$ and $\bar h$, we find that the variation can be written as
\begin{equation}
{\delta n^I_\mu = - v^{(I}n^{J)\nu}n_{\mu J}\delta \tau_\nu + \frac{1}{2}n^{\rho J} n_{\mu J} n^{\nu I}\delta h_{\rho \nu}- {n}^I_\nu u^a_\mu\delta u^\nu_a~~,}\label{eq:NormalVarFromNull}
\end{equation}
in agreement with the result \eqref{eq:NormalVar} (up to a local $\mathfrak{so}(d-p)$ transformation that we ignored).

With this at hand, we rederive \eqref{eq:VarOfK} using the method of null reduction. The relativistic result reads \cite{Armas:2017pvj}
\begin{equation} \label{eq:theexpression}
\delta_{\hat X}\hat K_{\hat a\hat b}{^I} = -\hat n_{\hat\mu}^I\hat{D}_{\hat{a}}\hat{D}_{\hat{b}}\hat{\xi}^{\hat\mu} + \hat n_{\hat\mu}^I\hat\xi^{\hat\lambda}  \hat{u}^{\hat\nu}_{\hat{a}}\hat u^{\hat\rho}_{\hat b} \hat{R}_{\hat\lambda\hat\nu\hat\rho}{^{\hat\mu}} +\hat{\lambda}^I{}_J \hat K_{\hat{a}\hat{b}}{^J}~~,
\end{equation}
where
\begin{equation}
\hat{\lambda}^{IJ}= \hat n^{\hat\mu[I}\hat n^{J]\hat\nu}\hat\xi^{\hat\rho}\D_{\hat\nu} \hat g_{\mu\hat{\rho}}= \hat n_{\hat\rho}^{[I}\hat n^{J]\hat\nu}\hat\Gamma^{\hat\rho}_{\hat\nu\hat\sigma}\hat\xi^{\hat\sigma}\,.\label{eq:NullRedThis}
\end{equation}
We keep the null direction fixed, so that
\begin{equation}
\hat\xi^{\hat\mu} = -\delta \hat X^{\hat\mu}~~,~~ \hat{\xi}^u=0~~.
\end{equation}
We are interested in $(\hat{a},\hat{b}) = (a,b)$ and since $\hat n^I_u = 0 = \hat u^u_a$, \eqref{eq:theexpression} reduces to
\begin{equation} \label{eq:var11}
\delta_{X} \hat{K}_{ab}{^I} =  - n_\mu^I\hat{D}_{a}\hat{D}_{b}{\xi}^\mu +  n_\mu^I\xi^\lambda  {u}^\nu_{a} u^\sigma_{b}\hat{R}_{\lambda\nu\sigma}{^\mu} +  \hat{\lambda}^I{}_J\hat{K}_{ab}{^J}~~,
\end{equation}
where $\hat{\xi}^\mu = \xi^\mu$ so that $\delta\hat X^\mu=\delta X^\mu$. In the absence of torsion, the null reduction of the Riemann tensor gives
\begin{equation}
\hat{R}_{\lambda\nu\sigma}{^\mu} = -\D_\lambda \hat\Gamma^\mu_{\nu\sigma} + \D_\nu \hat\Gamma^\mu_{\lambda\sigma} - \hat \Gamma^\mu_{\lambda \hat\rho}\hat \Gamma^{\hat\rho}_{\nu\sigma} + \hat\Gamma^\mu_{\nu \hat\rho}\hat\Gamma^{\hat\rho}_{\lambda\sigma}= R_{\lambda\nu\sigma}{^\mu}~~.
\end{equation}
Since in the absence of torsion $\hat{D}_w \xi^\mu=0$ and $\hat{D}_b \xi^\mu={D}_b \xi^\mu$, we find that
\begin{equation}
\hat{D}_{a}\hat{D}_{b}{\xi}^\mu = {D}_{a} {D}_{b}{\xi}^\mu~~,
\end{equation}
while the null reduction of \eqref{eq:NullRedThis} gives  $\hat{\lambda}^{IJ}= n_\rho^{[I}n^{J]\nu}\Gamma^\rho_{\nu\sigma}\xi^\sigma$ and so we obtain \eqref{eq:VarOfK}, as expected.

\subsubsection{Note on the reduction of the Lorentzian action}
The variational principle for NC surfaces in section \ref{sec:EoMs} can be obtained from null reduction of the relativistic variational principle \cite{Armas:2013hsa}, namely
\begin{equation} \label{eq:varrel}
\delta S = \int_\Sigma \diff^{p+1}\sigma\sqrt{-\hat\gamma}\left(\frac{1}{2}\hat T^{\hat a\hat b}\delta \hat\gamma_{\hat a\hat b} +\hat{\mathcal{D}}^{\hat a\hat b}{_I}\delta\hat K_{\hat a\hat b}{^I} \right)~~.
\end{equation}
The null reduction formulae of the previous section, for instance \eqref{eq:var11}, imply that the null reduction of \eqref{eq:varrel} will include a dependence on variations of $\hat K_{wa}{}^I=-\frac{1}{2}\tau_{Ib}$. Such torsion dependent terms were not included in \eqref{eq:NCact}. The reason, as mentioned throughout the paper is that we have assumed to be working without torsion, that is $\tau_{\mu\nu}=0$ at the expense of only being able to extract the divergence of the energy current instead of the energy current itself.

\subsection{Perfect fluid from null reduction}\label{sec:PerfectFluidNullRed}
In this section, we consider the null reduction of the equilibrium partition function of a relativistic space-filling perfect fluid, that is a fluid that is not living on a surface. The case in which the fluid is confined to the surface (i.e. a fluid membrane) considered in section \ref{sec:Fluids} is a straightforward modification of this analysis. The result provides us with the hydrostatic partition function of a Galilean-invariant perfect fluid.

We begin with the null reduction of the unit normalised relativistic fluid velocity $\hat u^{\hat\mu}$, which obeys $\hat g_{\hat\mu\hat\nu}\hat{u}^{\hat\mu}\hat{u}^{\hat\nu}=-1$. We define the non-relativistic fluid velocity $u^\mu$ as follows \cite{Hartong:2016nyx}, 
\begin{equation}\label{eq:defu}
u^\mu = \frac{\hat{u}^\mu}{\hat{u}_u}~~,
\end{equation}
where $\hat u_u=\hat g_{u\hat\mu}\hat u^{\hat\mu}=\tau_\mu\hat u^\mu$. This implies that $\tau_\mu u^\mu = 1$ which is the standard normalisation of the contravariant velocity of a non-relativistic fluid. The relativistic condition
\begin{equation}
\hat g_{\hat\mu\hat\nu}\hat{u}^{\hat\mu}\hat{u}^{\hat\nu}=\bar h_{\mu\nu}\hat u^\mu \hat u^\nu+2\tau_\mu \hat u^\mu \hat u^u=-1~~,
\end{equation}
can be used to solve for $\hat u^u$, leading to
\begin{equation}\label{eq:uup}
  \hat u^u=-\frac{1}{2\hat{u}_u}-\frac{1}{2}\hat{u}_u\bar h_{\mu\nu}u^\mu u^\nu~~.
\end{equation}
We still need to find a lower-dimensional interpretation of $\hat u_u$. This can be achieved as follows. Let $\hat T^{\hat\mu}{}_{\hat\nu}$ be the energy-momentum tensor of the higher-dimensional relativistic theory. For a perfect fluid this is $\hat T^{\hat\mu}{}_{\hat\nu}=\left(\hat E+\hat P\right)\hat u^{\hat\mu}{}\hat u_{\hat\nu}+\hat P\delta^{\hat\mu}_{\hat\nu}$. The mass current of the null reduced theory is given by $\hat T^\mu{}_u$ (see for example \cite{Hartong:2016nyx}). In the lower-dimensional theory, this is equal to $nu^\mu$, where $n$ is the mass density. Comparing the two expressions yields
\begin{equation}\label{eq:udown}
\hat u_u^2=\frac{n}{\hat E+\hat P}~~.
\end{equation}
We will later find expressions for $\hat E$ and $\hat P$ in terms of the non-relativistic energy and pressure.

In the hydrostatic partition function approach for a relativistic fluid, one identifies the intensive fluid variables such as temperature and velocity with a timelike Killing vector of an otherwise arbitrary Lorentzian curved background geometry. By varying the metric while keeping the Killing vector fixed, one extracts the fluid energy-momentum tensor. This approach has been applied to non-relativistic fluids on a NC background in \cite{Jensen:2014ama, Banerjee:2015uta} and here we will show how this follows from null reduction. In the higher-dimensional Lorentzian geometry, we assume the existence of a Killing vector $\hat{k}^{\hat\mu}$ such that
\begin{equation}
\hat{k}^{\hat{\mu}} = \hat\beta \hat{u}^{\hat\mu}~~,
\end{equation}
where $\hat{\beta}$ is the relativistic (inverse) temperature, and $\hat{u}^{\hat\mu}$ the relativistic fluid velocity. Just like in the Lorentzian setting, we will introduce a Newton--Cartan Killing vector $k^\mu$ that is proportional to the non-relativistic fluid velocity $u^\mu$ and that is timelike, where $\tau_\mu k^\mu$ relates to the non-relativistic temperature. Hence we write
\begin{equation}
k^\mu=\beta u^\mu~~,
\end{equation}
where $\beta = \tau_\mu k^\mu$ is the non-relativistic (inverse) temperature. The null reduction of $\hat{k}^{\hat{\mu}}$ is just $\hat{k}^{\hat{\mu}}=(\hat k^u\,,k^\mu)=\beta\left(\hat \mu, u^\mu\right)$, where we write $\hat k^u=\beta\hat\mu$ with $\hat\mu$ a parameter to be determined. This means that
\begin{equation}
\beta u^\mu = \hat{\beta}\hat u^\mu~~.
\end{equation}
Now, since $\hat{k}^{\hat\mu}$ is a Killing vector, we have that
\begin{equation}
\pounds_{\hat{k}} \hat{g}_{\hat\mu\hat\nu} = 0~~,
\end{equation}
which, after null reduction, turns into the statements
\begin{equation}
{\pounds_k\tau_\mu = 0~~,~~ \pounds_k \bar h_{\mu\nu}= -2\tau_{(\mu}\D_{\nu)}\hat k^u~~.}
\end{equation}
In a NC geometry a Killing vector is defined by setting to zero the transformations in \eqref{eq:TNCgaugetrafos} (and thus also implying that the variations in \eqref{eq:sigmatrafo} give zero). Here $\hat k^u$ is thus a specific $U(1)$ gauge transformation parameter that is associated with the existence of a Killing vector.

The relativistic hydrostatic partition function at ideal order in derivatives is an integral of the pressure which depends on the intensive variables, i.e. scalar quantities built from the Killing vector. One of these is the norm of $\hat k^{\hat\mu}$ which relates to the relativistic temperature. However, in the case of null reduction we actually have, besides $\hat k^{\hat\mu}$, another Killing vector which is $U^{\hat\mu}=\left(\frac{\partial}{\partial u}\right)^{\hat\mu}$. Since $U^{\hat\mu}$ is null, we can form only one other scalar, namely
\begin{equation}
\hat{g}_{\hat\mu\hat\nu}  U^{\hat\mu} \hat k^{\hat\nu} = \tau_\mu k^\mu = \beta~~,
\end{equation}
which is the non-relativistic (inverse) temperature. The other scalar is of course
\begin{equation}
-\hat\beta^2= \hat{g}_{\hat\mu\hat\nu} \hat k^{\hat\mu} \hat k^{\hat\nu} = \beta^2\left(2\hat\mu +\bar h_{\mu\nu} u^\mu u^\nu \right)~~.
\end{equation}
This determines the proportionality between the relativistic and non-relativistic temperatures. We define
\begin{equation}\label{eq:tildemu}
\mu =\hat\mu +\frac{1}{2}\bar h_{\mu\nu} u^\mu u^\nu\,.
\end{equation}
We will see below that $\mu$ is a chemical potential related to the mass conservation, which is a consequence of the null Killing vector and we note that its definition implies $\mu<0$. In the grand canonical ensemble for a system at rest, the partition function is of the form $\mathcal{Z}=\text{Tr}\, e^{-\beta H+\beta\mu N}$, where $H$ is the Hamiltonian and $N$ the conserved mass of the system.

\subsubsection{Null reduction of the hydrostatic partition function}
At the end of section \ref{sec:NullRed}, we discussed the role of the $U(1)$ transformation from the null reduction point of view, and we showed that such a transformation corresponds to a diffeomorphism generated by $\hat \xi^{\hat\mu}=-\sigma\delta^{\hat\mu}_u$. Applying this to our Killing vector $\hat k^{\hat\mu}$, we learn that under $\delta_{\hat\xi}\hat k^{\hat\mu}=\pounds_{\hat\xi}\hat k^{\hat\mu}$, the NC Killing vector $k^\mu$ is left inert and that $\hat k^u$ transforms as
\begin{equation}
\delta_\sigma \hat k^u = k^\mu \D_\mu\sigma\,.
\end{equation}
Since $\tau_\mu$ is also invariant it follows that $\beta$ also does not transform. Hence, using $\hat k^u=\beta\hat\mu$ and $k^\mu=\beta u^\mu$, we can write
\begin{equation}
\delta_\sigma \hat\mu= u^\mu \D_\mu\sigma\,.
\end{equation}
It then follows that $ \mu$ defined in equation \eqref{eq:tildemu} is $U(1)$ invariant, making $\mu$ together with $\beta$ the two parameters on which the lower dimensional pressure in the hydrostatic partition function should depend. 

In a $d+1$-dimensional theory, the hydrostatic partition function is given by
\begin{equation} \label{eq:hydroP}
S = \int \diff^{d+1}x~eP(T,{\mu})~~,
\end{equation}
where $P$ is the fluid pressure. Next, we vary $S$ keeping the Killing vector fixed, i.e. $\delta k^\mu = 0 = \delta \hat k^u$. The variation of the temperature is then given by
\begin{equation}
\delta T = \delta (\tau_\mu k^\mu)^{-1} = -(\tau_\nu k^\nu)^{-2} k^\mu\delta \tau_\mu = -T u^\mu\delta \tau_\mu~~,
\end{equation}
while the variation of the chemical potential reads
\begin{equation}
\delta \mu = \delta\hat\mu + \frac{1}{2}u^\mu u^\nu \delta\bar h_{\mu\nu} + \bar h_{\mu\nu} u^\nu \delta u^\mu = \hat\mu\frac{\delta T}{T} + \frac{1}{2}u^\mu u^\nu \delta \bar h_{\mu\nu} + \bar u^2\frac{\delta T}{T}~~.
\end{equation}
This allows us to compute
\begin{align}
\delta P =& 	\left(\frac{\D P}{\D T}\right)_{\mu}\delta T + 	\left(\frac{\D P}{\D {\mu}} \right)_T \delta \mu = s \delta T +n\delta\mu \\
=& -(s T+ n \mu + \frac{1}{2} n \bar u^2   )u^\mu\delta \tau_\mu + \frac{1}{2}n u^\mu u^\nu \delta \bar h_{\mu\nu}\,,
\end{align}
where $s$ is the entropy density and $n$ the mass density. Thus, combining our findings, we obtain
\begin{eqnarray}
\delta S & = & \int\diff^{d+1}x~e\left(\mathcal{T}^\mu \delta\tau_\mu + \frac{1}{2}\mathcal{T}^{\mu\nu}\delta \bar h_{\mu\nu}\right)\label{eq:varS}\\
&=& \int \diff^{d+1}x~e\left[\frac{1}{2}\left(P h^{\mu\nu} + nu^\mu u^\nu \right)\delta \bar h_{\mu\nu} - P\hat v^\mu \delta \tau_\mu - \left(s T + n \mu + \frac{1}{2} n\bar u^2 \right) u^\mu \delta \tau_\mu \right]~,\nn
\end{eqnarray}
leading us to identify the energy current and the Cauchy stress--mass tensor as
\begin{align}
\mathcal{T}^\mu =& -P \hat v^\mu - (s T + n \mu + \frac{1}{2}n\bar u^2)u^\mu = -P \hat v^\mu - (\mathcal{E} + P + \frac{1}{2}n\bar u^2 )u^\mu~~,\label{eq:energy}\\
\mathcal{T}^{\mu\nu} =& P h^{\mu\nu} + nu^\mu u^\nu~~,
\end{align}
where we defined $\mathcal{E}$, the internal energy, via the relation $\mathcal{E} + P=s T + n\mu$. This matches the results of \cite{Hartong:2016nyx}, where these equations were obtained by directly null reducing the expression for the relativistic energy--momentum tensor.

The relation between the relativistic and non-relativistic currents can be found from
\begin{equation}
\frac{1}{2}\sqrt{-\hat g}\hat T^{\hat\mu\hat\nu}\delta\hat g_{\hat\mu\hat\nu}=e\left(\mathcal{T}^\mu \delta\tau_\mu + \frac{1}{2}\mathcal{T}^{\mu\nu}\delta \bar h_{\mu\nu}\right)~~.
\end{equation}
Hence the energy current is given by $\mathcal{T}^\mu=\hat T^{u\mu}$.  For a perfect fluid, this is $\mathcal{T}^\mu=\left(\hat E+\hat P\right)\hat u^\mu\hat u^u-\hat P\hat v^\mu$. Comparing this with \eqref{eq:energy} implies that we have the identification $\hat P=P$, as well as
\begin{equation}
\mathcal{E} + P + \frac{1}{2}n\bar u^2=-\left(\hat E+\hat P\right)\hat u_u\hat u^u=\frac{1}{2}\left(\hat E+\hat P\right)+\frac{1}{2}n\bar u^2\,,
\end{equation}
where $\bar u^2=\bar h_{\mu\nu}u^\mu u^\nu$ and where we used \eqref{eq:defu}, \eqref{eq:uup} and \eqref{eq:udown}. Hence we conclude that, since $\hat P=P$, we have $\hat E=2\mathcal{E} +P$. Finally, we note that equation \eqref{eq:udown} can be obtained from comparing $\hat T^{\mu\nu}=\mathcal{T}^{\mu\nu}$. Replacing $P$ in \eqref{eq:hydroP} by $\chi$ and confining the fluid to a surface leads to \eqref{eq:freesurface} upon Wick rotation.

\section{Classes of Newton-Cartan geometries} \label{app:NCtypes}
As mentioned in section \ref{sec:abs}, while it is not necessary to work with torsion for relevant systems, it is nevertheless formally necessary to introduce it in order to obtain the correct variational calculus (see discussion around \eqref{eq:torsion2}). Thus it is instructive to briefly mention other types of Newton-Cartan geometry for which different conditions on $\tau_\mu$ are considered. In the most general version of NC geometry, “torsional Newton--Cartan geometry” (TNC geometry \cite{Christensen:2013lma,Christensen:2013rfa,Hartong:2014oma}), the clock 1-form is completely unconstrained. A more moderate version, referred to as “twistless torsional Newton--Cartan geometry” (TTNC geometry), requires that the clock 1-form be hypersurface-orthogonal (i.e. it satisfies the Frobenius integrability condition $\tau\wedge\diff\tau = 0$). We summarise these different notions in table \ref{table:NCGeoms} below. 
\begin{table}[H]
\centering
\caption{The three classes of Newton--Cartan geometries and their properties.}
\label{table:NCGeoms}
\begin{tabular}{|c|c|c|c|}
\hline
Geometry & Constraint on $\tau$ & Causality  & Torsion\\
\hline
TNC & None  & acausal & yes \\
TTNC & $\tau\wedge\diff\tau = 0$ & surfaces of absolute simultaneity & yes \\
NC & $\diff\tau = 0$ & absolute time  & no \\
\hline
\end{tabular}
\end{table}
In fact, these conditions are intimately linked with torsion. In particular if $\tau$ is closed ($\diff\tau=0$), there is no
torsion, but if $\tau$ is hypersurface-orthogonal ($\tau\wedge\diff\tau=0$) the \textit{twist} vanishes, $\omega^2=h^{\mu\rho}h^{\nu\sigma}\omega_{\mu\nu}\omega_{\rho\sigma}=0$, where the twist tensor is given by $\omega_{\mu\nu}=h^{\rho\sigma}h_{\sigma\mu}h^{\lambda\kappa}h_{\kappa\nu}\tau_{\rho\lambda}$. Finally, if the clock 1-form is completely unconstrained, so is the torsion.

When there is no constraint on $\tau_\mu$, it was shown in \cite{Geracie:2015dea} that the spacetime becomes acausal in the sense that given a point $P$ there exists a neighborhood of $P$ such that all points in the neighborhood are separated from $P$ by curves that are spacelike, i.e. their tangent vectors are orthogonal to $\tau_\mu$. When $\tau_\mu$ is hypersurface orthogonal, the spacetime admits a foliation in terms of constant time slices. At different points on such a hypersurface clocks may tick at a slower or faster rate as time evolves, although all observers on such a constant time slices agree that they are simultaneous with each other. When there is no torsion (and $\tau$ is exact) the rate at which time evolves is the same for all points on the constant time slices and we are dealing with absolute time. In this case the interval between two events $P$ and $Q$ connected by a curve $\gamma$ joining $P$ and $Q$, i.e. $\int_\gamma\tau$, is independent of the choice of $\gamma$.

\section{Connections on the submanifold}\label{App:ConnectionsAreImportant}

The purpose of this appendix is to find the relation between the NC connections of the ambient spacetime and the submanifold as described in section \ref{sec:CovDandExtK}.

Consider first the projection of the submanifold covariant derivative acting on a vector $V^\nu$, that is
\begin{align}
u^\mu_a u^b_\nu  \nabla_\mu V^\nu =& u^\mu_a u^b_\nu \left(\D_\mu V^\nu + {\Gamma}^\nu_{\mu\rho} V^\rho \right) = \D_a(u_\nu^b V^\nu) - V^\nu\D_a u^b_\nu + u^\mu_a u^b_\nu\Gamma ^\nu_{\mu\rho} V^\rho\nn \\
=& \D_a V^b - V^\sigma(u^\nu_c u_\sigma^c + n_I^\nu n_\sigma^I)\D_a u^b_\nu + u^\mu_a u^b_\nu\Gamma ^\nu_{\mu\rho}(u^\rho_c u_\sigma^c + n_I^\rho n_\sigma^I) V^\sigma\nn\\
=& {\D_a V^b +  \Gamma^b_{ a c} V^c - V^c u^\nu_c \D_a u^b_\nu } - V_I h^{bc}\tilde{K}_{ac}{^I}~~,
\end{align}
where we defined
\begin{equation}
\Gamma ^b_{c a} = u^\mu_a u^b_\nu u_c^\rho\Gamma^\nu_{\rho\mu}~~.
\end{equation}
Now, if the vector is a pushforward of a submanifold vector as in $V^\mu = u^\mu_a V^a$, the last term in the expression above vanishes, which leads us to define
\begin{equation}
{ \gamma ^b_{a c} = \Gamma ^b_{ac} - u^\mu_c\D_a u^b_\mu ~~.}\label{eq:gamma2}
\end{equation}
The connection on the submanifold is also given by \eqref{eq:indcon} which we can write using the ambient structures as
\begin{align}
-\hat{v}^c \D_a\tau_b =& -u^c_\mu u^\nu_a u^\rho_b \hat{v}^\mu \D_\nu \tau_\rho - { \hat{v}^c u^\nu_a \tau_\rho \D_\nu u^\rho_b}~~,\\
 h^{cd}\D_a \bar h_{b d} =& u^c_\mu u^d_\nu u^\rho_a u^\lambda_b u^\sigma_d \D_\rho \bar h_{\lambda\sigma} + h^{cd}\bar h_{b \lambda}\D_a u^\lambda_d + h^{c\lambda}\bar h_{\lambda\sigma}\D_a u^\sigma_b~~.
\end{align}
Substituting these back into \eqref{eq:indcon}, we find that
\begin{align}
{\gamma}^c_{ab} - \Gamma^c_{ab} =& - { \hat v^c \tau_\rho \D_a u^\rho_b} + \frac{1}{2}h^{cd}\bar h_{b \lambda}\D_a u^\lambda_d + \frac{1}{2}h^{c\lambda}\bar h_{\lambda\sigma}\D_a u^\sigma_b+\frac{1}{2}h^{cd}\bar h_{a \lambda}\D_b u^\lambda_d\nn \\
&+ \frac{1}{2}h^{c\lambda}\bar h_{\lambda\sigma}\D_b u^\sigma_a
-\frac{1}{2}h^{cd}\bar h_{a \lambda}\D_d u^\lambda_b - \frac{1}{2}h^{cd}\bar h_{\lambda b}\D_d u^\lambda_a\nn\\
=& u^c_\sigma\D_a u^\sigma_b~~,
\end{align}
obtaining the result \eqref{eq:gamma}.

\section{Gauss--Bonnet \textit{\&} $(2+1)$-dimensional membranes}\label{sec:GaussBonnet}
For a closed co-dimension one surface embedded in flat $(3+1)$-dimensional Newton--Cartan geometry, the Gauss-Codazzi equation \eqref{eq:GaussCodazzi} relates $K^2$ and $K\cdot K$ according to 
\be
K^2 - K\cdot K = { \mathcal{R}}~~,\label{eq:GaussCodazzi2}
\ee
where ${ \mathcal{R}}$ is the spatial Ricci scalar ${ \mathcal{R}}=h^{ab} {\mathcal{R}}_{ac b}{^c}$.
This is the Ricci scalar of a 2-dimensional spatial metric on constant time slices of $\Sigma$. This can be seen from the perspective of gauging the Bargmann algebra (see e.g. \cite{Andringa:2010it,Bergshoeff:2014uea,Festuccia:2016awg}) as we will briefly review.

In this section we will denote surface tangent space indices as $\bar a,\bar b,\ldots = 1,2$. It is well known that $(2+1)$-dimensional Newton--Cartan geometry arises as a gauging of $\mathfrak{barg}(2,1)$, which is generated by $(H,P_{\bar{a}},G_{\bar{a}},J_{\bar{a}\hspace{0.5pt}\bar{b}},N)$ with the following non-vanishing brackets
\begin{align}
&[H,G_{\bar a}] = P_{\bar a}~~,~~ [J_{\bar a\bar b},G_{\bar{c}}] = 2\delta_{\bar{c}[\bar{a}}G_{\bar{b}]}~~,~~ [J_{\bar{a}\bar{b}},P_{\bar{c}}] = 2\delta_{\bar{c}[\bar{a}}P_{\bar{b}]}~~,\nn\\
&[J_{\bar{a}\bar{b}},J_{\bar{c}\bar{d}}] = 4\delta_{[\bar{a}[\bar{d}}J_{\bar{c}]\bar{b}]}~~,~~ [P_{\bar{a}},G_{\bar{b}}] = N\delta_{\bar{a}\bar{b}}~~.\label{BargmannAlg}
\end{align}
The gauging procedure then proceeds as follows. We introduce a Lie algebra valued connection
\begin{equation}
\mathcal{A}_a = H\tau_a+P_{\bar{a}} e^{\bar{a}}_a+Nm_a+G_{\bar{a}}\omega_\mu{^{\bar{a}}}+\frac{1}{2}J_{\bar{a}\hspace{0.5pt}\bar{b}}\omega_{a}{^{\bar{a}\hspace{0.5pt}\bar{b}}}~~,
\end{equation}
with an associated curvature two-form $\mathcal{F}=\diff \mathcal{A}+\mathcal{A}\wedge\mathcal{A}$ whose Lie algebra expansion is given by
\begin{equation}
\mathcal{F}_{ab} = HR_{ab}(H)+P_{\bar{a}} \bar{\mathcal{R}}_{ab}{^{\bar{a}}}(P)+N\bar{\mathcal{R}}_{ab}(N)+G_{\bar{a}}\bar{\mathcal{R}}_{ab}{^{\bar{a}}}(G)+\frac{1}{2}J_{\bar{a}\bar{b}}\bar{\mathcal{R}}_{ab}{^{\bar{a}\bar{b}}}(J)~~.
\end{equation}

In  \cite{Hartong:2015zia} it is shown that the Riemann tensor is related to the curvatures appearing in the gauging procedure as follows,
\begin{equation}
{\mathcal{R}}_{abd}{^c}=e^c_{\bar{a}}\tau_d\bar{\mathcal{R}}_{ab}{^{\bar{a}}}(G) - e_{d {\bar{a}}}e^c_{\bar{b}}\bar{\mathcal{R}}_{ab}{^{\bar{a}\bar{b}}}(J)~~.
\end{equation}
The curvature of the spatial rotations $\bar{\mathcal{R}}_{ab}{^{\bar{a}\bar{b}}}(J)$ is the curvature 2-form of the constant time slices which for (twistless torsional) NC geometry is Riemannian.
In $(2+1)$-dimensional Newton--Cartan geometry, therefore, the spatial Ricci scalar $\mathcal{R}$ only depends on the curvature two-form $\bar{\mathcal{R}}_{ab}{^{\bar{a}\bar{b}}}(J)$ and we have the usual identities from 2-dimensional Riemannian geometry for the spatial projections of ${\mathcal{R}}_{abd}{^c}$. For example the vanishing of the 2-dimensional Einstein tensor would read 
\begin{equation}\label{eq:no2dEinstein}
h^{ac}h^{be}{\mathcal{R}}_{abc}{^d}-\frac{1}{2}\mathcal{R}h^{de}=0\,.
\end{equation}

In the case of torsionless NC geometry the (2+1)-dimensional integration measure $e$ is just the integration measure on the constant time slices (since the time direction has a trivial measure when we are dealing with absolute time). The Gauss--Bonnet theorem then tells us that 
\begin{equation}
\int_\Sigma \diff^3\sigma~e{\mathcal{R}}=4\pi \int d\sigma^0\chi(\Sigma_s)~~,
\end{equation}
where $\chi(\Sigma_s)$ is the Euler characteristic of the constant time slices $\Sigma_s$. Hence, the Gauss--Codazzi equation \eqref{eq:GaussCodazzi2} gives us a relation between the coefficients $a_2,a_3$ of \eqref{eq:model}, allowing us to set either $a_2$ or $a_3$  equal to zero (but only when both $a_2$ and $a_3$ are constant). In \eqref{eq:reducedmodel}, we have chosen to set $a_3$ to zero.

\addcontentsline{toc}{section}{\refname}

%

\end{document}